\documentclass[letterpaper]{ar-1col}
\usepackage{graphicx}
\usepackage{amsmath}
\usepackage{amssymb}
\usepackage{color}
\usepackage{ARAstroBib}
\usepackage{deluxetable}

\newcommand{\mcore}{M_{\rm core}}
\newcommand{\sigz}{\Sigma_{\rm solids}}
\newcommand{\sigg}{\Sigma_{\rm gas}}
\newcommand{\gcms}{{\rm g}{\rm cm}^{-2}}
\newcommand{\au}{{\rm AU}}
\begin{document}
 \input psfig.sty

\jname{Annu.\ Rev.\ Astron.\ Astrophys.}
\jyear{}
\jvol{}
\doi{10.1146/((please add article doi))}

 \title{Origins of Hot Jupiters}

\author{
 Rebekah I.\ Dawson$^1$ and
  John Asher Johnson$^2$
\affil{$^1$Department of Astronomy \& Astrophysics, Center for Exoplanets and Habitable Worlds,The Pennsylvania State University,
    University Park, PA 16802; email:~{\tt rdawson@psu.edu}}
\affil{$^2$ Department of Astronomy, Harvard University,
  Cambridge, MA 02138}}


\begin{abstract}
Hot Jupiters were the first exoplanets to be discovered around main sequence stars and astonished us with their close-in orbits. They are a prime example of how exoplanets have challenged our textbook, solar-system inspired story of how planetary systems form and evolve. More than twenty years after the discovery of the first hot Jupiter, there is no consensus on their predominant origin channel. Three classes of hot Jupiter creation hypotheses have been proposed: in situ formation, disk migration, and high eccentricity tidal migration. Although no origin channel alone satisfactorily explains all the evidence, two major origins channels together plausibly account for properties of hot Jupiters themselves and their connections to other exoplanet populations. 
\end{abstract}

\maketitle

\section{INTRODUCTION}
\label{sec:intro}

Prior to the discovery of the first exoplanets, our conception of planetary systems was generally informed by an observational sample of one: our own Solar System. In the origins story informed by this limited yet well-studied sample, planets formed from a cloud of gas and dust that collapsed into a disk. Beyond the snow line, where feeding zones are large and solid icy materials are abundant, rocky cores grew quickly and accreted massive gaseous atmospheres before the gas disk dissipated. In the inner disk, where feeding zones are small and ices absent, rocky cores were too small to accrete gas and had to wait until after the dissipation of the gas disk to grow by giant impacts. The planets maintained the orbits on which they formed: circular and coplanar with rocky planets inside and gas giants outside.

The discovery of the first gas giants outside of the Solar System immediately upended this Solar System-centric formation picture. The first exoplanet discovered around a main sequence star, 51\,Peg\,b  \citep{mayo95}, has an astonishingly close-in orbit.\footnote{Although hot Jupiters surprised the modern astronomical community, their existence, discovery via radial velocity, and propensity to transit were proposed decades ago by \citealt{struv52}.}  Instead of orbiting beyond several AU like our Solar System's gas giants in the region we expected giant planets to form, 51\,Peg\,b orbits 10 times closer to its star than Mercury to the Sun. 51\,Peg\,b belongs to the class of planets known as hot Jupiters, which we define here as gas giants with masses greater than or equal to 0.25 Jupiter masses (0.83 Saturn masses) and orbital periods shorter than 10 days (for comparison, Mercury's orbital period is 88 days). It was immediately clear from the discovery of 51 Peg b -- and from the other hot Jupiters that followed -- that theories of planet formation needed revision. In parallel, the discovery of the Kuiper belt has given us hints that even our own Solar System's history was more dramatic than we once assumed (e.g., \citealt{malh93,thom99}). 

The discovery of the first hot Jupiters not only sparked a revolution in planet formation theory but kick-started the field of exoplanet discovery and characterization. If the Solar System paradigm for gas giants was the rule throughout the Galaxy, the progress of the field of exoplanetary science would have been glacially slow.  Radial velocity surveys require the observation of at least one and ideally multiple orbits, which would take multiple decades for a Jupiter analog. In contrast, early planet hunters could observe multiple hot Jupiter orbits within a single week. Hot Jupiters also cause a larger reflex motion of the star than a Jupiter analog, allowing observers to characterize the planets with fewer measurements necessary to beat down the noise. Rapid discovery and confirmation of hot Jupiters led to the build-up of a large sample of planets over a decade. The first statistical studies of exoplanets, including the earliest measurements of the planet mass distribution with its power-law rise toward less massive planets \citep{marc00} and the planet-metallicity correlation \citep{gonz97}, were informed primarily by giant planets orbiting within 1 AU.

The rising discovery rate of exoplanets revealed that hot Jupiters, while playing a major role in the field of exoplanets, are relatively rare in the Galaxy. Only ~1\% of Sun-like stars host one, and the occurrence rate falls off around the most numerous stars, the M dwarfs \citep{john10}. However, this rare class of planets has played an out-sized role in our understanding of the internal structure,  atmospheric composition, and orbital architecture of giant planets outside the Solar System. These insights, like the sample of planets discovered from Doppler surveys, are a direct benefit of the unexpected orbital architecture of hot Jupiters. The probability of a transit scales with the inverse of a planet's semi-major axis, so a hot Jupiter at 0.05 AU is 100 times more likely to transit than a planet at 5 AU. What was statistically impractical for a true Jupiter analog became a statistical inevitability for the growing population of hot Jupiters discovered by radial velocity surveys. Photometric monitoring of these hot Jupiters led to the first discovery of a  transiting planet, hot Jupiter HD 209458 b \citep{char00,henr00}. When a planet transits, we can measure its size and combine the size and mass to derive a bulk density. We can use secondary eclipses and transmission spectroscopy to study its atmosphere, a step toward ultimately characterizing smaller, potentially habitable planets. We can also measure the sky-projected stellar obliquity -- the angle between the planet's orbital axis and the star's spin axis -- via the Rossiter-McLaughlin effect \citep{ross24,mcla24}.

The discovery of hot Jupiters paved the way for modern exoplanetary science by inspiring extensive theoretical work on several physical processes that were missing from our pre-exoplanet story of how planetary systems form and evolve. In this review we provide an overview of the various theories currently invoked to explain the origins of hot Jupiters (\S 2).  In \S 3, we summarize how properties of hot Jupiters themselves---including their eccentricities, host star obliquities, radii, semi-major axes, host star ages, and atmospheric compositions---square with these theories. In \S 4, we synthesize tests of origin theories that involve connecting hot Jupiters to other exoplanet populations. We conclude in \S 5 that there are two formation channels that together are most consistent with observations, and we outline future observational and theoretical studies that would clarify this picture.

\section{OVERVIEW OF HOT JUPITER ORIGIN THEORIES}

Here we provide a theoretical overview of the three main classes of hot Jupiter origin theory: in situ formation (\S\ref{subsec:form}), disk migration (\S\ref{subsec:disk}), and high eccentricity tidal migration (\S\ref{subsec:hem}).

\begin{figure}
\includegraphics[width=4.4in]{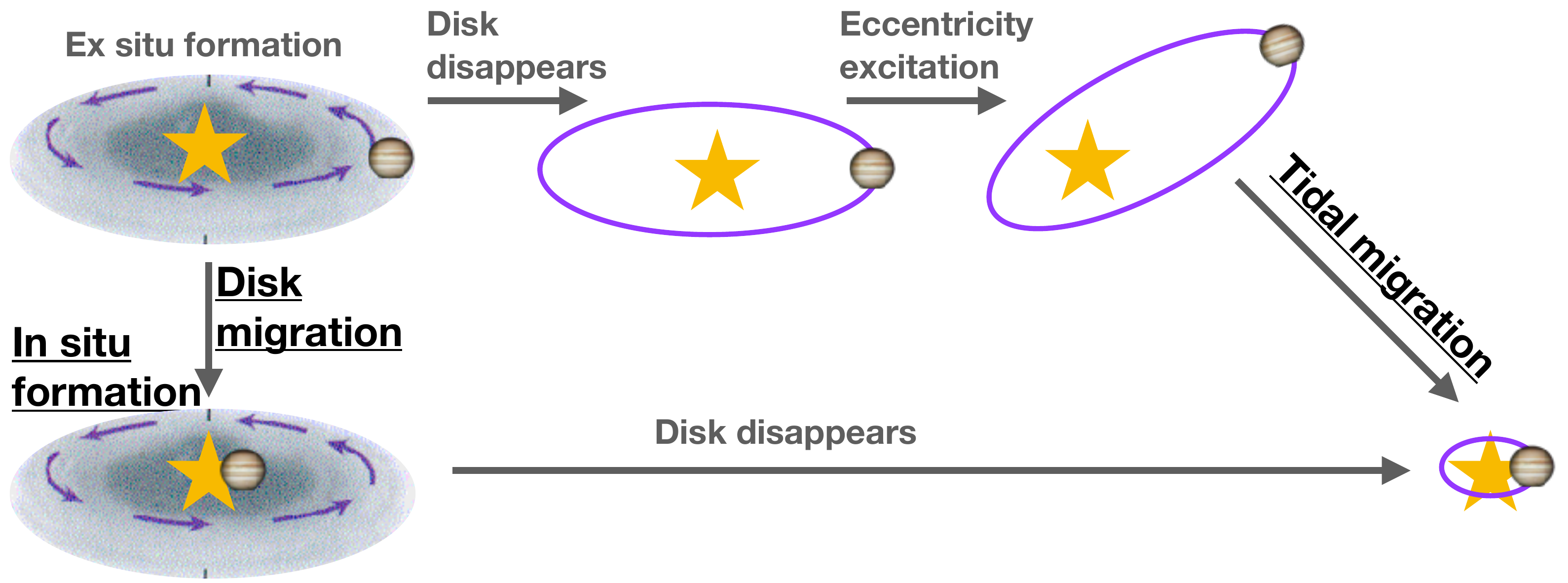}
\caption{Three origins hypotheses for hot Jupiters: in situ formation (\S\ref{subsec:form}), disk migration (\S\ref{subsec:disk}), and tidal migration (\S\ref{subsec:hem}). 
}
\label{fig:cartoon}
\vspace{-5mm}
\end{figure}

\subsection{In situ formation}
\label{subsec:form}
A major open question is whether hot Jupiters can form at their present day short orbital periods. In situ formation is feasible if one or both of the two mechanisms proposed for giant planet formation can operate close to the star: gravitational instability, in which part of the proto-planetary disk fragments into bound clumps (e.g., \citealt{boss97}; see \citealt{duri07} for a review), or core accretion, in which a rocky proto-planet core accretes many times its mass in gas from the proto-planetary disk (e.g., \citealt{peri74,poll96}; see \citealt{chab14} for a review). Until recently, it was widely believed that neither gravitational instability nor core accretion could operate at hot Jupiters' close in locations \citep{rafi05,rafi06} and hence hot Jupiters must have formed further from their stars and migrated to their present-day orbits (\S \ref{subsec:disk}--\ref{subsec:hem}). Here we review the feasibility of in situ formation of hot Jupiters by either mechanism. 

\subsubsection{Gravitational instability: not plausible}

A region of the proto-planetary disk will be susceptible to gravitational instability if the free-fall time due to self gravity is sufficiently rapid to overcome Keplerian sheer. The {\bf criterion for gravitational instability instability} is parameterized as \citep{toom64}
\begin{equation}
\label{eqn:q}
    Q = \frac{2 \sqrt{kT/\mu}}{ G P \sigg} = 130 \left(\frac{T}{1500 \rm K}\right)^{1/2} \left(\frac{2.3 m_H}{\mu}\right)^{1/2} \left(\frac{3 \rm~day}{P}\right)\left(\frac{2 \times 10^5 ~ \gcms}{\sigg}\right) \lesssim 1 ,
\end{equation}
\noindent where $T$ is the temperature, $\mu$ is the mean molecular weight, $m_H$ is the mass of a hydrogen atom, $k$ is the Boltzmann constant, $P$ is the orbital period, $G$ is the universal gravitational constant, and $\sigg$ is the gas surface density. $T\sim 1500$ K and $\sigg \sim 2 \times 10^5 ~ \gcms$ are plausible disk conditions at $P=$ 3 day.

It remains generally accepted that nebular conditions at hot Jupiters' present day locations could not meet $Q \lesssim 1$ during the disk stage. At short orbital periods (denominator of Eqn. \ref{eqn:q}), fast rotation supports the local gas against gravitational collapse . At high temperatures (numerator of Eqn. \ref{eqn:q}), thermal pressure supports the local gas against gravitational collapse. For a disk heated by starlight, temperature increases as orbital period decreases as $T \propto P^{-1/3}$, so temperatures are higher closer to the star. Although gas surface density (denominator of Eqn. \ref{eqn:q}) is higher close to the star, typical gas surface density profiles are not steep enough to compensate for hotter temperature and faster rotation close in the star. For example in the minimum mass solar nebula, the gas surface density is constructed to scale as $\sigg \propto P^{-1}$ (e.g., \citealt{haya81}). In observed disks -- at least at the wide separations we can observe -- the density scales even more weakly with $P$ (see \citealt{andr15} Fig. 12, Section 5.2.2, references therein). Only a density profile steeper than $\sigg \propto P^{-7/6}$ would compensate for the fast rotation and high thermal pressure close to the star.

More importantly, even if the gas were dense enough for disk fragmentation, at short orbital periods fragments would shear out before they could cool and contract. Following \citet{rafi05} with an optical depth of unity (most efficient possible cooling) and adiabatic index of 7/5, the criterion for fragments to cool before they are rotationally sheared is
\begin{equation}
\xi = \frac{5\pi k \sigg}{\sigma \mu P T^3} = 2200 \left(\frac{\sigg}{2 \times 10^5 ~ \gcms} \right)\left(\frac{3 \rm~day}{P}\right)  \left(\frac{1500 \rm K}{T}\right)^3 \lesssim 1,
\end{equation}
\noindent where $\sigma$ is the Stefan-Boltzmann constant.

To meet both criteria simultaneously (gravitational instability, $Q \lesssim 1$, and cooling before shearing, $ \xi \lesssim 1$) requires implausibly high temperatures and gas surface densities:
\begin{eqnarray}
\frac{T}{1500\rm K} &\sim& 150 \left(\frac{3 \rm~day}{P}\right)^{4/5}   \left(\frac{2.3 m_H}{\mu}\right)^{1/5}, \nonumber \\
\frac{\sigg}{2 \times 10^5 ~ \gcms} &\sim& 1600 \left(\frac{3 \rm~day}{P}\right)^{7/5}   \left(\frac{2.3 m_H}{\mu}\right)^{3/5}. \nonumber
\end{eqnarray}
\noindent As argued by \citet{rafi05}, such high temperatures would unbind the gas from the star. Therefore we do not expect the conditions for gravitational instability close to the star.

\subsubsection{Core accretion: requires huge build up of solids}
\label{subsec:core}
The challenge for forming hot Jupiters close to their stars via core accretion lies not in the accretion but in the core. In the core accretion hypothesis (e.g., \citealt{poll96}), giant planets form when a large, solid $\sim 10$ Earth mass core\footnote{Until recently, the existence of these cores had not been proven, even for Jupiter and Saturn (e.g., \citealt{guil05}). Using Juno gravity field measurements, \citealt{wahl17} found a core mass of 6--25 M$_\oplus$ for Jupiter. Hot Jupiter HAT-P-13b has a core mass of $\sim 11 M_\oplus$, \citealt{buhl16}. } accretes gas from the nebula. Hot nebular gas can be accreted by a massive core almost as easily as cool gas, because the accretion efficiency depends primarily on the conditions at the radiative-convective boundary deep in the accreting proto-planet's atmosphere, not on the conditions in the nebula (e.g., \citealt{stev82,lee14,piso15}). Throughout the proto-planetary disk, a giant planet core, if formed, will accrete gas. The challenge lies in growing a sufficiently massive core before the gas disk dissipates.

In gaseous proto-planetary disks, embryos grow by coagulation from planetesimals in their reservoir of planet material. To grow into a core adequately large for gas accretion ($\sim 10 M_\oplus$; \citealt{rafi06,lee14,piso15b}), the disk must satisfy two requirements (e.g., \citealt{poll96}):

\begin{enumerate}
\item The timescale to grow the core is shorter than the gas disk lifetime, so that the gas accretion stage can take place. Typical gas disk lifetimes are a few Myr \citep{fede10,bare16}.
\item The amount of mass in the feeding zone where the core grows is sufficiently large to build up a $\sim 10 M_\oplus$ core.
\end{enumerate}
\begin{figure}
\includegraphics[width=4in]{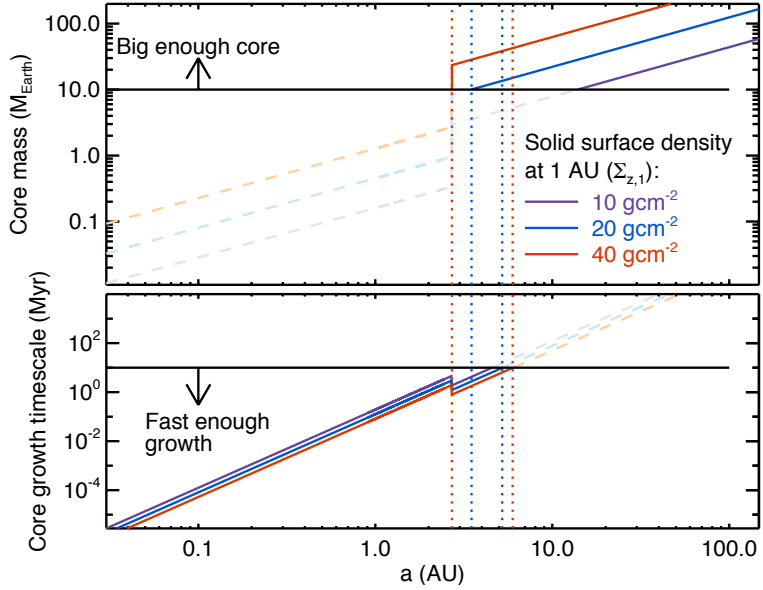}
\caption{Intermediate semi-major axes facilitate conditions for giant planet formation in high solid surface density disks. Top: Maximum core mass (set by the reservoir of planet formation material, not disk lifetime; Eqn. \ref{eqn:mcore}) vs. semi-major axis. Bottom: core growth timescale vs. semi-major axis (Eqn. \ref{eqn:tgrow}). The solid surface density profile is $\sigz = \Sigma_{z,1} \left(\frac{a}{\au}\right)^{-3/2} $ for three normalizations $\Sigma_{z,1} = 10, 20, 40 \gcms$; the normalization is set to increase by a factor of 4 at the ice line, 2.7 AU. The factor of 4 and ice distance of 2.7 AU are still under investigation (e.g., \citealt{leca06}) and taken as illustrative values. Under these assumptions, at small semi-major axes (i.e., where hot Jupiters are observed today), the maximum core mass is too small to undergo runaway accretion, whereas at large semi-major axes, the core cannot grow massive enough before the gas disk dissipates. When the disk solid surface density is sufficiently high (red, blue), both conditions can be met at intermediate semi-major axes (within the vertical dotted lines), but not if the solid surface density is too low (purple). Values are illustrative to demonstrate the tension between the core mass criterion and the core timescale criterion. The tension can potentially be resolved by transporting solids to the inner disk, increasing the effective solid surface density and hence the core mass, and by speeding up the core growth timescale at wider separations through the mechanism of pebble accretion (see \citealt{joha17} for a review).
}
\label{fig:miso}
\vspace{-3mm}
\end{figure}

Figure \ref{fig:miso} displays these competing requirements. The first requirement is easier to meet close to the star due to the short orbital timescales. We can estimate the coagulation timescale of a 10 Earth mass core ($M_{\rm core} = 10 M\oplus$) as \citep{gold04,chia13}
\begin{equation}
\label{eqn:tgrow}
t_{\rm coag} \sim \frac{\rho_{\rm core}^{2/3} M_{\rm core}^{1/3}}{\sigz \mathcal{F} } P =  10^5 {\rm yr} \left(\frac{\rho_{\rm core}}{\rm 8 g cm^{-3}}\right)^{2/3}\left(\frac{M_{\rm core}}{\rm 10 M_\oplus}\right)^{1/3} \left(\frac{\sigz}{10^{3} \rm \gcms }\right)\left(\frac{P}{3 \rm day}\right)\mathcal{F}^{-1}
\end{equation}
where $\mathcal{F} > 1$ is the gravitational focusing factor, $\sigz$ is the solid surface density, and $\rho_{\rm core}$ is the bulk density of the core. Gravitational focusing and gas drag on planetesimals (e.g., \citealt{rafi04}; \citealt{cham16} and references therein) can speed up this process further. In Fig. \ref{fig:miso}, we use 10 Myr as generously long timescale for the dissipation of the gas disk.

Unfortunately for forming giant planets close to their stars, short orbital periods also translate to tiny feeding zones and hence tiny core masses. Forming sufficiently massive cores is more difficult close to the star, because 
\begin{equation}
\label{eqn:mcorepre}
\mcore = 2\pi \sigz a \Delta R_H .
\end{equation}
\noindent where $a$ is the semi-major axis and the dimensionless quantity $\Delta$ is the width of the feeding zone in units  Hill radii, $R_{\rm H}$, given by
\begin{equation} 
\label{eqn:rh}
R_{\rm H} = a \left( \frac{ \mcore}{3 M_\star} \right)^{1/3},
\end{equation}
 where $M_\star$ is the host star mass. Eqn. \ref{eqn:mcorepre} and \ref{eqn:rh} constitute a joint set of equations for $\mcore$ and $R_{\rm H}$. By solving for $\mcore$ and substituting $P$ for $a$ using Kepler's law, we obtain
\begin{eqnarray}
\mcore &=& G \sigz^{3/2} \left(\frac{M_\star}{6\pi}\right)^{1/2} \Delta^{3/2} P^2 \nonumber\\&=& 0.005 M_\oplus \left(\frac{\sigz}{10^3 \gcms} \right)^{3/2} \left( \frac{M_\star}{M_\odot} \right)^{1/2} \left( \frac{\Delta}{7}\right)^{3/2} \left( \frac{P}{3 \rm day}\right)^2.
\label{eqn:mcore}
\end{eqnarray}
\noindent where $\Delta = 7$ is typical (e.g., \citealt{gree90}). The core mass is tiny, about half a Moon mass. To grow a large core of $\sim 10 M_\oplus$ would require a factor of 160 enhancement in $\sigz$ or $\Delta$. Although the core can grow quickly near the star, its growth \emph{appears} to stall at a low mass. (See below for a discussion of how radial drift can greatly increase the effective $\sigz$ by replenishing the planet formation reservoir.)

Recent studies have revisited the plausibility of forming hot Jupiters in situ (e.g., \citealt{lee14,lee16,baty15,bole16}),  motivated by the abundance of super-Earths, planets with masses between that of Earth and Neptune, at short orbital periods \citep{howa10,mayo11}. Many super-Earths are $\sim > 10 M_\oplus$, large enough to be the cores of giant planets. However, there are several important caveats to this connection. First, super-Earths have a different orbital period distribution than giant planets, with their occurrence rate drops steeply with orbital period within 10 days (see \citealt{lee17} and references therein), as we will discuss further in \S\ref{subsec:small}. The relative occurrence rates can only be reconciled if it much easier for cores grow in gas giant planets very close to the star than further from the star, which is not what we expect from planet formation theory (\S \ref{subsec:form}). Second, in situ formation tends to produce multiple nearby planets (e.g., \citealt{hans13,beck15}), as we will discuss further in \S\ref{subsec:comp}. Finally, increasing $\Delta$ beyond 7 requires mergers of multiple cores via giant impacts. However, damping from the gas disk prevents orbits from crossing until after the gas disk dissipates. \citet{lee16} highlight this argument as the strongest against forming hot Jupiters in situ. Furthermore, $\Delta$ has a hard upper limit set by the escape velocity (e.g., \citealt{schl14}), which limits the self-stirring growth of cores' eccentricities for orbit crossing. This results in a limit of 
\begin{equation}
\Delta_{\rm max} \lesssim \frac{2v_{\rm esc}}{R_{\rm H} 2 \pi P^{-1}} \simeq 13\left(\frac{P}{3 \rm day}\right)^{1/3} \left(\frac{\rho}{{\rm 8 g\, cm}^{-3}}\right)^{1/6}.
\end{equation}
Therefore, given that we cannot make $\Delta$ arbitrarily large, forming a $~\sim 10 M_\oplus$ close to the star can only be achieved by greatly enhancing the solid surface density beyond the minimum mass solar nebula (e.g., \citealt{chia13}). Enhancing the entire disk mass (gas plus solids) by two orders of magnitude would destabilize the disk (e.g., \citealt{schl14}).

However, the effective $\sigz$ can be greatly increased by radial drift of solids. Highly efficient transport and pile-up of solids in the inner disk could perhaps deliver 10 $M_\oplus$ of solids to a tiny feeding zone. The mechanism of pebble accretion, in which cores grow by accreting millimeter-to-centimeter-sized pebbles that drift from the outer disk, allows cores to grow more quickly (see \citealt{joha17} for a review). However, core growth stales when the core carves a gap, staunching the flow of pebbles, which build up at a pressure bump. For a variety of assumptions about the disk properties, the limiting core mass is $<1 M_\oplus$ at typical hot Jupiter separations. Therefore, the material for hot Jupiter cores cannot come solely in the form of pebbles. A better understanding of how solids are transported through the disk -- as dust, pebbles, planetesimals, embryos, or proto-cores -- may  resolve the open question of whether hot Jupiters can form in situ by core accretion. 

\subsection{Gas disk migration}
\label{subsec:disk}

In gas disk migration, torques from the gaseous proto-planetary disk can shrink a giant planet's semi-major axis from several AU to hundredths of an AU (e.g., \citealt{gold80,lin86,lin96,ida08}; see \citealt{baru14} for a comprehensive review). The planet exchanges angular momentum with the disk by perturbing nearby gas onto horseshoe orbits (via corotation torques) and deflecting more distant gas (via Lindblad torques). The net Lindblad torque tends to be inward  (\citealt{ward97}; see also \citealt{armi13} for a pedagogical overview of migration torques). The strength and sign of corotation torques depend on the disk's turbulent viscosity, opacity, and radial entropy profile (e.g., \citealt{paar06,duff15}); for disk conditions, corotation torques can drive a net outward migration. As the planet deflects gas, it can clear a gap; the resulting reduction in its migration speed depends on the disk viscosity and scale height. It was once believed that a sufficiently massive planet could staunch the flow of gas across the gap and that the planet would move inward on the timescale of disk's viscous accretion onto the star (``type II migration''). Staunching gas flow would have guaranteed inward migration at or above the disk's viscous accretion rate.  However, recent simulations have shown that a significant amount of material passes through the gap on horseshoe orbits (e.g., \citealt{duff14,durm15}), and the type II migration rate can be slower than viscous rate. The contribution of gas disk migration to the origin of hot Jupiters remains ambiguous because the magnitude and sign of the migration rate are highly sensitive to disk conditions.

When migration is faster than the disk lifetime, a giant planet risks migrating past the hot Jupiter region and being tidally disrupted or engulfed by the star. When proposing disk migration for the origin of the first discovered hot Jupiter 51 Peg b \citep{mayo95}, \citet{lin96} invoked two mechanisms for halting 51 Peg b's migration: angular momentum transfer from the star and a ``magnetocavity" in the inner disk. In the former mechanism, an inward migrating hot Jupiter can sustain its close-in orbit and avoid infall during the disk lifetime by extracting angular momentum from its star via tides \citep{tril98}. The planet may lose some mass through Roche lobe overflow but remains giant. This tidal mass loss can contribute to reversing a hot Jupiter's orbital decay (e.g., \citealt{vals15}). In the magnetocavity mechanism, the stellar magnetic field creates a cavity in the innermost disk. Planet migration driven by the disk can shut off when the hot Jupiter reaches the 2:1 resonance with the inner disk edge (e.g., \citealt{rice08,chan10}). Another hypothesis is that hot Jupiters halt at the dust sublimation radius \citep{kuch02}. However, \citet{eisn05} find that dust sublimation radii in observed disks are too large to be consistent with hot Jupiters' short orbital periods. In \S \ref{subsec:semi}, we will review the consistency the semi-major axis distribution with hot Jupiter origins hypotheses. If hot Jupiters arrive in the inner disk via migration, the magnetospheric cavity, mass loss, and tidal interactions with the star likely all play a role in setting its final location (e.g., \citealt{chan10}).

Disk migration -- at least in the single planet case -- changes a planet's semi-major axis while keeping its eccentricity low, as depicted in Fig. \ref{fig:evsa}. Planet-disk interactions can excite eccentricities $e$ under some disk conditions \citep{gold03,tsan14a,duff15}. However, \citet{duff15} find the excitation is limited to small values corresponding to epicyclic velocities approximately equal to the sound speed within the disk. Above this {\bf maximum disk eccentricity excitation value}, the planet tends to undergo collisions with the gap edge that damp its eccentricity:
\begin{eqnarray}
\label{eqn:cse}
   e_{\rm disk} \lesssim \frac{\sqrt{kT/\mu}}{2\pi a/P} &= 0.015 \left(\frac{T}{1500 \rm K}\right)^{1/2} \left(\frac{2.3 m_H}{\mu}\right)^{1/2} \left(\frac{P}{3 \rm day}\right)^{1/3}\nonumber \\&= 0.05 \left(\frac{T}{400 \rm K}\right)^{1/2} \left(\frac{2.3 m_H}{\mu}\right)^{1/2} \left(\frac{P}{600 \rm day}\right)^{1/3}.
\end{eqnarray}

A migrating giant planet may capture planets into mean motion orbital resonances (e.g., \citealt{malh93,lee02,raym06}). We  summarize what the properties of hot Jupiters' companions reveal about their origins in \S 4.3.

\subsection{High-eccentricity tidal migration}
\label{subsec:hem}
High-eccentricity tidal migration is another proposed mechanism to move a giant planet from several AU to several hundredths of an AU. Turning a cold Jupiter into a hot Jupiter requires reducing its orbital angular momentum by a factor of 10 and its orbital energy by a factor of 100. Unlike disk migration (\S \ref{subsec:disk}), in which the gravitational back reaction from the disk changes the planet's energy and angular momentum simultaneously, high eccentricity migration can often be approximated as a two step process: reducing the planet's orbital angular momentum and then reducing its energy. (See \S \ref{subsec:deco} for exceptions.) During the first step, a perturber extracts orbital angular momentum from the Jupiter by perturbing the Jupiter onto a highly elliptical orbit. During the second step, the Jupiter tidally dissipates its orbital energy through interactions with the central star. At periapse the Jupiter undergoes close passages to its host star, which raise tides on the Jupiter. The Jupiter dissipates energy as it stretches, changing shape to conform to the rapidly changing tidal potential. During this tidal dissipation step, the hot Jupiter's angular momentum is roughly conserved as it circularizes to a final, close-in semi-major axis,
\begin{equation}
    a_{\rm final} = a (1-e^2).
\end{equation}
\noindent For example, if the Jupiter begins the tidal dissipation stage at $a=4$ AU (1 AU) and circularizes to $a=0.04$ AU, the perturber must have raised the Jupiter's eccentricity to $e=0.995$ (0.98), corresponding to a periapse of 0.02 AU. Although these eccentricities are quite high, they are plausible for eccentricity excitation mechanisms (\S\ref{subsec:excite}) and in line highly elliptical planets observed (e.g., HD 80606b has e=0.93, \citealt{naef01}; Fig. \ref{fig:evsa}).

\subsubsection{Eccentricity excitation: reducing the Jupiter's angular momentum}
\label{subsec:excite}

Several mechanisms have been proposed as sinks for Jupiter's orbital angular momentum.

\begin{figure}
\includegraphics[width=4.9in]{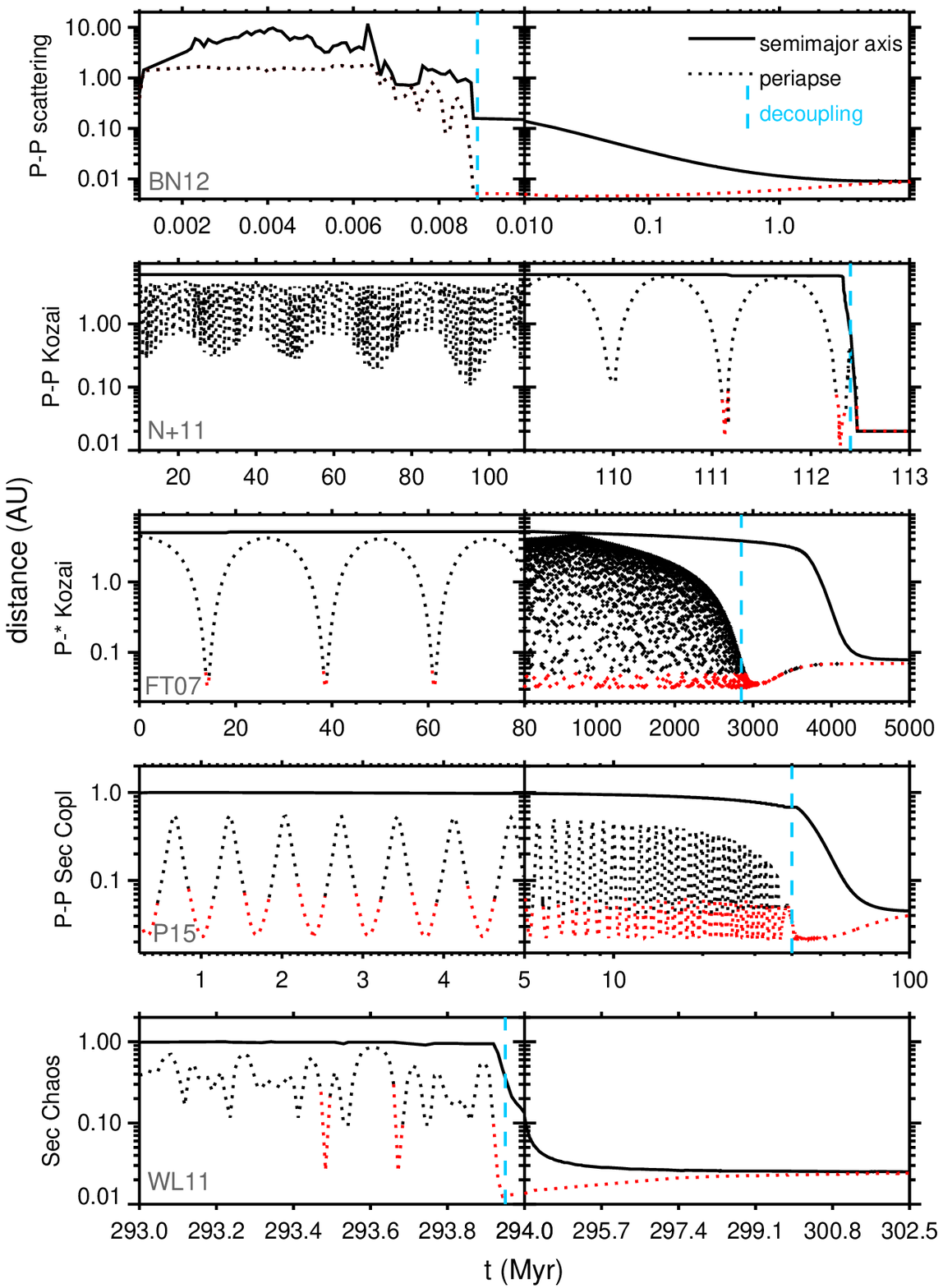}
\vspace{-.3in}
\caption{Dynamical histories leading to hot Jupiter: planet-planet scattering \citep{beau12}, planet-planet Kozai \citep{naoz11}, planet-star Kozai \citep{fabr07}, planet-planet coplanar secular \citep{petr15}, and secular chaos \citep{wu11}. The planet loses angular momentum/is perturbed to a small periapse (dotted line), i.e., high eccentricity. When the planet gets close enough to the star (red), it tidally circularizes, i.e., loses orbital energy, and its semi-major axis (solid line) shrinks. Eventually the planet decouples from its perturber (blue vertical dashed line). Simulation data was extracted from each cited article. See text for a discussion of timescales.
}
\label{fig:dynamics}
\end{figure}

{\bf Planet-planet scattering: } Planet-planet scattering converts Keplerian sheer (i.e., differences in angular velocity between planets with different semi-major axes) into angular momentum deficit, triggering high eccentricity migration  (e.g., \citealt{rasi96,weid96,ford06,chat08}). An example of planet-planet scattering is shown in the top panel of Fig. \ref{fig:dynamics}. Planet-planet scattering can take place in systems that form tightly packed (e.g., \citealt{juri08}) or when high eccentricities are generated by stellar fly-bys (e.g., \citealt{shar16}). Although planet-planet scattering also alters planets' semi-major axes, it is highly unlikely that a hot Jupiter can reach its present day short orbital period solely by scattering. Since the system's total orbital energy is conserved, a Jupiter would need to eject 100 planets of its own mass to reduce its semi-major axis by 100. (The hot Jupiter can reduce its semi-major axis through tidal dissipation in the next step.) In planet-planet scattering, the planet's eccentricity typically grows over a series of multiple close encounters. {\bf Eccentricity growth via scattering is limited} to an epicyclic velocity corresponding to the escape velocity from the surface of the planet (e.g., \citealt{gold04,ida13,petr14}):
\begin{eqnarray}
\label{eqn:vesce}
   e_{\rm scatter} \lesssim \frac{\sqrt{2GM_p/R_p}}{2\pi a/P} 
   &= \left(\frac{M_p}{M_{\rm Jup}}\right)^{1/2}\left(\frac{R_{\rm Jup}}{R_p}\right)^{1/2} \left(\frac{P}{50 \rm day}\right)^{1/3}\nonumber \\
   &= 0.2 \left(\frac{M_p}{0.5 M_{\rm Jup}}\right)^{1/2}\left(\frac{2 R_{\rm Jup}}{R_p}\right)^{1/2} \left(\frac{P}{3 \rm day}\right)^{1/3}
\end{eqnarray}
\noindent Once the eccentricity exceeds $e_{\rm scatter}$, the cross section for collisions exceeds the cross section for scattering and planets merge rather than scatter during close encounters. At $\sim$AU separations, the $e_{\rm scatter}$ exceeds 1 and Jupiters can be scattered onto highly elliptical orbits.

{\bf Secular interactions: } Secular interactions are slow exchanges of angular momentum between widely separated planets. Secular interactions take place over thousands or even millions of years, depending on the separation and mass of the planets involved. Through secular interactions, a Jupiter can deposit its angular momentum into other planets or stars in the system on a timescale of many orbits. Planets can swap angular momentum periodically (e.g., \citealt{petr15}) or chaotically (e.g., \citealt{wu11,hame16}), driving the Jupiter's eccentricity to large values. The latter scenario requires three or more planets and is known as secular chaos. (Chaos can occur in hierarchical two planet systems as well, but requires large eccentricities and inclinations; see \citealt{wu11} for a discussion.) Kozai-Lidov cycles are a type of periodic angular momentum exchange that also trade off mutual inclination and eccentricity \citep{koza62,lido62,naoz16}. Kozai-Lidov cycles can also occur in an initially coplanar system when the outer body is on a highly elliptical orbit \citep{li14a}.  Kozai-Lidov cycles driven by a star (e.g., \citealt{wu03,fabr07,katz11,naoz12,petr15a}) or planet (e.g., \citealt{naoz11}; see \citealt{tesy13} for distinctions between planet-planet and planet-star Kozai) have been widely hypothesized to trigger high eccentricity tidal migration of hot Jupiters. In another scenario, multiple secular frequencies within a system coincide, creating a secular resonance that elevates planets' eccentricities (e.g., \citealt{mint11,xu16}). Examples of several types of secular excitation interactions are shown in Fig. \ref{fig:dynamics}.

Secular interactions conserve the system's total orbital angular momentum. The proto-hot Jupiter loses its angular momentum by taking a share of the system's angular momentum deficit, the difference between the system's actual angular momentum and the angular momentum it would have if all the bodies had circular, coplanar orbits. The angular momentum deficit could have originally been created from Keplerian sheer (i.e., via scattering as discussed above). In the case of Kozai-Lidov cycles generated by a stellar perturber, a widely separated binary may naturally form in a different plane than the planet-primary, resulting in a mutual inclination that can drive Kozai-Lidov. 

{\bf Timescales for exciting eccentricity:} In planet-planet scattering, eccentricities are excited on a synodic timescale at planets' conjunctions. Each encounter is a random kick, and the eccentricity grows as a random walk. In Figure \ref{fig:dynamics}, first row, the periapse shrinks to a value sufficient for tidal circularization on a timescale of thousands of years.

Secular excitation of eccentricities occur on secular timescales of thousands of orbits. The timescale is primarily set by the perturber's mass and orbit. Rows 2--4 of Fig. \ref{fig:dynamics} feature eccentricity excitation by secular cycles. In row 3, a 1 $M_\odot$ Jupiter at 1000 AU causes $\sim 20$ Myr Kozai-Lidov eccentricity oscillations of a mutually inclined coplanar Jupiter at 5 AU \citep{fabr07}.
In row 4, a coplanar, 3.3 $M_{\rm Jup}$ Jupiter at 8 AU causes $\sim 1$ Myr eccentricity oscillations of a coplanar Jupiter at 1 AU \citep{petr15}. In row 2, a 3 $M_{\rm Jup}$ Jupiter at 61 AU causes eccentricity oscillations of a mutually inclined Jupiter at 6 AU on two timescales: a shorter Kozai-Lidov timescale of $\sim 1$ Myr and longer envelope of $\sim 20$ Myr caused by the perturber's eccentricity \citep{naoz11}. See \citet{li14a,anto15} for Kozai-Lidov eccentricity excitation timescales in the regime where the perturber is nearby and/or eccentric. In secular chaos, the proto-hot Jupiter's eccentricity diffusively randomly walks over many secular timescales. In the example shown in row 5 of Fig. \ref{fig:dynamics}, the eccentricity reaches a value large enough for tidal circularization after 300 Myr \citep{wu11}. Timescales can range from millions of years to many billions of years (e.g., the Solar System, \citealt{lask08}).

\subsubsection{Tidal dissipation: reducing the hot Jupiter's orbital energy}

Once the Jupiter's orbit is sufficiently elliptical, tidal dissipation in the planet shrinks and circularizes its orbit. Tidal dissipation in the planet decreases the planet's orbital energy but keeps its angular momentum constant. Once the hot Jupiter is gravitationally decoupled from the perturber(s) that originally removed the hot Jupiter's angular momentum, its orbit evolves according to 
\begin{equation}
\label{eqn:afinal}
    a_{\rm final} = a(t) [1-e(t)^2].
\end{equation}
\noindent where $a(t)$ and $e(t)$ are the time evolution of the semi-major axis and eccentricity respectively. The link between $a$ and $e$ lets us rewind a hot Jupiter's tidal evolution history. For example, a hot Jupiter observed today on a low eccentricity orbit at 0.04 AU that underwent tidal migration had $e=0.9$ at $a=0.2$ AU and $e=0.99$ at 2 AU. Tidal evolution tracks of constant angular momentum are included in Fig. \ref{fig:evsa}.

Regardless of the details of the tidal physics, the tidal evolution timescale has a strong dependence on $a_{\rm final}$. For example, in the constant tidal time lag model (e.g., \citealt{eggl98})
\begin{equation}
\label{eqn:adot}
a/\dot{a} \propto a_{\rm final}^8.
\end{equation} \noindent When the initial orbit at the beginning of tidal circularization is highly elliptical, the initial periapse $a(t=0)[1-e(t=0)] \approx \frac{1}{2} a_{\rm final}$ because tracks of constant angular momentum are defined by $a_{\rm final} = a (1-e^2)$ and $a (1-e^2) \approx 2 a (1-e)$ for $e \rightarrow 1$. The tidal evolution timescale is highly sensitive to the initial periapse and spans many orders of magnitude. A planet that takes 1 Myr to circularize to $a_{\rm final}=0.03$ AU (initial periapse 0.015 AU) would take longer than the age of the universe to circularize to $a_{\rm final}=0.1$ AU (initial periapse 0.05 AU). Figure \ref{fig:dynamics} features tidal circularization on a range of timescales from $\sim 0.1$ Myr (row 2) to Gyrs (row 3).

It is important to remember that high eccentricity tidal migration is caused by tides raised on the planet by the star, not tides raised on the star by the planet. Tides raised on the star by the planet are typically much less efficient and much more sensitive to $a$ than to $a_{\rm final}$. If and when a hot Jupiter arrives at a small $a_{\rm final}$, its tides raised on the star may drive orbital evolution on a longer timescale (e.g., \citealt{vals15}). If the planet arrives via tidal dissipation in the planet at an orbital period longer (shorter) than the star's rotational period, its orbit will expand (shrink) during this final stage driven by tidal dissipation in the star.

\subsubsection{Decoupling from the perturber}
\label{subsec:deco}

The hot Jupiter's tidal interactions with the central star eventually decouple it from the gravitational perturbations that originally raised its eccentricity. In the case of planet-planet scattering, the planet decouples when its orbit shrinks sufficiently to avoid subsequent close encounters with its scatterer (Fig. \ref{fig:dynamics}, top panel). In the case of secular interactions, the planet decouples as tides shrink its semi-major axis and precession from general relativity or tides exceeds the precession caused by the secular perturber. Figure \ref{fig:dynamics} features a range of decoupling timescales. In the top panel, we see a quick decoupling triggered by a kick and subsequent tidal circularization. In panel 3, we see a gradual decoupling as the semi-major axis shrinks and eccentricity oscillations are gradually quenched.

Some binary companions are too weak to even compete with general relativity in the first place and high eccentricity migration is never triggered, as we will discuss further in \S 4.3. In order for a perturber companion to excite the eccentricity of the proto-hot Jupiter, the precession caused by the perturber must dominate over the precession caused by the central star, including due to general relativity , the tidal distortion of the planet, and the oblateness of the star (e.g., \citealt{fabr10}, Eqn. 15--17). \citet{dong14} derived {\bf quenching criterion} (their Eqn. 4). In the limiting case of initially circular orbits for the proto-hot Jupiter and perturber and a proto-hot Jupiter with an orbital period much larger than a hot Jupiter's, this criterion is approximately 
\begin{eqnarray}
\label{eqn:grkoz}
\left(\frac{m_{\rm perturber}}{0.1 M_\odot}\right) \left(\frac{800 \rm yr}{P_{\rm perturber}}\right)^2 > \left(\frac{M_\star}{M_\odot}\right)^{5/3} \left(\frac{1 \rm yr}{P}\right)^{7/3} \left(\frac{3 \rm day}{P_{\rm HJ}}\right)^{1/3}
\end{eqnarray}

Figure \ref{fig:dynamics} depicts examples of decoupling in cases of planet-planet scattering and secular interactions. In some cases  -- particularly when the perturber is a massive, nearby planet  \citep{dong14} -- a proto hot Jupiter only decouples at the end of its tidal evolution and can undergo eccentricity oscillations throughout its journey (e.g., Fig. \ref{fig:dynamics}, row 3). In that case, Eqn. \ref{eqn:afinal} does not strictly apply: the planet's angular momentum is not constant but changing due to continued interactions with the perturber. However, the angular momentum averaged over an eccentricity cycle is roughly conserved throughout tidal evolution (e.g., as argued by \citealt{socr12}). 

\section{TESTING HOT JUPITER ORIGIN THEORIES USING PROPERTIES AND CORRELATIONS}
\label{sec:prop}

The observed properties of hot Jupiters and their host stars---and correlations among those properties---can be used to test theories of the origins of hot Jupiters. In this section we summarize how hot Jupiter properties square with their origin via in situ formation (\S\ref{subsec:form}), disk migration (\S\ref{subsec:disk}), and different varieties of high eccentricity tidal migration (\S \ref{subsec:hem}). We divide the properties into intrinsic properties (this section) and connections of hot Jupiters to other populations (\S \ref{sec:conn}).

\subsection{Eccentricities of hot Jupiters}   
\label{subsec:ehj}

The present-day eccentricities of giant planets’ are relics of their origins. Each orbital evolution process we described in \S 2---including planet-disk interactions, planet-planet scattering, secular interactions, and high eccentricity tidal migration---plays out differently in the parameter space of eccentricity and semi-major axis. We illustrate how different processes populate the (semi-major axis, eccentricity) space and compare to the observed population in Fig. \ref{fig:evsa}. In this subsection, we focus on the eccentricities of hot Jupiters themselves ($P<10$ days in Fig. \ref{fig:evsa}). (We will assess the predictions of hot Jupiter origin theories for longer period giant planets’ eccentricities in \S \ref{sec:conn}.) A hot Jupiter that gets close to its host star can collide with the star (Fig. \ref{fig:evsa}, dark yellow region), be tidally disrupted (light yellow region), or tidally circularize (red region). In the observed distribution (Fig. \ref{fig:evsa}), hot Jupiters are absent from the stellar collision and tidal disruption regions. 

Out to $\sim$3 day orbital periods, most hot Jupiter orbits are consistent with circular. Jupiters at the shortest orbital periods likely have such fast tidal circularization timescales  (Eqn \ref{eqn:adot}) that we are unlikely to catch them on elliptical orbits. The circularization limit for hot Jupiters can be used to calibrate theories of tidal dissipation inside planets (e.g., \citealt{hans10, socr12}). An outstanding issues is that the tidal dissipation required to circularize these hot Jupiters is stronger than expected. \citet{socr12} found that the tidal efficiency of hot Jupiters must be at least 10 times higher than the tidal efficiency of our own Solar System's Jupiter. Current tidal models of dissipation by inertial waves do not predict such strong dissipation. Other processes -- such as the elliptical instability (e.g., \citealt{bark16a}) -- may enhance the dissipation efficiency.
\clearpage
\begin{figure}
\includegraphics[width=5in]{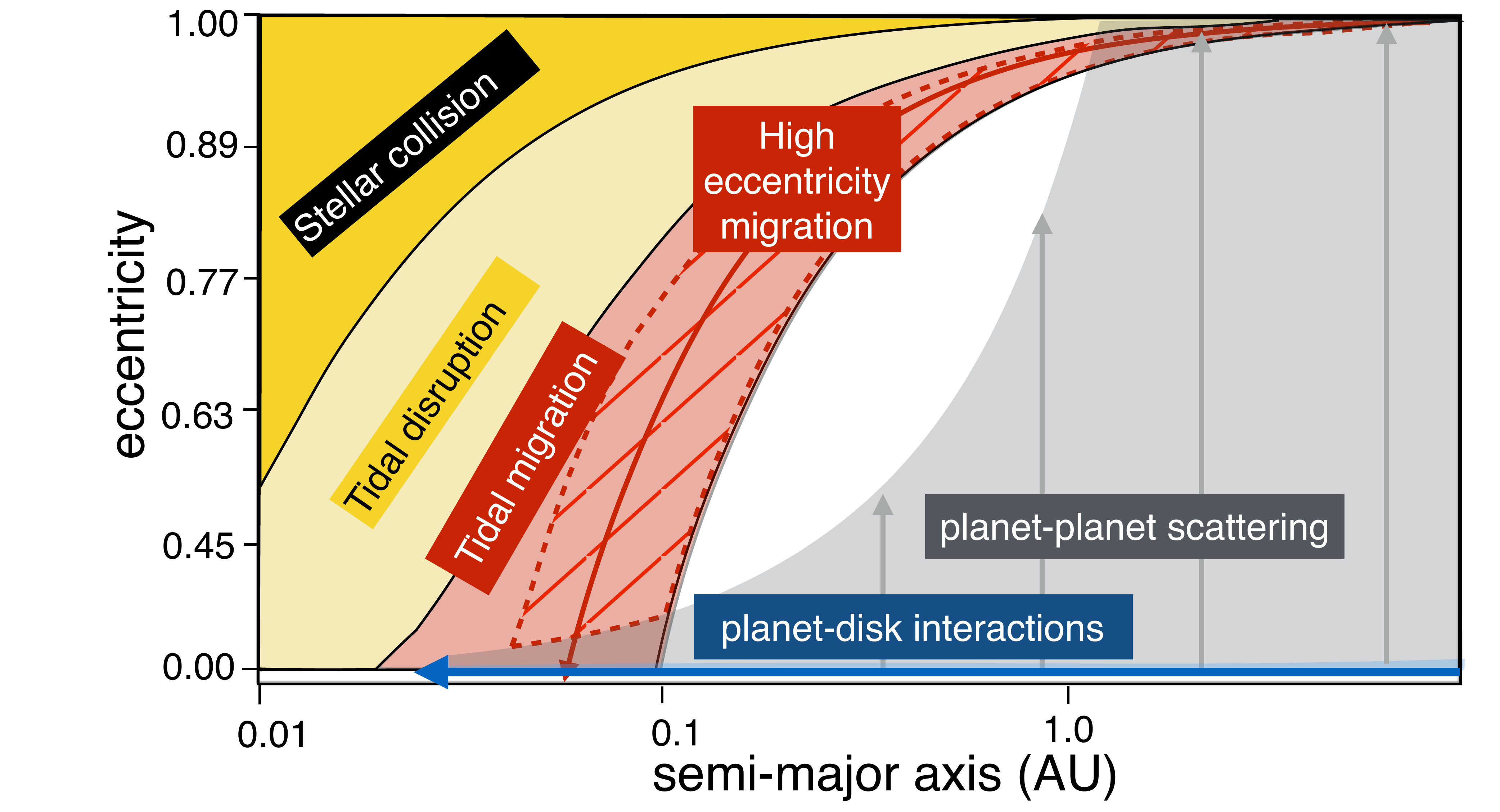}
\includegraphics[width=5in]{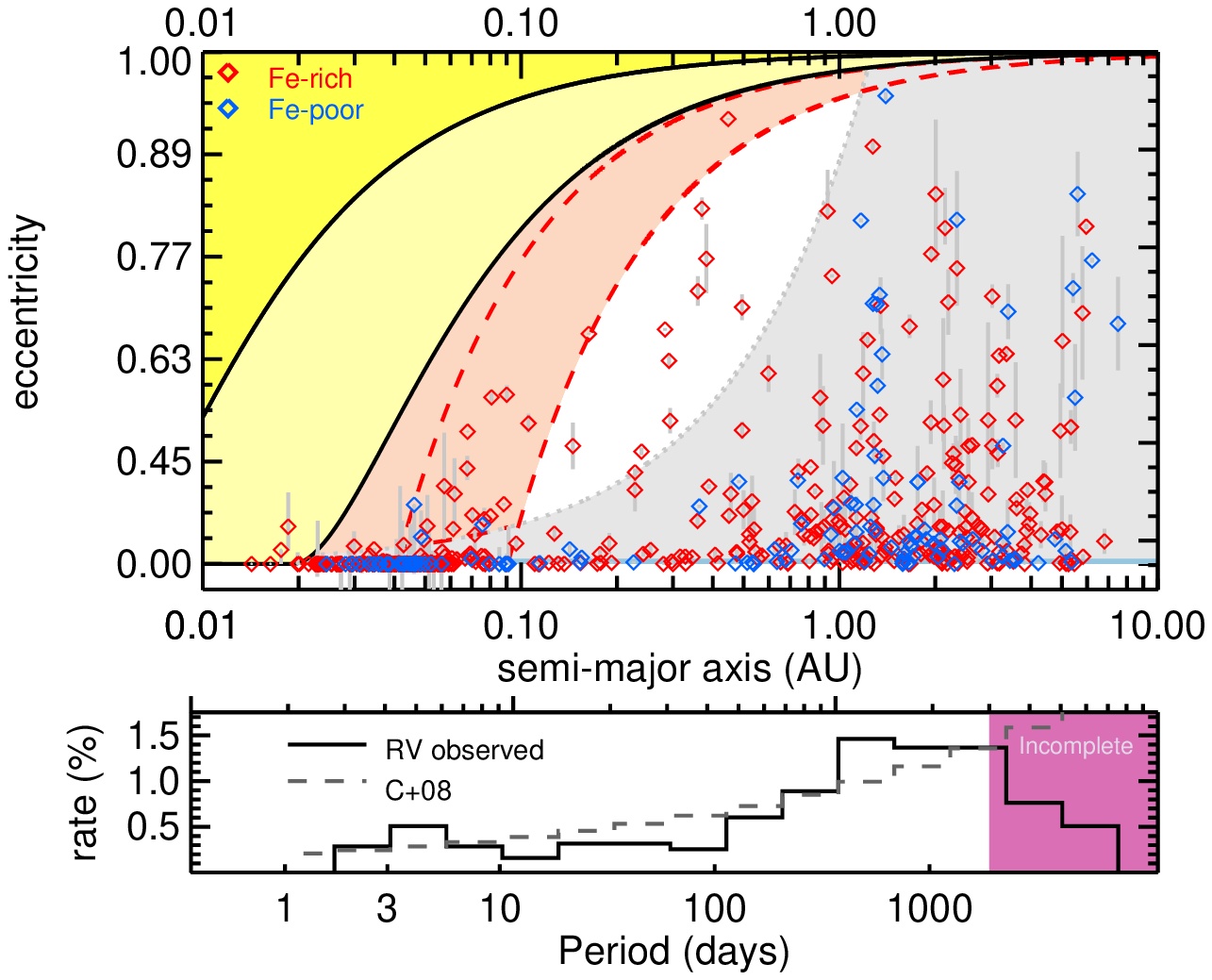}

\end{figure}

\begin{figure}
\caption{Regions of parameter space (eccentricity vs. semi-major axis) populated by different mechanisms of orbital evolution. Note that the y-axis is linear in $1-e^2$. Eccentricity and semi-major refer to a planet's instantaneous orbit: planets move through the diagram as their orbits evolve. For example, a planet could form at ($a=$5 AU, $e=0$), have its eccentricity excited by secular interactions with a companion and enter the red region, and tidally circularize a short orbital period. Top: Schematic. Middle: Observed radial-velocity planets with masses $M \sin i > 0.3 M_{\rm Jup}$, where $i$ is the sky-plane orbital inclination ($90^\circ$ is edge on). Planets disturbed into the yellow region would be tidally disrupted. Planets in the red region have small enough angular momentum for tidal circularization and those in the red dashed-outlined striped region could have started at high eccentricities (large semi-major axes) without having been tidally disrupted. Planet-planet scattering excites eccentricities to values limited by the planets' capacity to scatter rather than merge (gray region). Planet-disk interactions can alter semi-major axes and damp or modestly excite eccentricities (blue region). The solid red, gray, and blue arrows are examples of orbit changes via high eccentricity migration, eccentricity excitation, and planet-disk scattering respectively. \emph{Illustrative parameters are chosen to delineate the approximate region for each mechanism}; for example, the slope of the gray region depends on the planet's mass. See \S\ref{subsec:warm} for a discussion of if and how these mechanisms can populate the unshaded white region of eccentric warm Jupiters. Observed giant planets are color-coded by host star metallicity: those greater than or equal to solar within one sigma are red and others are blue. Bottom: histogram of semi-major axes of observed radial-velocity planets and occurrence rates inferred by \citet{cumm08} (C+08). Planets compiled from {\tt exoplanets.org} \citep{wrig12} and {\tt exoplanets.eu} \citep{schn11}. Eccentricities and upper limits taken from \citet{bono17} where available. The histogram is computed according to semi-major axis; the orbital period axis is for a Sun-like star. The observed histogram is normalized to total to 7.5\% for planets between 0.03--3 AU and 0.3--10 Jupiter masses (see \citealt{dong16} 5.1 discussion of \citealt{cumm08} results).
}
\label{fig:evsa}
\end{figure}

\clearpage

In the 3--10 day orbital period range, some hot Jupiters occupy moderately elliptical orbits ($0.2 < e < 0.6$).  If they originate through in situ formation (\S \ref{subsec:form}) or disk migration (\S \ref{subsec:disk}), their eccentricities would need to be excited at their present day close-in locations. However, as discussed in \S 2, interactions with the gas disk (Eqn. \ref{eqn:cse}) and planet-planet scattering in situ (Eqn. \ref{eqn:vesce}) cannot account for the moderate eccentricities observed. Possibly an outer planet could secularly force the eccentricity, but the perturber would need to be massive and/or nearby to overcome precession from general-relativity and tides. In the limit of Laplace-Lagrange secular excitation (which applies for low-to-moderate eccentricities and inclinations), an approximate criterion for the precession caused by the perturber (e.g., \citealt{fabr10}, Eqn. 21) to exceed general-relativistic precession is
\begin{eqnarray}
\label{eqn:grll}
    \left(\frac{m_{\rm perturber}}{ M_{\rm Jup}}\right) \left(\frac{37 \rm day}{P_{\rm perturber}}\right)^2 \gtrsim \left(\frac{M_\odot}{M_\star}\right)^{5/3} \left(\frac{3 \rm day}{P}\right)^{8/3} 
\end{eqnarray}
As we will discuss further in \S\ref{subsec:comp}, such companions have been ruled out for many moderately eccentric hot Jupiters. Similar criteria can be derived for the perturber precession to overcome the precession caused by the tidal distortion of the planet or rotation of the star.

In contrast, high eccentricity tidal migration (\S\ref{subsec:hem}) provides a natural explanation for these hot Jupiters on moderately eccentric orbits. According to this theory, moderately elliptical hot Jupiters are in the process of tidal circularization.  However, the occurrence of these moderately eccentric Jupiters may not be compatible with all tidal migration theories. \citet{petr15a} found that high eccentricity tidal migration spurred by stellar binary Kozai-Lidov cannot account for the number of moderately eccentric Jupiters relative to hot Jupiters: they predict only 1 moderately eccentric hot Jupiter for every 300 hot Jupiters on circular orbits. Among the 228 currently known hot Jupiters ($m > 0.1 M_{\rm Jup}$) in the 3--10 day orbital period range, 31 have eccentricities constrained at two sigma to be below $e<0.1$ (i.e., that could have been excited in situ), 10 have eccentricities constrained at two sigma to be $e>0.2$ (i.e., that most likely could not have been excited in situ and are evidence of high eccentricity tidal migration), and the rest have intermediate or poorly constrained eccentricities. Because the tidal circularization timescale is shorter for more less planets among planets of similar radii, less massive planets should be circularized out to larger host star separations \citep{pont11}. In a large sample of hot Jupiters with uniformly derived eccentricities and upper limits, \citet{bono17} found that this trend is present, supporting high eccentricity migration.

Why do we observe both circular and eccentric hot Jupiters at the same orbital periods? Under the high eccentricity migration hypothesis, circular hot Jupiters with 3--10 day orbital periods have completed their tidal circularization. Possibly circular hot Jupiters began their migration earlier than eccentric hot Jupiters at similar orbital periods (a possibility we will return to in \S\ref{subsec:sage}) or have more efficient tidal dissipation properties.  Another possibility is that multiple hot Jupiter formation channels are at work and some low eccentricity hot Jupiters originated via disk migration or in situ formation. \citet{daws13} suggested two formation channels for hot Jupiters---one of which is high eccentricity tidal migration---based on trends with host star metallicity. Eccentric hot Jupiters (red diamonds, Fig. \ref{fig:evsa}) orbit metal rich stars, while circular hot Jupiters orbit both metal rich and metal poor stars. (See also \citealt{shab16}). The correlation with host star metallicity may indicate that high eccentricity migration was spurred by some type of planet-planet gravitational interaction (\S\ref{subsec:hem}), as giant planet occurrence is correlated with host star metallicity \citep{gonz97,sant01,sant03,sant04,fisc05,sous11}. 

In summary, the existence of moderately eccentric hot Jupiters is evidence at least a fraction hot Jupiters underwent high eccentricity tidal migration.  Furthermore, although we do not know exactly where they began their tidal circularization, they must have originated far enough from the star for planet-planet scattering and/or secular excitation to be effective. We will return to the possibility that multiple origins channels contribute substantially to the hot Jupiter population throughout the review. We will tabulate which properties of hot Jupiters the three origins hypotheses explain or fail to explain in Table \ref{tab:evid}. 

\begin{deluxetable}{p{1.6in}|ccc}
\tabletypesize{\small}
\tablewidth{0pt}
\tablecaption{Evidence for origins hypotheses of hot Jupiters (HJ), including links to warm Jupiters (WJ) \label{tab:evid}}

\tablehead{
\colhead{Evidence}&
 \colhead{In situ formation} &
 \colhead{Disk migration} &
 \colhead{Tidal migration} \\
 &
 \colhead{(\S\ref{subsec:form})} &
 \colhead{(\S\ref{subsec:disk})} &
 \colhead{(\S\ref{subsec:hem})} 
}

\startdata
Elliptical HJ (\S\ref{subsec:ehj})&X&X&\checkmark\\
HJ obliquitities (\S\ref{subsec:obli}) & O & O & O\\
Inflated HJ radii (\S\ref{subsec:infl})&\checkmark & \checkmark & \checkmark\\
HJ semimajor axes (\S\ref{subsec:semi})&\checkmark&\checkmark&\checkmark\\
T Tauri HJs  (\S\ref{subsec:sage}) & \checkmark & \checkmark & X\\
Host star ages (\S\ref{subsec:sage}) & O & O & O\\
Atmospheres (\S\ref{subsec:atmo}) & T & T & T\\
Occurrence rates (\S\ref{subsec:occu}) & T &\checkmark &X\\
Companions (\S\ref{subsec:comp}) & X & X & \checkmark \\
HJ vs. WJ occurrence (\S \ref{subsec:warm}) & X & \checkmark & X \\
Circular WJs (\S \ref{subsec:warm}) & \checkmark & \checkmark & X \\
Elliptical WJs (\S \ref{subsec:warm}) & X & X & \checkmark \\
Nearby WJ companions (\S \ref{subsec:warm}) & \checkmark & T & X \\
Small planets (\S \ref{subsec:small}) & X & X & T \\
Hoptunes (\S \ref{subsec:small}) & X & T & \checkmark\\
\enddata
\tablecomments{\checkmark: consistent, X: inconsistent, T: no clear prediction from theory yet, O: additional or complementary observations needed}
\end{deluxetable}

\subsection{Obliquities of hot Jupiters' host stars}
\label{subsec:obli}

Hot Jupiter host star obliquities were once pursued as the Rosetta stone of hot Jupiters' origins. The host star obliquity refers to the angle between the star's spin angular momentum vector and the hot Jupiter's orbital angular momentum vector. The stellar obliquity can be measured in projection for individual stars using the Rossiter-McLaughlin effect (\citealt{mcla24,ross24}; see \citealt{tria17} for a review), the rotational broadening of stellar spectral lines (e.g., \citealt{schl10}), star spot crossings (e.g., \citealt{sanc11}), Doppler tomography (e.g., \citet{coll10}), gravity darkening (e.g., \citealt{barn09}), or asteroseismology (e.g., \citealt{hube13}) and for an ensemble based on star spot modulation amplitudes (e.g., \citealt{maze15,li16}). Many hot Jupiters appear well-aligned with their host stars' spin, but others are dramatically misaligned, even polar and retrograde (e.g., \citealt{albr12} and references therein).

We might naively expect planets that form in an accretion disk to be on orbits aligned with the star's rotational plane. In this simple picture, hot Jupiters that originate in situ (\S\ref{subsec:form}) or via disk migration (\S\ref{subsec:disk}) would maintain aligned orbits (even as they interact gravitationally with the gas disk, e.g., \citealt{bits13}). In contrast, the gravitation interactions that reduce the magnitude of the Jupiter's orbital angular momentum--- triggering high eccentricity tidal migration (\S\ref{subsec:hem})---would commonly change the direction of the angular momentum vector. Different mechanisms for generating the Jupiter's eccentricity (planet-planet scattering, Kozai-Lidov cycles driven by a binary, etc.) would result in different distributions of stellar obliquities (e.g., \citealt{fabr07,chat08,naoz11,tesy13,li14}). The theoretically predicted obliquity distribution could be compared to the observed distribution to identify the predominant eccentricity generating mechanism and even tease out the contributions of multiple pathways for hot Jupiter origins (e.g., \citealt{fabr09,mort11,naoz12}). Tides raised on the planet, which erase hot Jupiters' eccentricities (\S \ref{subsec:ehj}), leave \emph{stellar} obliquities intact. The hot Jupiter's orbital angular momentum is comparable to a star's spin angular momentum and timescale for realigning the star is (in this simple picture) comparable to tidal decay of the hot Jupiter. Although tides raised by the planet on the star could erase the stellar obliquity, this process would require complete tidal decay and disruption of the hot Jupiter (e.g., \citealt{hut81}). In this naive picture, stellar obliquities would be a powerful, unambiguous indicator of hot Jupiters' origin channel.

\begin{figure}
\includegraphics[width=5.06in]{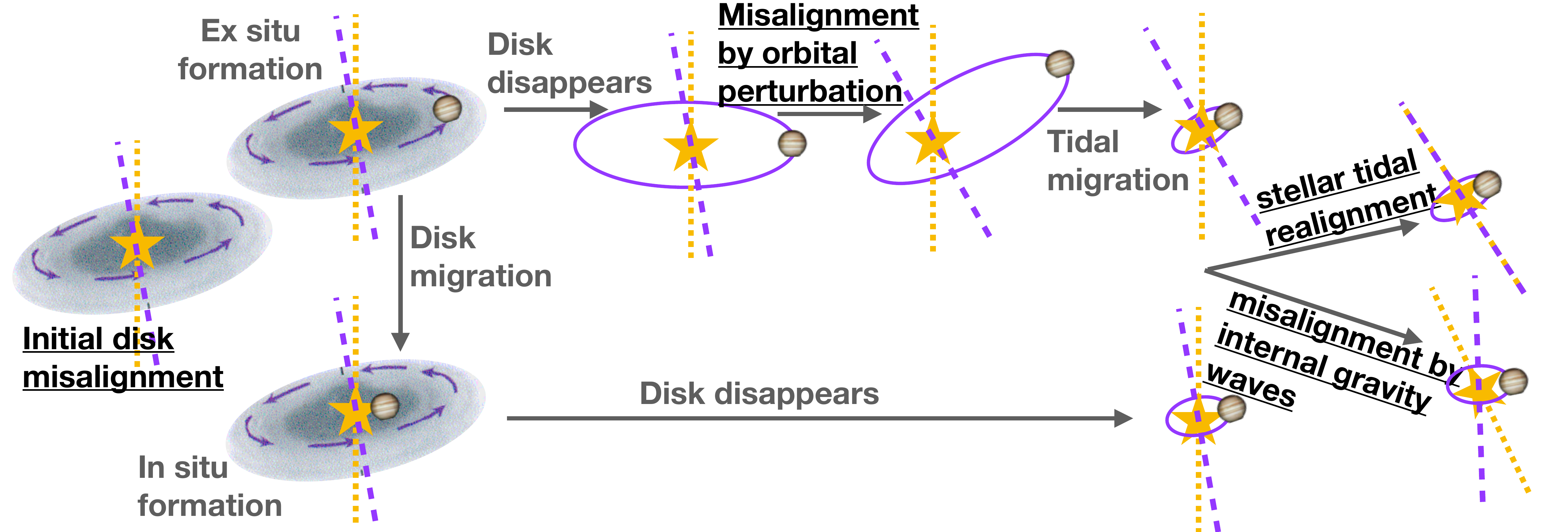}
\caption{Misalignments between a hot Jupiter's orbital angular momentum vector (purple dashed) and its host star's spin axis (yellow dotted) are influenced by a variety of physical processes including primordial misalignment of the disk the planet forms from, misalignment of the planet's orbit by a perturber, and realignment via tides raised on the star. In the absence of primordial misalignment, the star spin (dashed yellow) is initially taken to be perpendicular to the disk. Additional misalignment pathways not shown include chaotic evolution of the host star's spin axis during tidal evolution with Kozai-Lidov cycles \citep{stor14a} and misalignment by injection of an earlier planet (e.g., \citealt{mats15}).
}
\label{fig:obl}
\end{figure}

Unfortunately, interpreting stellar obliquities has proved to be far more challenging in reality. The following complications weaken our expectations that in situ formation and disk migration result in low obliquitities and that high eccentricity tidal migration should go hand in hand with spin-orbit misalignments:
\begin{itemize}
    \item Despite expectations to the contrary, spin-orbit alignments may be erased by tides raised on the star by the planet. As the sample of hot Jupiter host star obliquities grew, correlations between these obliquities and stars' tidal properties raised the alarm that the observed obliquity distribution, rather than purely reflecting the hot Jupiters' origins, has been sculpted by tidal realignment. \citet{schl10} and \citet{winn10} found that among hot Jupiter hosts, only hot stars' spins are misaligned with their hot Jupiters' orbits; cool stars may be more easily tidally realigned \citep{winn10,albr12}. The temperature cut-off coincides with the Kraft break \citep{kraf67}, implicating stellar spin-down in the realignment process (e.g., \citealt{daws14a}). Empirical evidence has emerged that planets can indeed influence stars' rotation (e.g., \citealt{cata07,kova14,popp14,mill15}). However, the interpretation that tidal realignment has erased primordial obliquities has a major unresolved theoretical problem: how can a mere planet realign an entire star's spin without the planet sacrificing all of its angular momentum (and without retrograde planets ending up perfectly anti-aligned; e.g., \citealt{lai12,dami15,lin17})? 
    \item Hot Jupiters originating via in situ formation or disk migration may be misaligned.  Mechanisms have recently been proposed for misaligning the disk or star itself, including a binary perturber (e.g., \citealt{baty12,spal15}) or stellar fly-by (e.g., \citealt{xian16}) tilting the disk; the star misaligning itself through internal gravity waves  (e.g., \citealt{roge12}); change in the spin of the proto-stellar cloud over the star formation timescale (e.g., \citealt{fiel15}); injection of an earlier planet (e.g., \citealt{mats15}); or chaotic evolution of the host star's spin axis during tidal evolution with Kozai-Lidov cycles \citep{stor14a} . Some of these mechanisms can plausibly account for the observed stellar temperature dependence without the planet needing to tidally realign the star. For example, \citet{roge12} found that misalignment through internal gravity waves only affects hot stars above the Kraft break. \citet{spal15} found that magnetic torques can realign the stellar spin-axes of lower-mass stars, which tend to be cooler, with the disk plane.
\end{itemize}
Figure \ref{fig:obl} schematically depicts how the above complications affect our interpretation of hot Jupiters' host star obliquities. 

The case remains open on whether hot Jupiters' host star obliquities implicate a particular origin scenario, require multiple origins scenarios, or primarily reflected physical processes unrelated to hot Jupiters' origins. A promising path is to compare hot Jupiters' host star obliquities to obliquities of stars hosting other classes of planets: we will discuss progress and prospects for comparative studies of obliquities in \S\ref{subsec:comp}.

\subsection{Hot Jupiter radius inflation}
\label{subsec:infl}

A number of hot Jupiters---including HD 209458b, the first hot Jupiter to be discovered to transit \citep{char00,henr00}---have radii larger than expected from internal structure models. Hot Jupiters' inflated radii require an additional heat source, such as tidal heating during high eccentricity migration (e.g., \citealt{bode01}), thermal tides caused by stellar irradiation (e.g., \citealt{arra10,socr13}), or deposition of stellar irradiation energy into the interior (e.g., \citealt{guil02,baty10,youd10}). All three classes of hot Jupiter origins hypotheses can be consistent with our current understanding of hot Jupiter inflation but the history of inflation plays out differently in each. In each case, the hot Jupiter begins its life hot and inflated from formation (e.g., \citealt{spie12}) and heat mechanisms must sustain inflation or re-inflate the hot Jupiter. Hot Jupiter radii are observed to be strongly correlated with stellar flux (e.g., \citealt{weis13}), supporting the interpretation that one or more stellar irradiation deposition mechanisms plays a role.

The scenarios of in situ formation and disk migration only require a mechanism to \emph{sustain} inflation. Disk migration---which by definition occurs early in the planet's lifetime--- naturally delivers the hot Jupiter close to its star before it cools on a $\sim$ 10 Myr timescale (e.g., \citealt{spie12,wu13}). One or more stellar irradiation mechanisms can subsequently keep the hot Jupiter inflated as it ages.

Tidal migration can also be compatible with inflated hot Jupiter radii. One possibility is that tidal heating re-inflates the proto-hot Jupiter after it cools (e.g., \citealt{igbu11}): in this case, the hot Jupiter does not need to arrive early in the star's lifetime. Tidal heating shuts off as the orbit nears circularization (e.g., \citealt{leco10,hans10}), so stellar irradiation mechanisms could take over at that point. A second possibility is that thermal tides or stellar irradiation deposition mechanisms may be capable of re-inflating a hot Jupiter after its arrival (e.g., \citealt{hart16,lope16}). An important caveat is that heating mechanisms that decay with depth, such as the commonly invoked stellar irradiation deposition mechanism of ohmic dissipation (e.g., \citealt{baty10}), have extremely long re-inflation timescales and are unable to re-inflate a Jupiter within the star's lifetime if it cools before arrival (e.g., \citealt{ginz16}; see also \citealt{wu13}). 

A final possibility is that the inflated hot Jupiters are those that tidally migrated early in the star's lifetime, before cooling. This possibility requires hot Jupiters to begin high eccentricity tidal migration at a young age, which is plausible because the dissipation of the gas disk may naturally trigger eccentricity excitation. When the gas dissipates, planets cushioned by the gas disk begin to scatter and secular excitation turns on as the precession from the gas disk shuts off. Hot Jupiters with smaller $a_{\rm final}$ migrate more quickly (Eqn. \ref{eqn:adot}), so this scenario is consistent with inflated hot Jupiters being closer to their stars (i.e., as argued by \citealt{wu07}). 

\subsection{Hot Jupiter semi-major axis distribution}
\label{subsec:semi}

The three origins channels in \S 2 each make different predictions for the distribution of hot Jupiter semi-major axes observed today (Fig. \ref{fig:evsa}). The small end of the semi-major axis distribution is affected by tidal disruption and large end by where formation and migration can effectively deposit hot Jupiters. A modest peak in the distribution occurs at $\sim 3$ days, a feature known as the three-day pile-up. We caution that this feature appears misleadingly large in plots that include hot Jupiters discovered by ground-based transit surveys, for which selection effects sculpt a prominent three-day pile-up (see \citealt{gaud05} for a discussion of these selection effects). Because the statistical significance of the three-day pile up has not been definitively established (i.e., whether it is a pile-up or simply a drop off interior to 3 days), here we focus on the inner edge of the hot Jupiter region.

\begin{figure}
\includegraphics[width=4in]{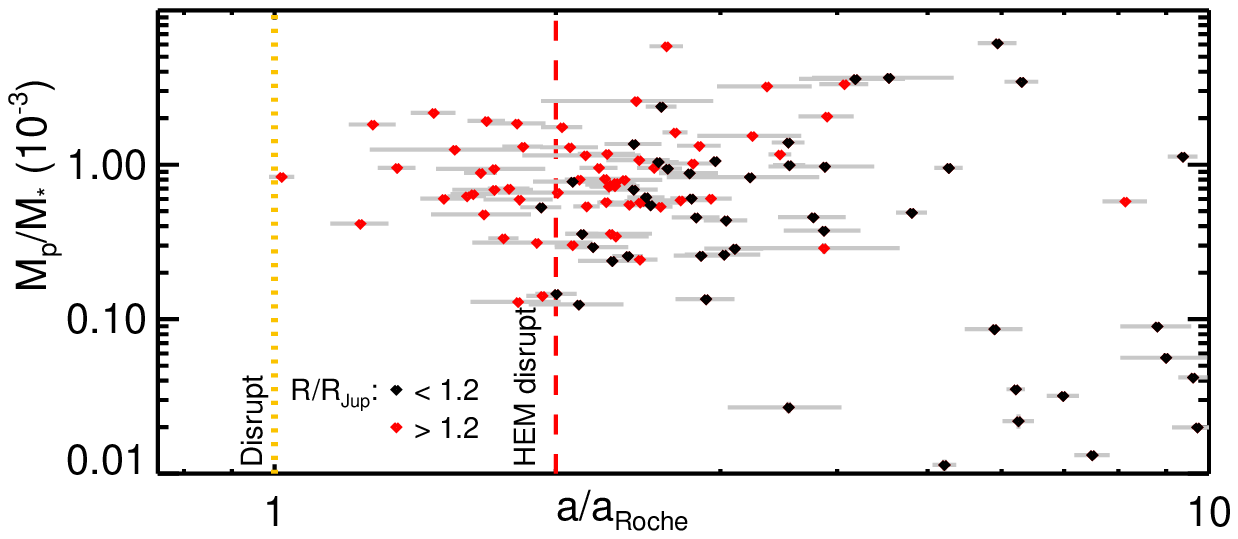}
\includegraphics[width=4in]{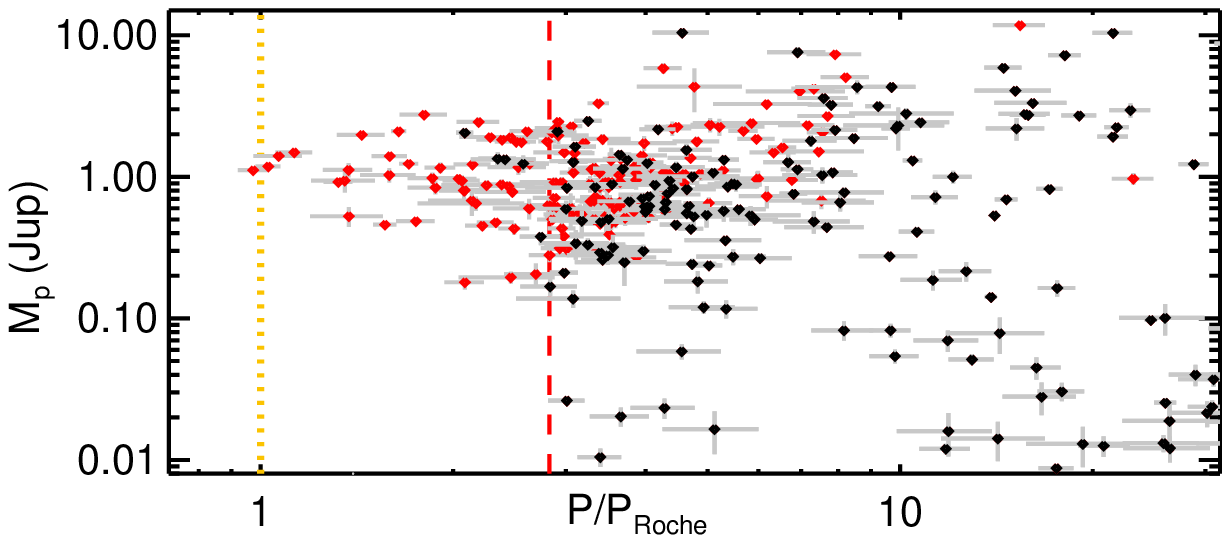}
\includegraphics[width=4in]{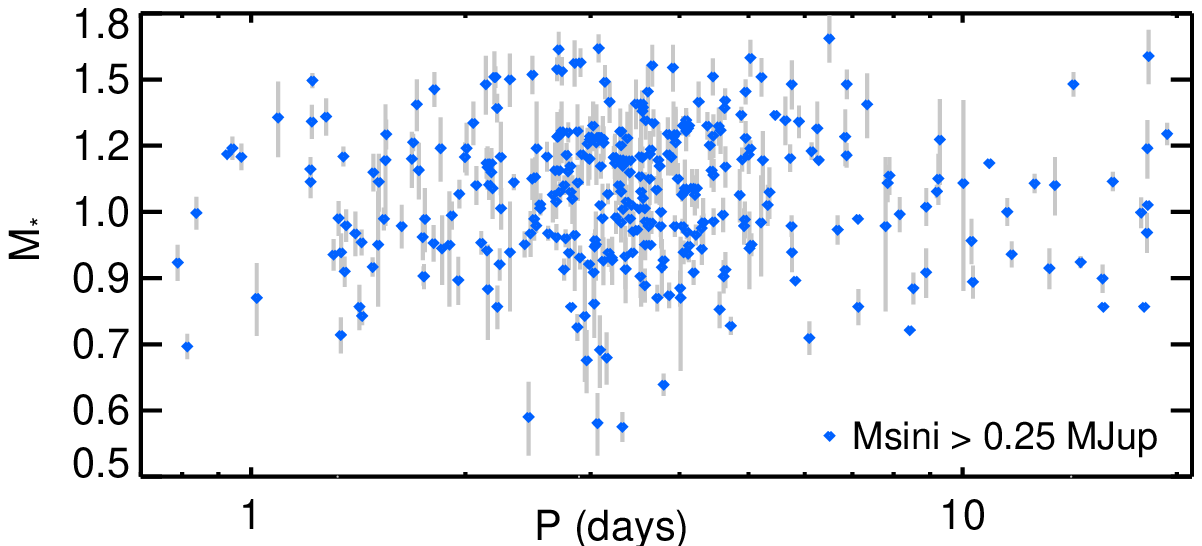}
\caption{The distribution of hot Jupiters' orbital periods and semi-major axes tests theories for their origin. All plotted planets have mass precision of 50\% or better. Yellow dotted line: tidal disruption limit. Red dashed line: tidal disruption limit during high eccentricity tidal migration (HEM) for $e \rightarrow 1$. Top: Planet-to-star mass ratio vs. semi-major axis scaled by the Roche radius for planets with 30\% precision or better on $a_{\rm Roche}$. Middle: Same using planet mass and Roche period (Eqn. \ref{eqn:roche}) for planets with 30\% precision or better on $P_{\rm Roche}$. Bottom: Stellar mass vs. orbital period of hot Jupiters (including non-transiting planets): no trend is evident, consistent with halting at the corotation radius. Planets compiled from {\tt exoplanets.org} \citep{wrig12} and {\tt exoplanets.eu} \citep{schn11}. 
}
\label{fig:roche}
\end{figure}

Hot Jupiters' present day orbits are consistent with tidal disruption limits (yellow regions, Fig. \ref{fig:evsa}). A hot Jupiter will be tidally disrupted inside the Roche limit, $a_{\rm Roche}$,
\begin{eqnarray}
\label{eqn:roche}
a_{\rm Roche} &\simeq& f_p R_p \left(\frac{M_\star}{M_p}\right)^{1/3} , \nonumber \\
P_{\rm Roche} &\simeq& \frac{2 \pi f_p^{3/2} R_p^{3/2}}{G^{1/2} M_p^{1/2} } = f_p^{3/2} \left(\frac{R_p}{\au}\right)^{3/2}\left(\frac{M_\odot}{M_p}\right)^{1/2} \nonumber \\&=& 0.79 {\rm~days} \left(\frac{f_p}{2.7}\right)^{3/2} \left(\frac{R_p}{1.3 R_{\rm Jup}}\right)^{3/2}\left(\frac{M_{\rm Jup}}{M_p}\right)^{1/2}, 
\end{eqnarray}
\noindent where $R_p$ is the planet radius, $M_p$ is the planet mass, $M_\star$ is the stellar mass, and $f_p$ is a dimensionless scale that depends on physical properties of the body. See the introduction of \citet{fabe05} for a pedagogical review of the Roche limit. The Roche limit is related to the Hill radius (Eqn. \ref{eqn:rh}): $R_H = R_p$ when $a = a_{\rm Roche}$ for $f_p = 3^{1/3}\approx1.44$. The limit $f_p = 3^{1/3}$ corresponds to the maximum distance at which a test particle on the surface of a perfectly spherical planet can remain at rest on the surface. For a planet subject to tidal disruption, $f_p$ depends on the material properties of the Jupiter. From  three-dimensional hydrodynamical simulations of tidal disruption of giant planets, \citet{guil11} found that $f_p \ge 2.7$.  In their simulations, planets at $f_p \sim 2.7$ are not immediately tidally disrupted but destroyed after mass loss and re-accretion over a number of close encounters. The re-accretion makes the planet puffier and easier to destroy in subsequent encounters. In Fig. \ref{fig:roche}, we plot $a/a_{\rm Roche}$of observed hot Jupiters (setting $f_p=2.7$). All hot Jupiters have $a/a_{\rm Roche} > 1$ today. (WASP-19b has $a/a_{\rm Roche}$ slightly less than 1 but consistent within the uncertainties.) However, some of these Jupiters may have been larger in the past at their time of formation. We color-code those smaller than $1.2 R_{\rm Jup}$ in black and see that almost all are beyond $a/a_{\rm Roche}>2$. These hot Jupiters would not have been in danger of tidal disruption even if they were inflated in the past (\S\ref{subsec:infl}).

In the case of in situ formation with no migration at all, we expect the inner semi-major axis of the hot Jupiter population to occur at the disk edge. The disk edge is thought to be set by the corotation radius, and therefore hot Jupiters are a factor of several closer to their stars than expected (e.g., \citealt{lee17}). We discuss expectations for the entire semi-major axis distribution of giant planets formed in situ in \S \ref{subsec:occu} and \S\ref{subsec:warm}. 

As described in \S\ref{subsec:disk}, disk migration may deliver hot Jupiters to half the corotation period (i.e., the 2:1 resonance with the disk inner edge). These shorter orbital periods of $\sim 5$ days are more consistent with hot Jupiters' observed orbital periods. See \S\ref{subsec:warm} for expectations for the occurrence rate of hot Jupiters (periods $<$ 10 days) relative to warm (periods 10--200 days) Jupiters from disk migration. 

In high eccentricity tidal migration, we expect to see surviving planets at or beyond $2 a_{\rm Roche}$ \citep{rasi96,mats10}. When planets begin their migration at high eccentricities, their initial periapses are approximately half to their final semi-major axes because $a_{\rm final}= a (1-e^2) \approx 2 a (1-e) $ for $e \rightarrow 1$ (\S \ref{subsec:hem}). Therefore planets at $2 a_{\rm Roche}$ today must have reached $a_{\rm Roche}$ during high eccentricity migration. Moreover, if the planets underwent high eccentricity migration before cooling and contracting, their $a_{\rm Roche}$ during migration would have been larger, so we would see them beyond $2 a_{\rm Roche}$ today. Although many hot Jupiters are beyond $2 a_{\rm Roche}$, we also see a population between 1--2 $a_{\rm Roche}$ that high eccentricity migration alone cannot easily account for.

However, in all origins scenarios, subsequent tidal evolution, in which planets raise tides on their stars, can further shrink hot Jupiters' semi-major axes. \citet{vals14} find that this subsequent evolution could account for hot Jupiters at 1--2 $a_{\rm Roche}$ for certain stellar tidal models and parameters. The paucity of very massive hot Jupiters in this region ($M_p > ~3 M_{\rm Jup}$, Fig. \ref{fig:roche}) may be due to orbital decay (e.g., \citealt{dami16}). It is possible that the inner edge hot Jupiters is set not by the origin channel but by $a_{\rm Roche}$. See \S \ref{subsec:usp} for a discussion of tidally stripped hot Jupiters than become super-Earths.

One possible way to distinguish among origins scenarios is to look for correlations between hot Jupiters' semi-major axes and other parameters. \citet{plav13} compared the observed distribution of hot Jupiter semi-major axis vs. stellar mass against the distribution expected from different origin scenarios. They found that the inner semi-major axis limit is consistent with scaling as $M_\star^{1/3}$, as expected if high eccentricity delivers hot Jupiters to $2 a_{\rm Roche}$ (Eqn. \ref{eqn:roche}). They ruled out scenarios in which hot Jupiters' inner semi-major axis is set by the magnetospheric cavity (i.e., in situ formation at the inner disk edge or disk migration halted at a 2:1 resonance with the inner disk edge), for which they expect the inner semi-major axis to scale as $M_\star^{1/7}$. However, there are two caveats to this conclusion. First, as discussed above, the inner edge of the Jupiter distribution may be sculpted primarily by tidal disruption rather than the formation or delivery location. Second, if the rotational periods of young stars are largely independent of stellar mass at the time of gas disk dispersal (\citealt{lee17},  references therein), the corotation radius would follow the same scaling as high eccentricity migration, $M_\star^{1/3}$. Panel 3 of Fig. \ref{fig:roche} shows a lack of trend in hot Jupiters' host star masses vs. hot Jupiters' orbital periods, consistent with the inner semi-major axis of the hot Jupiter distribution scaling with $M_\star^{1/3}$.

Hot Jupiters' semi-major axis distribution may have contributions from multiple origins channels. Recently \citet{nels17} modeled the semi-major axis distribution of hot Jupiters as resulting from two migration channels, disk migration and high eccentricity migration. They found that the Kepler and radial velocity sample of hot Jupiters can be accounted for by high eccentricity migration alone but that hot Jupiters discovered by WASP and HAT survey -- which are limited to shorter orbital periods -- need a $\sim35\%$ contribution from disk migration. 

In summary, the semi-major axes of the closest hot Jupiters appear most consistent with disk migration, which we expect to deliver to hot Jupiters interior to young stars' corotation radii. They appear less consistent with in situ formation, which we expect to deliver hot Jupiters beyond young stars' corotation radii, and high eccentricity tidal migration, which we expect to deliver hot Jupiters to beyond $2 a_{\rm Roche}$. However, we cannot definitively rule out in situ formation or high eccentricity tidal migration because tides raised on the star could move hot Jupiters to shorter orbital periods for certain tidal parameters.

\subsection{Ages of hot Jupiter hosts}
\label{subsec:sage}

If hot Jupiters form in situ or arrive via gas disk migration, they should be in place by time the gas disk dissipates. In contrast, hot Jupiters can arrive through high eccentricity tidal migration throughout a star's lifetime. Measuring host star ages may help distinguish among origin scenarios. However, this approach has not yet made any definitive breakthroughs in identifying hot Jupiters' predominant origins channel. Here we summarize the theoretical and observational challenges to identifying hot Jupiters' origins using stellar ages. 

\subsubsection{Expected age distinctions}
\label{subsec:eage}
Although hot Jupiters can complete their high eccentricity migration throughout a star's lifetime, we expect the bulk of them to arrive early. Once the mechanisms for generating high eccentricities turn on, they tend to work quickly. Planet-planet scattering can be triggered by dissipation of the gas disk. If planets remain stable through the dissipation of the gas disk, their subsequent instability timescales are drawn from a log distribution set by the planets' spacings (e.g., \citealt{cham96}): only particular special spacings result in the system going unstable on a timescale of order the stellar age. Timescales for eccentricity excitation via secular interactions typically range from thousands to millions of years depending on the perturber's mass and distance. Once the proto-hot Jupiter attains its high eccentricity orbit, the tidal circularization timescale scales as $a_{\rm final}^8$ (Eqn. \ref{eqn:adot}). Therefore the possible circularization timescales span many orders of magnitude and only a very special value for the proto-hot Jupiter's initial eccentricity will enable it to tidally migrate on a timescale similar to the star's lifetime. Otherwise, if it tidally migrates at all, it will do so quickly. The expectation that most hot Jupiters will arrive early via high eccentricity migration weakens the distinction between high eccentricity tidal migration and other origins scenarios. Only with a sample of very young stars or with a very large sample of main sequence stars can we hope to distinguish between high eccentricity migration vs. disk migration or in situ formation.

\subsubsection{Constraints from young stars}

Although the distinction in timescales between the scenarios is not as dramatic as we might hope, very young stars have the potential to powerfully distinguish between disk (in situ formation, disk migration) vs. post-disk (high eccentricity migration) mechanisms. If T Tauri stars, which still have their gas disks, host hot Jupiters, we can conclude that at least some form in situ or arrive via disk migration. However, it is challenging to detect and confirm planets orbiting these active young stars. Recently two hot Jupiters have been discovered orbiting Tauri stars using spectropolarimetry \citep{dona16,yu17}, which are very challenging to explain via high eccentricity migration. Ongoing surveys will help constrain the occurrence rate of planets orbiting Tauri stars, allowing us to evaluate whether all hot Jupiters might be delivered during the gas disk stage.

Next to T Tauri stars, young clusters are the best population to survey for young hot Jupiters. Hot Jupiters have recently been discovered in young clusters, including two hot Jupiters with 2 and 4 day orbital periods in the metal-rich 800 Myr Beehive cluster \citep{quin12} and HD 285507 b, an 6-day hot Jupiter with $e=0.09 \pm 0.02$ in the 600 Myr metal-rich Hyades cluster \citep{quin14}. All three hot Jupiters could be plausibly explained by any of three origins scenarios; the stars are old enough and the orbital periods are short enough that high eccentricity tidal migration may have operated. In that case, HD 285507 b may be at the end of its journey, still tidally circularizing. However, its eccentricity is low enough to have been excited in situ by scattering (Eqn. \ref{eqn:vesce}, secular perturbations by a nearby companion (if a sufficiently massive one is nearby, e.g., Eqn. \ref{eqn:grll}), or possibly even the disk (Eqn. \ref{eqn:cse}). Most observable open clusters are not young enough to definitively rule out high eccentricity migration for a given planet.

However, even if we cannot determine the origin of a particular hot Jupiter discovered in a young cluster, we may be able to distinguish among origins theories by comparing the occurrence rate of hot Jupiters in young clusters vs. in the field. If high eccentricity tidal migration is at work, the occurrence rate of hot Jupiters will be lower in young clusters (i.e., because planets with eccentricity excitation timescales and/or tidal circularization timescales of $\sim 1-10$ Gyr will not yet have arrived). Unfortunately, another factor complicates this expectation: the dynamical environment of the cluster. Encounters with other stars in the cluster can disturb the planetary system, triggering  high eccentricity tidal migration (e.g., \citealt{hao13,shar16}; \citealt{chat12} find that this would not be a common channel in NGC 6791). \citet{bruc17} find the occurrence rate for hot Jupiters in the solar-metallicity, solar-age open cluster M67 is actually higher than in the field (though the discrepancy is only marginally statistically significant). The addition of hot Jupiters from stellar fly-bys in open clusters could compensate for missing young hot Jupiters that have yet to undergo migration.

Hot Jupiters have also been discovered orbiting about half a dozen A stars (to date), which have main sequence lifetimes $~\sim 1$ Gyr and thus tend to be younger. For example, hot Jupiter WASP-33b, discovered via Doppler tomography, orbits a star younger than 400 Myr \citep{coll10}. However, comparing the properties of hot Jupiters orbiting A stars to those orbiting older FGK stars is complicated by the differences in stellar mass and stellar tidal properties.

\subsubsection{Constraints from middle age field stars}
 
Combining Gaia parallax measurements \citep{gaia16} with rotation periods and/or asteroseismology from TESS \citep{camp16} and PLATO \citep{raue14} should expand the sample of hot Jupiters with stellar age estimates. The vast majority of field stars are not young enough for us to check whether the hot Jupiter arrives while the proto-planetary disk is present. However, a large sample of field stars with stellar ages would allow us to test for trends. 

Two key age trends to investigate are the hot Jupiter occurrence rate and period distribution. From disk migration or in situ formation we expect the occurrence rate and period distribution to have no dependence on stellar age. In contrast, if high eccentricity tidal migration is the predominant channel, hot Jupiter occurrence rates should increase with age and the orbital period distribution should extend to longer orbital periods with age. As we discussed in \S \ref{subsec:eage}, we expect most hot Jupiters formed by high eccentricity migration to arrive early, so a large sample size is necessary to catch the late comers and identify differences in their distribution of orbital periods.

Furthermore, high eccentricity tidal migration leads to different eccentricity vs. semi-major axis distributions over time. Older stars should have circular hot Jupiters out to wider separations because the hot Jupiters have longer to tidally circularize. \citet{quin14} compared stellar ages to hot Jupiters' tidal circularization timescales and found that those with ages longer than the circularization timescale had significantly larger eccentricities. However, since the tidal circularization timescale is very sensitive to $a_{\rm final}$ ( \S\ref{subsec:hem}), the observed trend between eccentricity and tidal circularization timescale might reflect an increase in eccentricity with semi-major axis (i.e., still be statistically significant ignoring the stellar age). A trend in eccentricity vs. semi-major axis may be caused by in situ eccentricity excitation (\S \ref{subsec:ehj}). Comparing eccentricity vs. stellar age within each semi-major axis interval could help distinguish between tidal circularization vs. in situ excitation.

We caution that robustly identifying trends in a main sequence sample is challenging. Eccentricity excitation and tidal circularization timescales are drawn from log distributions, so the main sequence sample (e.g., with a typical $\sim1-10$ Gyr age range) is only a small fraction of the dynamic range of timescales. Moreover, trends with stellar age can be difficult to distinguish from trends with other stellar properties. For example, \citet{tria11} reported a trend of host star obliquity decreasing with stellar age, but the trend may instead be with host star effective temperature (e.g., \citealt{winn10}).

\subsubsection{Summary}
As of today, stellar ages provide no conclusive evidence regarding the origins of hot Jupiters. However, investigating how the occurrence rates and properties of hot Jupiters change with stellar age is an interesting area for future study. We recommend investigations of field stars with larger sample sizes of hot Jupiters discovered by CHEOPS \citep{broe13}, TESS \citep{rick15}, and PLATO \citep{raue14} and constraints on stellar properties and ages from TESS \citep{camp16}, Gaia \citep{gaia16}, and PLATO; continued searches for hot Jupiters orbiting T Tauri stars; efforts to disentangle the effects of age vs. cluster environment on the occurrence rate of hot Jupiters in open clusters; searches for longer period hot Jupiters in open clusters; and the development of statistical approaches to better distinguish whether a trend is due to stellar age or a different stellar property.

\subsection{Atmospheric properties of hot Jupiters}
\label{subsec:atmo}

The species present in a hot Jupiter's atmosphere are clues to where and how the hot Jupiter formed (e.g., \citealt{madh14}). The composition of gas and grains in a proto-planetary disk varies radially as the disk temperature drops and volatiles condense (e.g., \citealt{ober11}). The chemical composition of the gas and solids the planet accretes changes across snow lines (volatile condensation fronts). For example, the formation location of a hot Jupiter relative to the water, carbon dioxide, and carbon monoxide snow lines affect the atmosphere's C/O ratio. A hot Jupiter formed in situ would have a composition characteristic of the inner disk -- where very few ices can exist -- while a hot Jupiter arriving through high eccentricity tidal migration would have a composition reflective of the outer disk. A hot Jupiter that underwent disk migration may have an intermediate composition if it accreted gas along the way (e.g., \citealt{alib05}).

However, disk dynamics and chemistry can cause ice line locations to vary by an order of magnitude depending on the disk conditions (e.g., \citealt{piso15}). Uncertainty in disk conditions makes it challenging to back out a planet's formation location from its atmospheric properties. For example, we may deduce from a planet's low C/O atmosphere that it formed within the water ice line but suffer from an order of magnitude uncertainty in where that water ice line was located. Ongoing observations of proto-planetary disks with ALMA may provide us with the better understanding of realistic disk conditions and parameters necessary to pin down typical snow line locations. More pessimistically, ALMA may reveal too much diversity in disk conditions to ever infer a typical snow line.

Another uncertainty is the extent to which planetesimals get mixed into giant planets' atmospheres. This uncertainty complicated the interpretation of \citet{sing16}'s hot Jupiter characterization survey. \citet{sing16} found that hot Jupiter atmospheric compositions are consistent with no primordial water depletion relative to nebular gas containing water in vapor form. One possibility is that hot Jupiters formed within the water ice line. A second possibility is that they formed beyond the ice line and underwent disk migration but accreted gas along the way. A final possibility is that they formed entirely beyond the snow line and arrived via migration but icy planetesimals were oblated in their atmospheres, replenishing the water fraction.

Another challenge to using atmospheric properties to test hot Jupiter origin hypotheses is the difficulty in measuring species in the atmospheres. The C/O ratio is one example of an atmospheric quantity strongly influenced by location relative to ice lines yet challenging to measure. It can be inferred from the abundance of carbon monoxide (depleted at high C/O ratio) and methane (enhanced at high C/O ratio). However, C/O inferences have proven to be sensitive to which datasets and analysis techniques were used  (e.g., \citealt{krei15}) and to which priors were imposed dictating chemically-possible atmospheres (e.g., \citealt{heng16}). Higher signal-to-noise spectra observed by the upcoming James Webb Space Telescope may resolve some of these discrepancies by shifting the analysis into a more data-driven regime.

\section{TESTING THE ORIGIN THEORIES BY CONNECTING HOT JUPITERS TO OTHER POPULATIONS}

\label{sec:conn}
Hot Jupiters' origins manifest not only in their intrinsic properties (\S \ref{sec:prop}) but also in their connections to other exoplanet populations. Here we review how the following connections square with hot Jupiters' origins: the occurrence rate of hot Jupiters vs. more distant giant planets (\S \ref{subsec:occu}), companions of hot Jupiters (\S \ref{subsec:comp}), properties of hot vs. warm Jupiters (\S \ref{subsec:warm}, and properties of hot Jupiters vs. smaller planets (\S \ref{subsec:small}).

\subsection{Hot Jupiter occurrence rates relative to wider separation giant planets}
\label{subsec:occu}

\subsubsection{Overall relative occurrence rates}

About one in ten giant planet systems contains a hot Jupiter \citep[][cf \citealt{wrig12} and \citealt{guo17} regarding the apparent discrepancy between RV surveys and {\it Kepler} HJ occurrence rates]{howa10,mayo11}, with the giant planet occurrence rate dropping sharply within 200 days (Fig. \ref{fig:evsa}). The occurrence rate of hot Jupiters relative to wider separation giant planets reflects the efficiency of their origins channel. If they formed in situ (\S\ref{subsec:form}), their occurrence rate is set by the propensity of disks to form giant planets close to their stars vs. at wider separations. If they formed ex situ, their occurrence rate is set by the efficiency of transporting giant planets close to their star by disk migration (\S\ref{subsec:disk}) or high eccentricity tidal migration (\S\ref{subsec:hem}). 

The in situ formation hypothesis (\S\ref{subsec:form}) currently lacks a clear prediction for hot Jupiters' relative occurrence rate. The ease of forming hot Jupiter depends on the inner disk's local solid surface density and, to a lesser extent, the gas opacity (e.g., \citealt{lee16}). Building a core capable of runaway gas accretion requires a sufficient amount solids close to the star. Equation \ref{eqn:mcore} dictates a factor 10,000 higher surface density to build a massive core at a 3 day orbital period than at a 3000 day orbital period; whether this factor is achievable depends on the disk's mass in solids and their radial distribution. The radial distribution in turn depends on the (still not well quantified) efficiency of radial transport of pebbles, planetesimals, and embryos and the extent to which solids can pile up in one location. As we will discuss in \S\ref{subsec:small}, a complementary avenue to evaluating the in situ formation hypothesis is to compare giant planet occurrence rate vs. orbital period to super-Earth occurrence rates vs. orbital period.

Disk migration (\S\ref{subsec:disk}) can deliver hot Jupiters at the observed relative rate for plausible migration parameters. For example, \citet{cole16} found good agreement between their simulations of giant planet migration and the observed occurrence rates. Jupiters' final locations depend on the migration timescale, which may span many orders of magnitude depending on disk conditions, and the remaining disk lifetime after the Jupiter's formation. If the migration timescale is much longer than the disk lifetime, the Jupiter will not stray far from its birthplace. If the remaining disk lifetime is much longer than the migration timescale, the Jupiter can become a hot Jupiter (assuming some mechanism halts its migration prior to tidal disruption, \S\ref{subsec:disk}). 

\begin{deluxetable}{cccp{2.1in}}
\tabletypesize{\small}
\tablewidth{0pt}
\tablecaption{Theoretical efficiencies of forming hot Jupiters from cold Jupiters via high eccentricity tidal migration. The observed ratio of hot to cold Jupiters is $\sim1:10$.\label{tab:hje}}

\tablehead{
\colhead{Mechanism}&
 \colhead{Study} &
 \colhead{HJ/J (\%)} &
 \colhead{Assumptions of study} 
}

\startdata
Stellar binary Kozai & a & $\sim$1--3 &System of single giant planet and binary companion \\
Planet secular coplanar & b&$\sim$3--5& System of two giant planet; HJ begins at 1 AU\\
Planet-planet Kozai & c&$\sim5$&System of two giant planets; mutual inclinations drawn from an isotropic distribution; HJ begins at 1 AU\\
Planet-planet scattering & d &$\sim$3--5&System of at least three giant planets; system goes unstable.\\
Secular chaos & e & ? & Not quantified due to strong dependence on uncertain initial conditions 
\enddata
\tablecomments{a: \cite{muno16}, b: \citet{petr15}, c: \citet{petr16}, d: \citet{beau12}, e: \citet{wu11}.}
\end{deluxetable}

In the high eccentricity tidal migration scenario (\S\ref{subsec:hem}), hot Jupiters' relative occurrence rate depends on the efficiency of the mechanism that raises the proto-hot Jupiter's eccentricity. Forming a hot Jupiter requires driving the eccentricity high enough for tidal dissipation to be effective but not so high that the proto-hot Jupiter hits the star or is tidally disrupted. Investigations of high eccentricity tidal migration have found low rates of hot Jupiter production inconsistent with observations, even with optimistic assumptions, as we summarize in Table \ref{tab:hje}. A few outstanding theoretical issues remain in understanding in the hot Jupiter efficiency from high eccentricity tidal migration. First, the secular excitation mechanisms (\S \ref{subsec:hem}) -- stellar binary Kozai, planet-planet secular coplanar, and planet-planet Kozai -- can easily be shut off by bodies between the proto-hot Jupiter and its perturber (e.g., an additional giant planet, a nearby small planet). Accounting for such planets would reduce the efficiency further. Second, more investigation of how the results depend on initial conditions is necessary. For example, beginning a proto-hot Jupiter at 1 AU (e.g., \citealt{petr16}) requires less eccentricity excitation than starting it further from the star. Third, more investigation is needed on whether hot Jupiters can be saved from tidal disruption, a major sink of hot Jupiters in high eccentricity migration models. For example, \citet{wu17} recently showed that giant planets can undergo rapid tidal circularization through f-mode dissipation. This rapid circularization may allow a hot Jupiter to safely migrate and decouple from its perturber before its eccentricity is raised high enough for tidal disruption. Finally, the interplay among all the different planet-planet excitation mechanisms should be explored further (e.g., \citealt{naga11}); it is unclear whether hot Jupiter formation rates from different planet-planet eccentricity excitation mechanisms should be summed.

\subsubsection{Relative occurrence rate vs. planet mass}

Hot Jupiters have lower masses than their more distant counterparts (Fig. \ref{fig:roche}); in particular, we observe a dearth of giant planets above $> \sim 3 M_{\rm Jup}$ at orbital separations between the Roche limit and twice the Roche limit. One possibility is that lack of massive hot Jupiters is not a result of hot Jupiters' origin channel but of hot Jupiters raising tides on their host stars. More massive planets can more quickly transfer their angular momentum to spin up their stars and undergo tidal decay, ultimately leading to tidal disruption.

Alternatively, hot Jupiters' lower masses may be primarily established by their origins channel. The in situ formation hypothesis may preferentially produce lower mass hot Jupiters (e.g., \citealt{baty15}). A giant planet's core can either form as an isolation mass from a narrow feeding zone or grow via collisions of smaller cores from a wider annulus as the gas surface density declines (e.g., \citealt{bole16}). The latter scenario can operate with a lower local solid surface density (helping mitigate an obstacle to in situ formation) but can stunt the hot Jupiter's growth by giving it less time to accrete gas and less gas to accrete (e.g., \citealt{lee16}).

Migration is also consistent with lower masses for hot Jupiters. Less massive hot Jupiters may migrate more efficiently because they are less effective at opening deep gaps in the disk that would slow their migration. \citet{mass03} found that ``runaway migration," in which a feedback caused by corotation torques leads to rapid inward migration,operates most effectively for planets just below Jupiter mass. Sub-Jovians are suspectible to runaway migration because they are less massive than the surrounding disk yet massive enough to open shallow gaps in the disk that facilitate the corotation torque.

It is currently unclear whether lower masses for hot Jupiters are consistent with high eccentricity tidal migration. Lower mass planets are more easily disturbed onto elliptical orbits by other planets in planet-planet mechanisms \citep{naoz11,wu11,petr15}. However, they are also more easily tidally disrupted \citep{ande16,muno16}.

\subsubsection{Relative occurrence rate vs. stellar properties}
\label{subsec:stel}
The overall giant planet occurrence correlates with stellar mass (\citealt{john10}; peaking at 2 $M_\odot$, \citealt{reff15}) and metallicity \citep{gonz97,sant01,sant03,sant04,fisc05,sous11,guo17}. These correlations are thought to reflect an underlying connection between giant planet occurrence rate and disk mass, with more massive, solid-rich disks spawning cores massive and quick-growing enough to accrete gas during the gas disk lifetime (\S \ref{subsec:form}). The disk mass has been observationally linked to the host star mass \citep{ansd16,ansd17,pasc16}; it is still unclear whether the disk mass in solids is correlated with host star metallicity \citep{moro15,gasp16}. 

Different origins channels have different predictions for whether the occurrence rate of hot Jupiters {\emph relative to other giant planets} should depend on stellar mass and metallicity. The efficiency of forming hot Jupiters in situ relative to more distant giant planets is strongly dependent on the local solid surface density (\S \ref{subsec:form}), which is controlled by both the total disk mass in solids and the solid transport efficiency. The former should produce a correlation between hot Jupiters and host star metallicity/mass. A higher solid surface density would allow planets to form more quickly in the outer disk. A faster formation time would give a Jupiter more time to migrate, increasing the ratio of hot to cold Jupiters. However, given the many orders of magnitude of possible migration speeds (i.e., migration speeds can be much longer or much shorter than the disk lifetime), the increase may not be significant.

In the high eccentricity tidal migration channel, the relative hot Jupiter occurrence rate vs. stellar mass/metallicity depends on the mechanism for raising the hot Jupiter's eccentricity. In the case of high eccentricity migration triggered by a stellar perturber, we do not expect a correlation with host star metallicity/mass. We may even expect an anti correlation because if a disk spawns too many giant planets, they may protect each other from the secular effects of the stellar companion. In contrast, high eccentricity migration triggered by a planetary perturber is likely to produce a correlation. A more massive and/or solid enriched disk can spawn giant planets at more locations (Fig. \ref{fig:miso}), allowing multiple giant planets to form.

Observationally, there is some evidence for a further enhancement of hot Jupiters relative to cold Jupiters at higher host star metallicities.  \citet{jenk16} found that giant planets with orbital periods less than 100 days orbit more metal rich stars than giant planets with orbital periods greater than 100 days. \citet{daws13} found a larger difference in occurrence rates for hot Jupiters orbiting metal rich vs. metal poor stars than for longer period planets orbiting metal rich vs. metal poor stars. The correlation with stellar mass is less certain. \citet{ober16} found the occurrence rate of hot Jupiters orbiting M-stars vs. FGK stars {\emph could} be the same, but for the former, they only have an upper limit.

\subsection{Companions of hot Jupiters}
\label{subsec:comp}

The hot Jupiters origin theories make different predictions for whether hot Jupiters are likely -- or required -- to be members of multi-planetary or binary systems and the mass and proximity of these companions. In situ formation tends to spawn hot Jupiters accompanied by nearby planets (e.g., \citealt{hans13,bole16}), which are not necessarily in or near orbital resonance. Disk migration tends to deliver hot Jupiters with resonant companions  \citep{malh93,lee02,raym06}, which may be giant or small. Pairs of small planets can plausibly escape from resonance \citep{gold14}, but orbital resonances involving one or more giant planets tend to persist. Additionally, small, non-resonant planets might be able to form in situ nearby after the gas giant's migration. In contrast, tidal migration wipes out small planets inside the Jupiter's initial orbit (e.g., \citealt{must15}) but requires a stellar or giant planet companion to trigger the high eccentricity. If planet-planet scattering produces the high eccentricity, the companion could be ejected but for secular mechanisms, the companion should still be present and have the necessary properties to have triggered high eccentricity migration. In summary, we expect nearby planets in the in situ formation scenario, nearby resonant planets in the disk migration scenario, and distant companions in the high eccentricity migration scenario.

\subsubsection{Distant companions of hot Jupiters}
\label{subsecdist}
\begin{figure}
\includegraphics[width=5in]{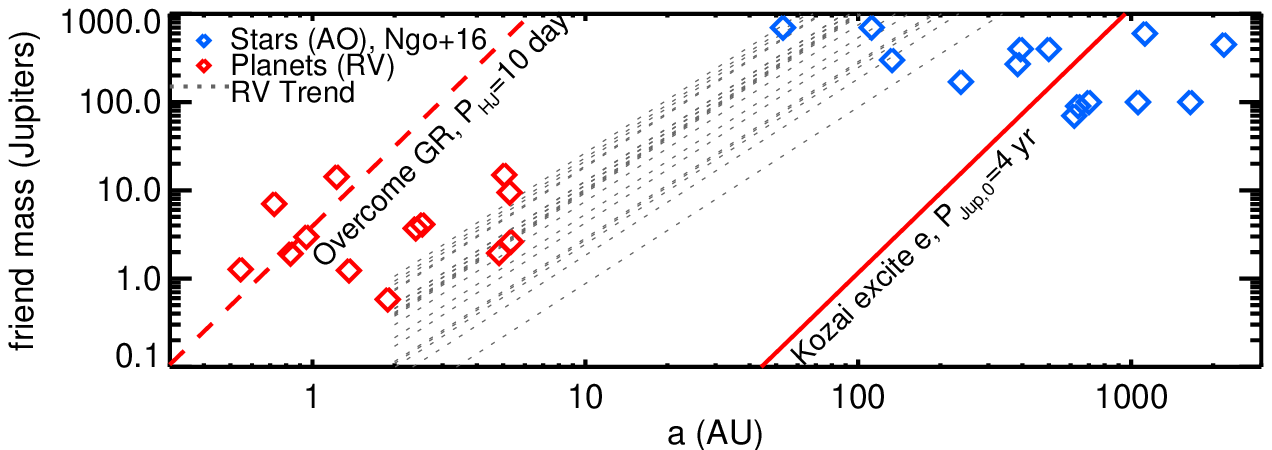}
\caption{Mass vs. semi-major axis of detected companions to hot Jupiters. Diamonds indicate hot Jupiter companions to detected via the radial velocity method (red; compiled from \citealt{knut14,wrig12}) or direct imaging (blue; \citealt{ngo16}). Dotted lines represent radial velocity trends. Companions to the top right the red dashed line are capable of overcoming GR precession to excite a hot Jupiter's eccentricity in situ at a 10 day orbital period (Eqn. \ref{eqn:grll}); those to the top right of the solid like are capable of exciting the eccentricity a proto-hot Jupiter on a 4 year orbital period high enough for the proto-hot Jupiter to circularize to a 5 day orbital period (Eqn. \ref{eqn:grkoz}). Most hot Jupiters do not have a companion capable of overcoming GR to raise their eccentricity through Kozai-Lidov at their present day short orbital periods (i.e., detection or upper limit is below red dashed line) or of triggering high eccentricity migration of a proto-hot Jupiter from beyond the ice line (i.e., points below the solid red line). See \S \ref{subsec:deco} for a discussion of other relevant timescales, such as tidal precession.
}
\label{fig:companions}
\end{figure}

Several follow-up studies of hot Jupiters have been conducted to search for both stellar and planetary companions (Fig. \ref{fig:companions}). The Friends of Hot Jupiters survey probed companions using long baselines of radial velocities \citep{knut14,brya16} sensitive to giant planet companions out to 10s of AU, infrared spectroscopy \citep{pisk15}, and direct imaging \citep{ngo16,ngo17} sensitive to low-mass stars at 10s to 1000s of AU. Other surveys for distant companions include \citet{endl14,evan16}, and \cite{neve16}. 

The properties of hot Jupiters' stellar companions are incompatible with high eccentricity migration triggered by stellar Kozai being the dominant channel of formation. \citet{ngo16} found an upper limit of 16$\pm 5\%$ on hot Jupiters with stellar companions capable of triggering high eccentricity tidal migration via Kozai--Lidov cycles (if the hot Jupiter formed between $\sim$1--5 AU). Most companions are not massive and nearby enough to overcome general relavistic (Eqn. \ref{eqn:grll}) and tidal precession. \citet{ngo16} find no correlation between binary companions and spin-orbit misalignments.  

In contrast, hot Jupiters' planet companions are generally nearby and massive enough to secularly trigger high eccentricity migration if they have the necessary eccentricity and/or mutual inclination. \citet{knut14,brya16} found that 70$\pm8\%$ of hot Jupiters have an outer planet between 1--20 AU and 1--13 Jupiter masses.  These outer planets are capable of having raised the eccentricities of proto-hot Jupiters formed ex situ between $\sim$1--5 AU, leading to high eccentricity tidal migration. How do these companions square with the population of eccentric hot Jupiters (\S\ref{subsec:ehj})? They are consistent with the theory that eccentric hot Jupiters are those finishing their high eccentricity migration. Most are not nearby and massive enough (Fig. \ref{fig:companions}) to having excited the hot Jupiter's eccentricity in situ (i.e., if the hot Jupiter formed in situ or underwent disk migration). No eccentric hot Jupiter has a \emph{known} companion -- or set of possible companions based on a linear radial-velocity trend -- capable of exciting the hot Jupiter's eccentricity in situ. However, for some eccentric hot Jupiters, the limits on companions are weak and more long baseline observations are needed to rule out the presence of such a companion.

\subsubsection{Nearby companions of hot Jupiters}

\begin{figure}
\includegraphics[width=5in]{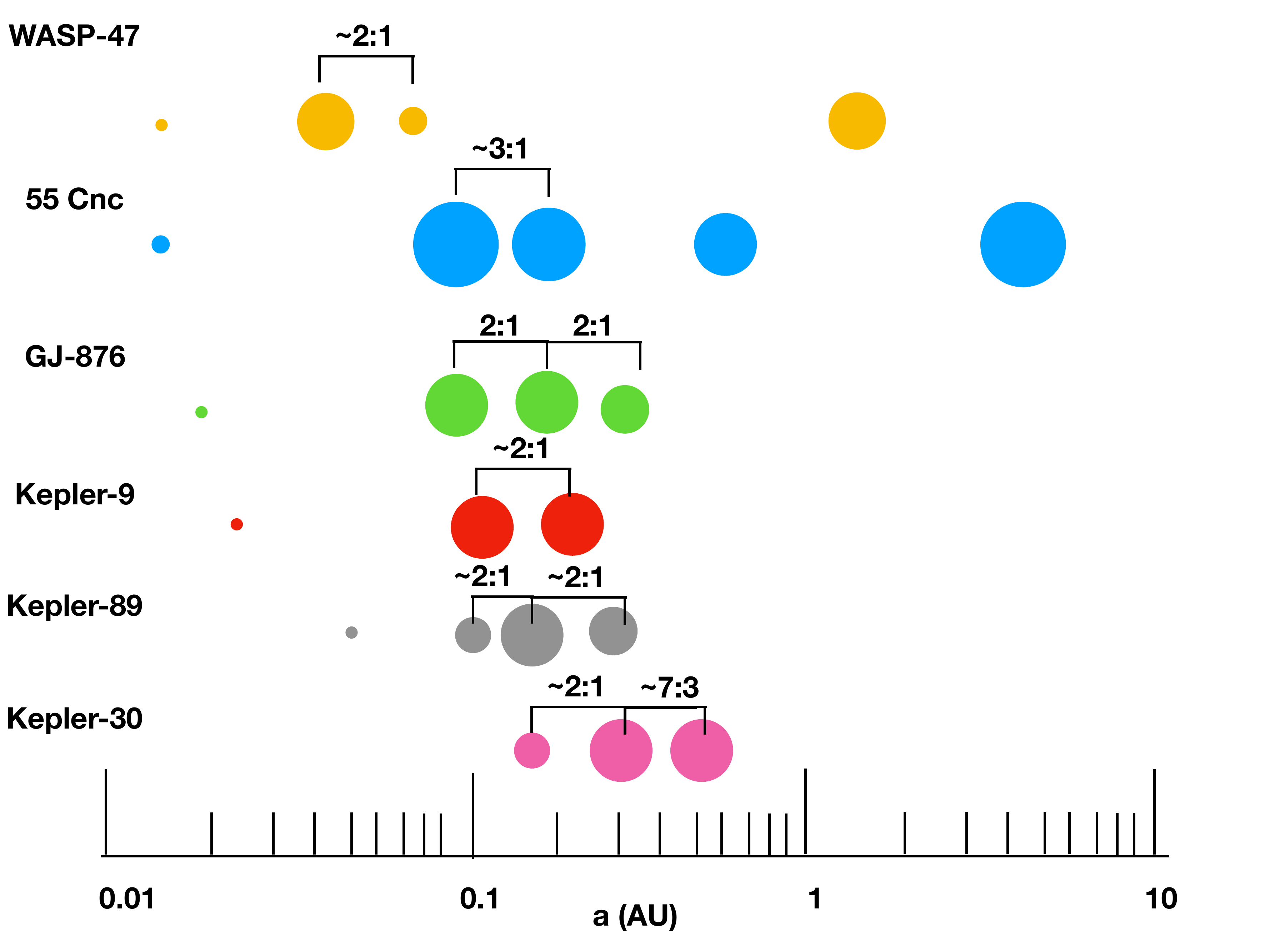}
\caption{Warm and Hot Jupiters for which the companions also have  semi-major axes less than 1 AU. Orbital resonances are labeled and sizes scale approximately with log planet mass. Hot Jupiter WASP-47b \citep{beck15} may be in the short period tail of a class of system featuring a close-in giant planet in orbital resonance with one or more neighbors. This architecture often also includes an ultra-short period super-Earth. 
 }
\label{fig:wasp}
\end{figure}

Nearby planets are generally absent in systems containing hot Jupiters. \citet{lath11} and \citet{huan16} found that hot Jupiters are much less likely to have other transiting planets in their systems. \citet{stef12} used the lack of transit timing variations of hot Jupiters to rule out nearby planets, even at low masses. However, there are exceptions to the trend that hot Jupiters lack nearby planets. \citet{mill16} found phase curve evidence for a non-transiting hot Jupiter in a system containing a 150 day orbital period candidate mini-Neptune. If confirmed, this system would be incompatible with high eccentricity migration of the hot Jupiter from beyond 1 AU (e.g., \citealt{must15}). The most striking exception is WASP-47b (Fig. \ref{fig:wasp}), a hot Jupiter near a 2:1 resonance with an outer Neptune \citep{beck15}. The WASP-47 system also has an ultra-short period super-Earth. WASP-47b bears resemblance to a number of systems featuring warm Jupiters with resonant neighbors, which are depicted in Fig. \ref{fig:wasp} and discussed further in \S \ref{subsec:warm}.

Although hot Jupiters' lack of nearby planets has generally been interpreted as evidence for high eccentricity migration, two sets of recent studies have argued that disk migration and in situ formation could be compatible with this observed trend. In the first set of studies, \citet{ogih13,ogih14} suggested that super-Earths cannot form near a hot Jupiter that migrated or formed in situ. Super-Earths would tend to drive the hot Jupiter into its star, provided that disk conditions did not enable hot Jupiters to open wide gaps. This theory does not predict a complete absence of nearby planets: low mass terrestrial planets could co-exist with the hot Jupiter. A more quantitative comparison of the surviving systems to the \citet{stef12} limits from transit timing variations would clarify whether low mass terrestrial planets would necessarily escape detection. Furthermore, in this scenario the hot Jupiters we observe today would be the exceptional survivors in systems where other planets never formed near the hot Jupiter or accompanied them on their migration. More work is necessary to explore whether a sizable population of hot Jupiters could form and remain lonely. 

In the second study, \citet{schl16} argued that the population statistics of hot Jupiter companions belay high eccentricity tidal migration from beyond the ice line. They found that the occurrence rate of longer period giant planet companions to hot Jupiters is consistent with that of companions to longer period Jupiters (periods $>$ 10 days). Their results hinge on two hot Jupiters with sub-AU companions discovered via radial velocity: upsilon Andromeda b (at 0.83 AU) and HIP 14810 (at 0.5454 AU). Companions in the \citet{schl16} sample are at $\sim$ 0.5--3 AU and are not incompatible with other studies investigating hot Jupiter companions at much closer \citep{lath11,stef12,huan16} or wider (e.g., Friends of Hot Jupiters) separations. \citet{stef12}, \citet{lath11}, and \citet{huan16} are sensitive to small, nearby planets that would not have been detectable by radial velocity. Conversely, the \citet{schl16} systems feature longer period planets that would be unlikely to transit and potentially too widely separated to cause TTVs. As part of the Friends of Hot Jupiters survey, \citet{brya16} found that the occurrence rate for hot Jupiter companions is larger than for warm Jupiter companions, but these companions are typically at wider separations than \citet{schl16}'s sample.

\citet{schl16}'s finding -- that hot Jupiters' companion rate is consistent with that of the overall population -- is not necessarily in tension with high eccentricity tidal migration. The efficiency of producing hot Jupiters through high eccentricity tidal migration is low and hot Jupiters are likely special outcomes of dynamical processes that affect many systems (\S \ref{subsec:occu}). Therefore we might expect the outer architectures of hot Jupiter systems to resemble those of giant planet systems without hot Jupiters. However, the question remains whether having a hot and warm Jupiter together in the same system is incompatible with high eccentricity tidal migration. Such an architecture would indeed seem to require that the hot Jupiter formed or migrated to the warm Jupiter region before the hot Jupiter high eccentricity tidal migration commenced; otherwise the warm Jupiter would be disturbed during hot Jupiter's tidal migration.

\subsubsection{Summary of hot Jupiters' companions}

Overall, the companions of hot Jupiters give most support to the origin hypothesis of high eccentricity tidal migration triggered by a planet companion. The majority of hot Jupiters are accompanied by long period planetary companions that are massive and nearby enough to have secularly triggered the hot Jupiter's high eccentricity tidal migration. (However, we do not yet know whether the companions have necessary eccentricities and inclinations. Also, such companions are not incompatible with disk migration or in situ formation.) In contrast, the majority of hot Jupiters are not accompanied by stellar companions capable of triggering their high eccentricity migration, implying that stellar Kozai triggered by high eccentricity migration is not a predominant channel. Most hot Jupiters are not accompanied by nearby planets, which we would expect to be present under the in situ formation and disk migration hypotheses and absent under the high eccentricity migration hypothesis. However, WASP-47b \citep{beck15} cannot be explained by high eccentricity tidal migration and would need to arrive via a different channel.

\subsection{Hot and Warm Jupiters}
\label{subsec:warm}

\emph{Warm} Jupiters are giant planets orbiting close to their star but at wider separations than \emph{hot} Jupiters. Their longer orbital periods, $\sim$10-200 days, made them challenging to discover in ground-based transit surveys (see \citealt{gaud05} for a discussion of ground-based surveys' selection effects), and hence they have received less attention than hot Jupiters.  Like hot Jupiters, warm Jupiters could have originated from in situ formation, disk migration, or high eccentricity migration. Also like hot Jupiters, their semi-major axes are too small for them to have been scattered directly from several AU (a factor of 10 change in energy; e.g., \citealt{dong14}). Fig. \ref{fig:evsa} shows how different mechanisms populate the warm Jupiter region. 

However, theoretical studies of the formation and orbital evolution of hot Jupiters have found it challenging to account for the occurrence rates, eccentricities, and other properties of warm Jupiters. This challenge may indicate a major problem in our understanding of hot Jupiters or that hot and warm Jupiters do not have a common origin. Distinguishing between these two possibilities is key to a complete understanding of hot Jupiters.

\subsubsection{Hot vs. Warm Jupiter occurrence rates}
\label{subsec:wjo}

The occurrence rate per log interval of giant planets dips in the warm Jupiter region by a factor of a few (Fig \ref{fig:evsa}). This feature is known as the Period Valley (\citealt{jone03,udry03,witt10,sant16}). See \citealt{witt10} for evidence of the statistical significance of the Period Valley, whose width, depth, and dependence on stellar properties are still being investigated (e.g., \citealt{sant16}). Although a Period Valley could be compatible with in situ formation if solids piled up in the innermost disk, we will discuss in \S\ref{subsec:small} that this explanation seems incompatible with super-Earth occurrence rates vs. orbital period. The Period Valley feature is plausibly compatible with disk migration. If a hot Jupiter's migration timescale is comparable to the disk lifetime after the hot Jupiter's formation, the Jupiter can get stranded mid-migration as a warm Jupiter. Therefore warm Jupiters are less common than hot or cold Jupiters because they require a special migration timescale (i.e., particular disk conditions). Another possibility is that photo-evaporation of the inner disk leaves later-forming giant planet stranded beyond $\sim 100$ days \citep{alex12}, unable to migrate. \citet{cole16} reproduced the period distribution of giant planets using a particular set of disk parameters. 

Although the occurrence rate of warm Jupiters per log interval is lower than for hot Jupiters, the total number of warm Jupiters is larger. Studies of high eccentricity tidal migration have severely under-produced warm Jupiters relative to hot Jupiters: in other words, they create a Period Valley that is far too deep. In the high eccentricity migration origins theory, warm Jupiters are proto-hot Jupiters in the midst of migration. Most high eccentricity migration mechanisms severely underproduce warm Jupiters (e.g., \citealt{wu11,beau12,petr14,petr15}). One way to enhance the number of warm Jupiters is to invoke perturbers massive and nearby enough to remain coupled to the proto-hot Jupiter (e.g., \citealt{dong14}). In this scenario, the proto-hot Jupiter's eccentricity can oscillate so that it spends less time at periapses small enough for effective tidal circularization, slowing down the migration. However, \citet{petr16} found that even this configuration leads to a ratio of warm Jupiters to hot Jupiters that is a factor of $\sim 5$ lower than observed. \citet{daws14b} found a dynamical configuration that can further extend the fraction of time spent as a warm Jupiter but did not quantify the likelihood of this configuration or its effect on the observed ratio of warm to hot Jupiters. The warm to hot Jupiter ratio is currently a major weakness in the high eccentricity tidal migration hypothesis, at least if it were the only channel.

\subsubsection{Warm Jupiters: too eccentric for comfort}
\label{subsec:circ}

Warm Jupiters have a wide range eccentricities. Their eccentricity distribution contains a low eccentricity component and a component with an approximately uniform distribution (e.g., \citealt{petr16}). The circular component, which we will discuss in \S\ref{subsec:supe}, is a challenge for the high eccentricity tidal migration hypothesis. The eccentric component is challenging for in situ formation or disk migration. Like for hot Jupiters (\S \ref{subsec:ehj}), in situ formation or disk migration leads to warm Jupiters on low eccentricity orbits that cannot be excited by subsequent scattering \citep{petr14}. \citet{ande17} advocate that warm Jupiters' eccentricities could be excited by an outer companion, which is feasible for warm Jupiters (e.g., \citealt{dong14}). However, \citet{ande17} find that this mechanism most easily explains moderately eccentric warm Jupiters, rather than high eccentricity warm Jupiters, because the fraction of time spent at high eccentricities for an initially circular warm Jupiter is low. \citet{must17} identified three additional mechanisms -- each of which involve several steps -- to generate warm Jupiters in systems of three or more planet. However, future work can quantify the efficiency of these more complex mechanisms.

\subsubsection{Warm Jupiters: too circular for comfort}
\label{subsec:supe}

Conversely, the high eccentricity tidal migration hypothesis has been unable to account for the low eccentricity component of the warm Jupiter population. Recall that under the high eccentricity migration hypothesis, all warm Jupiters are in the midst of tidal migration. If the perturber is decoupled from the migrating proto-hot Jupiter, its periapse should be close enough to the star for tides to operate (red region of Fig. \ref{fig:evsa}). With a couple exceptions, such as HD 80806b ($e=0.93$, \citealt{naef01,wu03}), elliptical warm Jupiters have eccentricities too high for in situ formation or disk migration (\S \ref{subsec:circ}) but too low to be undergoing tidal migration (i.e., as denoted by the white region in Fig. \ref{fig:evsa}).

\citet{dong14,daws14b}, and \citet{petr16} modeled warm Jupiters as undergoing tidal circularization but still coupled to a companion on a mutually inclined orbit. We observe warm Jupiters currently at eccentricities currently too low for tidal circularization but they periodically reach higher eccentricities. \citet{petr16} successfully reproduced the uniform eccentricity distribution of the eccentric component of warm Jupiters using planet-planet Kozai-Lidov driven high eccentricity tidal migration.

However, in most cases we do not have the constraints on the three-dimensional architectures of warm Jupiter systems necessary to test whether they have the requisite mutually inclined companions. Recently mutual inclinations have been measured for a handful of warm Jupiter systems using transit time and duration variations, and results have been mixed. \citet{masu17} found that the three-dimensional architecture of the Kepler-693 system is capable of driving high eccentricity migration. \citet{daws14} found that the massive nearby planet companion to eccentric (e=0.81; \citealt{daws12a}) warm Jupiter Kepler-419b is co-planar and not capable of driving high eccentricity tidal migration. \citet{daws14b} argued that half a dozen systems of warm Jupiters with nearby massive companions had the requisite mutual inclinations based on indirect evidence from clustering in the separations of their argument of periapse. There is currently debate over whether high eccentricity migration could have taken place in such systems. \citet{anto16,masu17} argue that rewinding the warm Jupiter's tidal migration to an initial semi-major axis $>$ 1 AU would result in a spacing too close to its companion for stability. In principle, such an instability may be what actually triggered the high eccentricity migration, but \citet{anto16} found in simulations that high eccentricity migration was an uncommon outcome of scattering in these systems.

Even if we can account for observed eccentricities too low for tidal circularization, a remaining problem is that the expected population of super-eccentric migrating proto-hot Jupiters is missing. Hot Jupiters are continuously being spawned throughout the galaxy as new stars are born, gas giants form, have their eccentricities excited, and tidally migrate. \citet{socr12} proposed that regardless of the mechanism of eccentricity excitation, we can test the origins theory of high eccentricity migration by searching for proto-hot Jupiters on highly elliptical orbits. If moderately eccentric hot Jupiters ($0.2<e<0.6$) (\S\ref{subsec:ehj}, Fig. \ref{fig:evsa}) are those finishing their tidal circularization, we expect the corresponding super-eccentric proto-hot Jupiters (i.e., those with $e>0.9$ and the same $a_{\rm final}$, ranging from 0.04--0.10 AU) dictated by the relative tidal evolution timescale. \citet{socr12} identified the Kepler sample (see \citealt{liss14} for a review of the Kepler Mission) as particularly well-suited for this search, because the transit probability of short and long period planets is the same for a given $a_{\rm final}$. Unlike most ground-based transit surveys, Kepler has the baseline and sensitivity to detect warm Jupiters. However, using an approach to identify super-eccentric Jupiters based on their transit shape and duration (e.g., \citealt{daws12}), \citet{daws15} found a paucity of super-eccentric proto-hot Jupiters in the Kepler sample, inconsistent with the expectation from high eccentricity tidal migration. \citet{wu17} recently proposed that f-mode tidal dissipation could cause hot Jupiters to migrate very quickly through the highly eccentric stage, causing us to miss super-eccentric Jupiters. However, this tidal dissipation mechanism requires $a{\rm final} \lesssim 0.04 $ AU and therefore can likely only account for a fraction of the missing super-eccentric Jupiters.

\subsubsection{Stellar obliquities of warm Jupiters' host stars}
\label{subsec:oblw}
If warm Jupiters have the same origin as hot Jupiters, the distribution of stellar obliquities of warm Jupiters' host stars could serve as a primordial distribution, unsculpted by tides raised on the star. Fig. \ref{fig:obl} depicts the processes affecting hot Jupiters' spin-orbit alignments. For warm Jupiters, we can escape the final confusing step in which the star can get realigned, erasing the distribution sculpted by hot Jupiters' origins mechanism. (Recall that tidally migrating warm Jupiters are experiencing tides raised on the planet. Tides raised on the star, sensitive to $a$ rather than $a(1-e^2)$, likely only become important when hot Jupiter-hood is achieved.)

However, the measurement and interpretation of warm Jupiters' stellar obliquities is an outstanding problem requiring a larger observational sample size. A few preliminary results have been ambiguous to interpret. \citet{li16} find that planets are more misaligned at longer orbital periods, consistent with the idea that the primordial distribution has been sculpted by tides. However, the trend of increasing misalignment vs. orbital period persists out to 50 days, a much larger distance than we expect tides to operate. Kepler-56, which hosts two transiting giant planets, is a misaligned cool star \citep{hube13}. However, the misalignment may be caused by a companion torquing the inner pair (e.g., \citep{li14c,otor16}), rather than a primordial disk misalignment. 

\subsubsection{Companions to warm Jupiters} 
Warm Jupiters have different companions than hot Jupiters, which may suggest they have a different origins channel. \citet{huan16} discovered a population of warm Jupiters with nearby super-Earths, which are incompatible with high eccentricity tidal migration \citep{must15}, in the Kepler sample. These systems exhibit a strong disparity from hot Jupiters, which generally lack nearby companions (\citealt{huan16}; see also \S \ref{subsec:comp}). Exceptional hot Jupiter systems like WASP-47 containing nearby planets (\S \ref{subsec:comp}, \citealt{beck15}) may be tail end members of this warm Jupiter population. Regarding longer period companions, \citet{brya16} found the companion rate to be lower for warm Jupiters than for hot Jupiters ($49\pm10\%$ vs. $75\pm5\%$). If these companions are necessary for high eccentricity migration, this finding may imply that fewer warm Jupiters came through this origins channel than hot Jupiters.

\subsubsection{Warm Jupiters: multiple origins channels?}

None of the three origins channels alone can account for all of warm Jupiters' observed properties. One promising option is to invoke
two origins channels, one for circular warm Jupiters and one for eccentric warm Jupiters. Distinctions in other properties between eccentric vs. circular warm Jupiters support this hypothesis. Low eccentricity warm Jupiters are less likely to have giant planet companions \citep{dong14,brya16} and warm Jupiters that have very nearby companions incompatible with tidal migration have lower eccentricities \citep{daws15}. Warm Jupiters orbiting metal-poor stars are confined to low eccentricities while those orbiting metal-rich stars exhibit a range of eccentricities (Fig. \ref{fig:evsa}, \citealt{daws13}; as discussed in \S\ref{subsec:ehj}). This trend may link the origins of eccentric warm Jupiters to planet-planet gravitational interactions.  Moreover, \citet{daws13} found that moderately eccentric \emph{hot} Jupiters orbit higher metallicity stars. If high eccentricity tidal migration is correlated with host star metallicity, the Kepler sample's lower metallicity may translate to fewer hot Jupiters originating from high eccentricity migration. A lack of high eccentricity migration could contribute to both the overall lower occurrence rate of hot Jupiters in the Kepler sample (\S \ref{subsec:stel}; but not entirely account for, see \citealt{guo17}) and the lack of super-eccentric Jupiters (\S \ref{subsec:supe}).
 
\subsection{Comparing and connecting hot Jupiters to smaller planets}
\label{subsec:small}

In the early 2000s, radial-velocity techniques reached the precision necessary to discover small planets (e.g., \citealt{sant04b}). In the 2010s, radial-velocity surveys amassed large enoughs sample to statistically characterize the properties of small planets (e.g., \citealt{howa10,mayo11}) and the Kepler Mission discovered an abundance of small transiting planets (see \citealt{liss14} for a review of Kepler results). Comparing hot Jupiters to smaller planets is another possible avenue to test theories for the origins of hot Jupiters. All three origins channels can produce hot planets of various masses and sizes, so whatever channel operates likely produces at least some hot super-Earths and Neptunes. However, the predominant origin channel for hot super-Earths and Neptunes may be different than for hot Jupiters. Here we review how different hot Jupiter origins channel contribute smaller planets and if and how we can compare the properties of hot Jupiters vs. smaller planets to test theories for hot Jupiters' origins.

\subsubsection{Small planet occurrence rates}

There are a number of features of super-Earth occurrence rates that are relevant to possible shared origins channels (or lack thereof) with hot Jupiters:

\begin{itemize}
    \item Super-Earths are more common than giant planets. $\sim$40\% of stars have at least one super-Earth within a fifty day orbital period \citep{howa10}, whereas only $\sim 10\%$ of stars have giant planets at \emph{any} orbital period (e.g., \citealt{cumm08,zeic13}).
    \item Super-Earths and Neptunes' occurrence rate is constant beyond ten day orbital periods and drops within 10 days (see \citealt{lee17} and references therein), whereas the giant planet occurrence rate drops within $\sim$ 3 days (\S \ref{subsec:semi}) and also exhibits a Valley from $\sim$ 10--100 days (\S \ref{subsec:wjo}). 
    \item The majority of super-Earths and Neptunes at sub year orbital periods are in multi-planet systems (e.g., \citealt{liss14,ball16,daws16}). However, hot Neptunes (2--6 Earth radii, $P<10$ days), recently termed Hoptunes, are most commonly in single transiting systems \citep{dong17}. This distinction between warm Neptunes and Hoptunes is reminiscent of that between warm Jupiters (accompanied by small nearby planets) vs. hot Jupiters (rarely accompanied by small nearby planets).
    \item The occurrence rates and host-star metallicity dependence of Hoptunes is similar to hot Jupiters \citep{dong17}; smaller planets at $<10$ day orbital periods are more common and less dependent on host star metallicity (e.g., \citealt{buch12}).
    \item The lack of Hoptunes orbiting M-dwarfs (e.g., \citealt{dres15}) also links them to hot Jupiters.
\end{itemize}
We now review how each of these features squares with different origins theories.

Under the in situ formation hypothesis, hot super-Earths form like hot Jupiters but fail to achieve runaway gas accretion. \citet{lee14} and \citet{lee16} considered why super-Earths formed in situ may fail to grow into hot Jupiters. They argued that if the local solid surface density is not high enough to form these bodies as isolation masses (\S\ref{subsec:form}), they may need to wait to grow through giant impacts when the supply of gas is nearly depleted. This explanation jives with the overall higher occurrence rate of super-Earths (vs. giant planets) across all orbital periods. This explanation also accounts for the metallicity dependence of gas giants (\S \ref{subsec:stel}) and of hot Neptunes with low mass gas envelopes (e.g., \citealt{daws15b}). However, no explanation has yet been posed for super-Earths' period cliff vs. giant planets' period valley. Nor does in situ formation satisfactorily explain the difference in nearby companions between hot vs. warm Jupiters.

Disk migration can deliver small planets to short orbital periods. Disk migration occurs more quickly for super-Earths, which (except in very low viscosity disks) are not able to open gaps that slow migration. Under the disk migration hypothesis, the overall higher occurrence rates of small planets and their metallicity dependence would reflect the efficiency of their formation ex situ. However, disk migration is difficult to reconcile with super-Earths' flat period distribution beyond 10 days (e.g., \citealt{lee17}). Early population studies of planetary migration predicted a planet desert of warm super-Earth/Neptunes (e.g., \citealt{ida08}), at odds with the observed high occurrence rate.

A key challenge for accounting for hot super-Earths and Neptunes via high eccentricity tidal migration is whether planetary systems have the requisite initial conditions to raise the eccentricities of these small bodies. If outer systems of smaller planets resemble inner ones, they are multi-planet systems of similar mass/size planets \citep{ciar13}. Close encounters among small planets lead to collisions rather than eccentricity excitation, even out to several AU (eqn. \ref{eqn:vesce}), making planet-planet scattering ineffective at generating highly elliptical orbits. Moreover, super-Earths in multi-planet systems keep each other safe from the secular eccentricity excitation by binary stars and giant planets. Secular chaos is a possible mechanism to generate large eccentricities but has not been explored in detail for systems of small planets.

However, although disk migration and tidal migration seem unlikely to be making a substantial contribution to the small planet population at large, they arguably play a role in Hoptunes' origins. Hoptunes are in danger of losing their atmospheres (e.g., \citealt{lope12}) while their stars are young and active. Young stars produce a relatively large amount UV and X-ray flux, which can photo-evaporate low mass planets' hydrogen and helium atmospheres. (Note: Hoptunes with close to 10 day orbital periods could potentially retain their hydrogen and helium atmospheres if their cores are sufficiently massive, e.g., \citealt{lope13}. However, none of the $\sim 10$ Hoptunes with mass measurements has a sufficiently high mass or long orbital period.) For smaller Hoptunes less than $4 R_\oplus$ (e.g., \citealt{lope14}), formation ex situ could endow Hoptunes with steam atmospheres that are less susceptible to photo-evaporation. Furthermore, high eccentricity tidal migration can occur later in the system's history, delivering Hoptunes after the star's $\sim 100$ Myr active stage. 

\subsubsection{Stellar obliquities of small planet host stars}
\label{subsec:obls}

Measuring stellar obliquities of small planets for comparison to hot Jupiters can potentially distinguish which physical processes are predominantly shaping the stellar obliquitities of hot Jupiter hosts (\S\ref{subsec:obli}; Fig.  \ref{fig:obl}). If spin-orbit misalignments are primarily caused by primordial misalignment of the disk (e.g., \citealt{baty12}), we would expect small planets and flat multi-planet systems to be misaligned around hot stars as well. Like warm Jupiters (\S\ref{subsec:oblw}), small planets are not effective at raising tides on their stars and we can study their obliquity distribution as one unsculpted by tidal realignment. In a study using the five compact multi-planet systems at the time with known obliquities, \citet{albr13} found that all are aligned with their star and the collection is inconsistent with being drawn from an isotropic obliquity distribution. However, all five stars had temperatures consistent with $<6250$ K. Most recently, \citep{winn17b}) found that most host stars hosting planets have low obliquities. Of the six possible misaligments they found, half were hot Jupiter hosts, whereas the vast majority of the sample hosted small planets.  This evidence supports the interpretation that hot Jupiters were misaligned after the gas disk stage and counters the interpretation that hot stars have high obliquitites independent of hot Jupiters.  The TESS \citep{rick15}, CHEOPS \citep{broe13}, and PLATO \citep{raue14} samples will provide more opportunities to compare obliquities of hosts of hot Jupiters vs. small planets (e.g., \citealt{quin16}).

\subsubsection{Ultra-short Period Planets}
\label{subsec:usp}

Ultra-short period planets (USPs) are planets on sub-day orbital periods. The overall occurrence rate of USPs is 0.5\% for USPs greater than 0.84 $R_\oplus$\citep{sanc14}; the majority of these USPs are Earths (0.84--1.25 $R_\oplus$), rather than super-Earths. USPS tend reside in compact multi-planet systems \citep{sanc14}. Some may be tidally stripped remnants of hot Jupiters (e.g., \citealt{vals15,jack16}). See Fig. 5 of \citet{vals15} for examples of planet trajectories in and out of the Roche limit as they evolve from hot Jupiters to super-Earths. However, most USPs are likely not tidally stripped Jupiters because, unlike most hot Jupiters, they reside in in compact multi-planet systems \citep{sanc14} and they lack hot Jupiters' host star metallicity dependence \citep{winn17}. \citet{lee17} showed that USPs can be accounted for by tidal evolution of compact multi-planet systems. All these lines of evidence suggest that the majority of USPs may have a different origin than hot Jupiters.

However, one observed feature does link USPs to giant planets: most of the special systems, like WASP-47, containing a hot or warm giant planet in orbital resonance with a neighbor also contain a USP (Fig. \ref{fig:wasp}). Perhaps this special type of giant planet system shares an origin with the more common compact, small multi-planet systems.

\subsubsection{Summary: small planets and hot Jupiters}

The links between small planets and hot Jupiters are still uncertain. However, the following possibility seems consistent with all the evidence. Perhaps the majority of super-Earths form in situ but some migration channel (disk or tidal) delivers some hot Jupiters and some planets as well. Perhaps these small planets generally blend in with the most abundant in situ super-Earth populations, except the distinctive Hoptunes \citep{dong17}, whose gas envelopes are unlikely to survive if Hoptunes formed in situ.

\section{SUMMARY AND DISCUSSION}

Despite thousands of observational and theoretical studies of hot Jupiters over the past twenty years, we still have no consensus on the predominant channel for their origin. However, in this review, we have attempted to demonstrate that our community's understanding of their origin has advanced from speculation and post-dictions to detailed comparisons between observations and theories and to testable predictions for upcoming missions and ongoing surveys. In \S 2, we described the three origins hypotheses for hot Jupiters: in situ formation, disk migration, and tidal migration. In \S 3, we synthesized which observed properties of hot Jupiters were consistent or inconsistent with these hypotheses. In \S 4, we summarized how connections between hot Jupiters and other exoplanet populations provide tests of hot Jupiters' origins. We tabulated which hot Jupiter properties the three origins hypotheses explain or fail to explain in Table \ref{tab:evid}. 

Although no hypothesis for hot Jupiters' origins can explain all the evidence, each piece of evidence is explained by at least one origin channel. Throughout the review, we have emphasized the power of two commonly operating origins channels to account for the diversity of hot Jupiter properties. High eccentricity tidal migration triggered by planet-planet Kozai-Lidov cycles is a strong contender for one of the two most prevalent origins channels (\S \ref{subsec:hem}). The supporting evidence is as follows. (We group related lines of evidence together, so the following list is not ordered by strength of evidence.)
\begin{itemize}
    \item Tidal migration accounts for a lack small planets near most hot Jupiters (\S\ref{subsec:comp}). 
    \item Tidal migration accounts for the similarities in occurrence rates, multiplicity, and stellar metallicity dependence between hot Jupiters and Hoptunes (\S \ref{subsec:small}).
    \item Moderately eccentric hot Jupiters cannot be explained by in situ formation or disk migration; they are most consistent with tidal migration (\S \ref{subsec:ehj}).
    \item These eccentric hot Jupiters orbit metal-rich stars, implicating planet-planet interactions in their origins (\S\ref{subsec:ehj}), rather than planet-stellar Kozai, because giant planet formation is strongly correlated with stellar metallicity but stellar multiplicity is not.
    \item Most hot Jupiters have long period giant planet companions capable of driving Kozai-Lidov cycles but few have stellar binary companions that are capable (\S\ref{subsec:comp}).
    \item Planet-planet Kozai is the only mechanism shown to produce a sizable population of eccentric \emph{warm} Jupiters (\S\ref{subsec:warm}). Moreover, these eccentric warm Jupiters orbit metal-rich stars (\S\ref{subsec:warm}) and are more likely to have outer giant planet companions (\S\ref{subsec:comp}), implicating planet-planet interactions.
\end{itemize}

We note that although we would summarize this type of dynamical history as tidal migration triggered by planet-planet Kozai (e.g., \citealt{naoz11,naga11,petr16}; \S \ref{subsec:hem}), other mechanisms may play important supporting roles. For example, planet-planet scattering may establish the requisite mutual inclinations for planet-planet Kozai, or disk migration may deliver the planets inside the ice line, if they begin their journey inside where they formed. We also caution that this hypothesis is our best attempt to synthesize the state of the field, rather than the universal consensus of the community. A major open question is whether giant planet systems have the mutual inclinations necessary for this mechanism to operate. Another open question is whether proto-hot Jupiters had nearby small planets at their formation locations that would quench secular interactions with more distant giant planets (though these may have been dislodged during the planet-planet scattering process).

That idea that two origins channels are prevalent has early roots in the distribution of hot Jupiter host star obliquities. \citet{fabr09} and subsequent studies identified two components to the obliquity distribution, one consisting of low obliquity (well-aligned systems) and another component with a broad distribution. The interpretation of hot Jupiter host star obliquities is currently ambiguous and no longer definitively linked to two origins channels (\S \ref{subsec:obli},\ref{subsec:oblw}, \ref{subsec:obls}). Instead, we invoke a second origins channel -- i.e., to supplement tidal migration triggered by planet-planet Kozai -- on following evidence:
\begin{itemize}
    \item The eccentric hot Jupiters orbiting metal rich stars are in contrast to the mixture of metal-rich and metal-poor stars hosting hot Jupiters (\S\ref{subsec:ehj}). This mixture of host star metallicities for hot Jupiters implies another origins channel that produces circular hot Jupiters orbiting lower metallicity stars.
    \item Two origins channels, correlated with host star metallicity, may contribute to the Kepler's sample lack of super-eccentric proto-hot Jupiters (i.e., generated from high eccentricity tidal migration) and overall lack of hot Jupiters  (\S\ref{subsec:supe}) .
    \item The high eccentricity tidal migration scenarios are not sufficiently efficient at producing hot Jupiters at the observed rates (\S \ref{subsec:occu}).
    \item High eccentricity tidal migration cannot easily account for circular warm Jupiters, which are also correlated with lower host star metallicity (\S \ref{subsec:warm}).
     \item High eccentricity tidal migration cannot account for warm Jupiters with nearby companions (\S \ref{subsec:warm}) or rare hot Jupiters with nearby companions, like WASP-47b (\citealt{beck15}, \S \ref{subsec:comp}).
    \item High eccentricity tidal migration cannot easily account for recent discoveries of hot Jupiters orbiting T Tauri stars (\S\ref{subsec:sage}).
    \item A number of hot Jupiters are observed on orbits within twice the Roche limit (\S \ref{subsec:semi}), where we don't expect high-eccentricity tidal migration to deliver them. A second channel could deliver hot Jupiters to $1--2 a_{\rm Roche}$ whether requiring subsequent evolution via tides raised on the star.
    \item \citet{beck17} recently found that detected planet companions in six hot Jupiters systems cannot have sufficient mutual inclination to have driven Kozai-Lidov high eccentricity tidal migration of the hot Jupiters.
\end{itemize}

The second origins channel could be either disk migration or in situ formation. For assessing both theories, the dominant uncertainty is whether disk conditions are suitable to produce the observed properties of hot Jupiters and their links to other planet populations. In situ formation, not feasible through gravitational instability, requires a local huge disk solid surface density for core accretion (\S \ref{subsec:form}). The solids likely need to be transported from further out in the disk. Disk migration has been more successful in accounting for the observed giant planet period distribution (\S \ref{subsec:occu}), including the Period Valley (\S\ref{subsec:wjo}), but requires tuned disk conditions. The inner limit of the hot Jupiter period distribution is nominally more consistent with disk migration (\S \ref{subsec:semi}), but subsequent tidal decay could deliver planets formed in situ at the disk edge to shorter orbital periods (\S \ref{subsec:semi}) for certain stellar tidal parameters. \citet{huan16} argued that nearby small planet companions of warm Jupiters are more consistent with in situ formation, but forming the nearby planets in situ following disk migration may be a plausible alternative. Better constraints on disk conditions and more detailed head-to-head comparisons between the period distribution and warm Jupiters' nearby companions would help distinguish whether situ formation vs. disk migration is the predominant second channel.

Kozai oscillations driven by a wide stellar companion, followed by tidal circularization, has been extensively investigated in the origins of hot Jupiters (e.g., \citealt{wu03,fabr07}). However, our interpretation of the current state of the field is that it cannot be a predominant channel, though it could still be operating in a handful of individual systems. Consistent with this mechanism not playing a major role in shaping the architectures of planetary systems, \citet{ngo17} found no significant differences in the mass, orbital eccentricity, and semi-major axis distribution of the innermost planet in the multi-stellar systems vs. single-star systems.

To resolve the outstanding issues of hot Jupiters' origins, we recommend follow-up observational and theoretical studies related to the following issues:

\begin{issues}[FUTURE ISSUES]
\begin{enumerate}
\item Eccentric hot Jupiters: several lines of evidence hinge on these planets, motivating continued, intensive radial-velocity follow up to map out the architecture of their systems. Do they all have giant planet companions that were capable of triggering their high eccentricity migration? And if so, were they protected from its influence by smaller planets in the system? Can we rule out for each eccentric hot Jupiter that a nearby eccentric giant planet excited their eccentricity in situ?
\item Many transiting hot Jupiters have limited radial-velocity follow-up designed only to confirm them as planets, not to constrain their eccentricities. Continued radial-velocity follow-up -- e.g., as pursued by the HARPS GAPS program (e.g., \citealt{bono17}) -- could increase the sample of valuable eccentric warm Jupiters. 
\item For hot Jupiters with giant planet companions, Gaia astrometry may diagnose whether the giant planet companion has the necessary mutual inclination to trigger high eccentricity migration using Kozai-Lidov cycles \citep{case08}. We recommend similar astrometric follow-up for companions of eccentric warm Jupiters.
\item Constraining the mutual inclination distribution of giant planets across all orbital periods will address the applicability of planet-planet Kozai. In addition to Gaia astrometry \citep{case08}, constraints from transit timing and duration variations of transiting Jupiters can contribute.
\item Theoretical and observational investigations should thoroughly constrain how often intervening small planets interfere with secular interactions between giant planets.
\item The biggest outstanding question for in situ formation is whether the requisite disk conditions are seen in nature. We can address this question with further observational and theoretical studies of solid transport within gas disks.
\item More theoretical studies are needed to address whether hot Jupiters' atmospheric properties can feasibly trace formation location or whether the diversity of disk conditions make this infeasible even for a large sample.
\item A larger sample of obliquitities of warm Jupiters hosts and small planets hosts will inform us of the extent to which the primordial distribution of hot Jupiter hosts has been altered by tides.
\item To harness the power of host star ages to distinguish among origins hypotheses (\S \ref{subsec:sage}), we need a large sample of stellar ages, ideally extending to very young ages. Gaia \citep{gaia16}, TESS \citep{camp16}, and PLATO \citep{raue14} will enlarge the sample of stellar age estimates, and ongoing surveys of T Tauri stars (e.g., \citealt{yu17b}) can assess how often hot Jupiters achieve their short orbital periods during the gas disk stage.
\end{enumerate}
\end{issues}

\section{DISCLOSURE STATEMENT}

The authors are not aware of any affiliations, sponsorships, funding, or financial holdings that might be perceived as affecting the objectivity of this review.

\section{ACKNOWLEDGMENTS}
We thank Eric Ford, Sivan Ginzburg, Chelsea Huang, Taisiya Kopytova, David Latham, Eve Lee, Gongjie Li, Henry Ngo, Ana-Maria Piso, Johanna Teske, and Yanqin Wu for helpful conversations. We are grateful to Simon Albrecht, Thomas Beatty, Chelsea Huang, Eve Lee, Smadar Naoz, Cristobal Petrovich, and Christopher Spalding for insightful comments on a manuscript draft. We thank Ewine van Dishoeck and two anonymous reviewers for helpful comments that improved the review. We thank Jose Manuel Almenara, Francesca Faedi, Michael Gillon, Guillaume H\'ebrard, and Rachel Street for providing us with upper limits on the eccentricities of certain hot Jupiters based on their published fits. This work was partially supported by funding from the NASA Exoplanets Research program (NNX16AB50G). The Center for Exoplanets and Habitable Worlds is supported by the Pennsylvania State University, the Eberly College of Science, and the Pennsylvania Space Grant Consortium.

\bibliography{ms.bib}

\begin{thebibliography}{}
\expandafter\ifx\csname natexlab\endcsname\relax\def\natexlab#1{#1}\fi

\bibitem[{{Albrecht} et~al.(2012){Albrecht}, {Winn}, {Johnson}, {Howard},
  {Marcy} et~al.}]{albr12}
{Albrecht} S, {Winn} JN, {Johnson} JA, {Howard} AW, {Marcy} GW, et~al. 2012.
\newblock \textit{\apj} 757:18

\bibitem[{{Albrecht} et~al.(2013){Albrecht}, {Winn}, {Marcy}, {Howard},
  {Isaacson} \& {Johnson}}]{albr13}
{Albrecht} S, {Winn} JN, {Marcy} GW, {Howard} AW, {Isaacson} H, {Johnson} JA.
  2013.
\newblock \textit{\apj} 771:11

\bibitem[{{Alexander} \& {Pascucci}(2012)}]{alex12}
{Alexander} RD, {Pascucci} I. 2012.
\newblock \textit{\mnras} 422:L82--L86

\bibitem[{{Alibert} et~al.(2005){Alibert}, {Mordasini}, {Benz} \&
  {Winisdoerffer}}]{alib05}
{Alibert} Y, {Mordasini} C, {Benz} W, {Winisdoerffer} C. 2005.
\newblock \textit{\aap} 434:343--353

\bibitem[{{Anderson} \& {Lai}(2017)}]{ande17}
{Anderson} KR, {Lai} D. 2017.
\newblock \textit{ArXiv e-prints} arXiv:1706.00084

\bibitem[{{Anderson}, {Storch} \& {Lai}(2016)}]{ande16}
{Anderson} KR, {Storch} NI, {Lai} D. 2016.
\newblock \textit{\mnras} 456:3671--3701

\bibitem[{{Andrews}(2015)}]{andr15}
{Andrews} SM. 2015.
\newblock \textit{\pasp} 127:961--993

\bibitem[{{Ansdell} et~al.(2017){Ansdell}, {Williams}, {Manara}, {Miotello},
  {Facchini} et~al.}]{ansd17}
{Ansdell} M, {Williams} JP, {Manara} CF, {Miotello} A, {Facchini} S, et~al.
  2017.
\newblock \textit{\aj} 153:240

\bibitem[{{Ansdell} et~al.(2016){Ansdell}, {Williams}, {van der Marel},
  {Carpenter}, {Guidi} et~al.}]{ansd16}
{Ansdell} M, {Williams} JP, {van der Marel} N, {Carpenter} JM, {Guidi} G,
  et~al. 2016.
\newblock \textit{\apj} 828:46

\bibitem[{{Antognini}(2015)}]{anto15}
{Antognini} JMO. 2015.
\newblock \textit{\mnras} 452:3610--3619

\bibitem[{{Antonini}, {Hamers} \& {Lithwick}(2016)}]{anto16}
{Antonini} F, {Hamers} AS, {Lithwick} Y. 2016.
\newblock \textit{\aj} 152:174

\bibitem[{{Armitage}(2013)}]{armi13}
{Armitage} PJ. 2013.
\newblock \textit{{Astrophysics of Planet Formation}}

\bibitem[{{Arras} \& {Socrates}(2010)}]{arra10}
{Arras} P, {Socrates} A. 2010.
\newblock \textit{\apj} 714:1--12

\bibitem[{{Ballard} \& {Johnson}(2016)}]{ball16}
{Ballard} S, {Johnson} JA. 2016.
\newblock \textit{\apj} 816:66

\bibitem[{{Barenfeld} et~al.(2016){Barenfeld}, {Carpenter}, {Ricci} \&
  {Isella}}]{bare16}
{Barenfeld} SA, {Carpenter} JM, {Ricci} L, {Isella} A. 2016.
\newblock \textit{\apj} 827:142

\bibitem[{{Barker}(2016)}]{bark16a}
{Barker} AJ. 2016.
\newblock \textit{\mnras} 460:2339--2350

\bibitem[{{Barnes}(2009)}]{barn09}
{Barnes} JW. 2009.
\newblock \textit{\apj} 705:683--692

\bibitem[{{Baruteau} et~al.(2014){Baruteau}, {Crida}, {Paardekooper}, {Masset},
  {Guilet} et~al.}]{baru14}
{Baruteau} C, {Crida} A, {Paardekooper} SJ, {Masset} F, {Guilet} J, et~al.
  2014.
\newblock \textit{Protostars and Planets VI} :667--689

\bibitem[{{Batygin}(2012)}]{baty12}
{Batygin} K. 2012.
\newblock \textit{\nat} 491:418--420

\bibitem[{{Batygin}, {Bodenheimer} \& {Laughlin}(2016)}]{baty15}
{Batygin} K, {Bodenheimer} PH, {Laughlin} GP. 2016.
\newblock \textit{\apj} 829:114

\bibitem[{{Batygin} \& {Stevenson}(2010)}]{baty10}
{Batygin} K, {Stevenson} DJ. 2010.
\newblock \textit{\apjl} 714:L238--L243

\bibitem[{{Beaug{\'e}} \& {Nesvorn{\'y}}(2012)}]{beau12}
{Beaug{\'e}} C, {Nesvorn{\'y}} D. 2012.
\newblock \textit{\apj} 751:119

\bibitem[{{Becker} et~al.(2017){Becker}, {Vanderburg}, {Adams}, {Khain} \&
  {Bryan}}]{beck17}
{Becker} JC, {Vanderburg} A, {Adams} FC, {Khain} T, {Bryan} M. 2017.
\newblock \textit{\aj} 154:230

\bibitem[{{Becker} et~al.(2015){Becker}, {Vanderburg}, {Adams}, {Rappaport} \&
  {Schwengeler}}]{beck15}
{Becker} JC, {Vanderburg} A, {Adams} FC, {Rappaport} SA, {Schwengeler} HM.
  2015.
\newblock \textit{\apjl} 812:L18

\bibitem[{{Bitsch} et~al.(2013){Bitsch}, {Crida}, {Libert} \& {Lega}}]{bits13}
{Bitsch} B, {Crida} A, {Libert} AS, {Lega} E. 2013.
\newblock \textit{\aap} 555:A124

\bibitem[{{Bodenheimer}, {Lin} \& {Mardling}(2001)}]{bode01}
{Bodenheimer} P, {Lin} DNC, {Mardling} RA. 2001.
\newblock \textit{\apj} 548:466--472

\bibitem[{{Boley}, {Granados Contreras} \& {Gladman}(2016)}]{bole16}
{Boley} AC, {Granados Contreras} AP, {Gladman} B. 2016.
\newblock \textit{\apjl} 817:L17

\bibitem[{{Bonomo} et~al.(2017){Bonomo}, {Desidera}, {Benatti}, {Borsa},
  {Crespi} et~al.}]{bono17}
{Bonomo} AS, {Desidera} S, {Benatti} S, {Borsa} F, {Crespi} S, et~al. 2017.
\newblock \textit{\aap} 602:A107

\bibitem[{{Boss}(1997)}]{boss97}
{Boss} AP. 1997.
\newblock \textit{Science} 276:1836--1839

\bibitem[{{Broeg} et~al.(2013){Broeg}, {Fortier}, {Ehrenreich}, {Alibert},
  {Baumjohann} et~al.}]{broe13}
{Broeg} C, {Fortier} A, {Ehrenreich} D, {Alibert} Y, {Baumjohann} W, et~al.
  2013.
\newblock In \textit{European Physical Journal Web of Conferences}, vol.~47 of
  \textit{European Physical Journal Web of Conferences}

\bibitem[{{Brucalassi} et~al.(2017){Brucalassi}, {Koppenhoefer}, {Saglia},
  {Pasquini}, {Ruiz} et~al.}]{bruc17}
{Brucalassi} A, {Koppenhoefer} J, {Saglia} R, {Pasquini} L, {Ruiz} MT, et~al.
  2017.
\newblock \textit{\aap} 603:A85

\bibitem[{{Bryan} et~al.(2016){Bryan}, {Knutson}, {Howard}, {Ngo}, {Batygin}
  et~al.}]{brya16}
{Bryan} ML, {Knutson} HA, {Howard} AW, {Ngo} H, {Batygin} K, et~al. 2016.
\newblock \textit{\apj} 821:89

\bibitem[{{Buchhave} et~al.(2012){Buchhave}, {Latham}, {Johansen}, {Bizzarro},
  {Torres} et~al.}]{buch12}
{Buchhave} LA, {Latham} DW, {Johansen} A, {Bizzarro} M, {Torres} G, et~al.
  2012.
\newblock \textit{\nat} 486:375--377

\bibitem[{{Buhler} et~al.(2016){Buhler}, {Knutson}, {Batygin}, {Fulton},
  {Fortney} et~al.}]{buhl16}
{Buhler} PB, {Knutson} HA, {Batygin} K, {Fulton} BJ, {Fortney} JJ, et~al. 2016.
\newblock \textit{\apj} 821:26

\bibitem[{{Campante} et~al.(2016){Campante}, {Schofield}, {Kuszlewicz},
  {Bouma}, {Chaplin} et~al.}]{camp16}
{Campante} TL, {Schofield} M, {Kuszlewicz} JS, {Bouma} L, {Chaplin} WJ, et~al.
  2016.
\newblock \textit{\apj} 830:138

\bibitem[{{Casertano} et~al.(2008){Casertano}, {Lattanzi}, {Sozzetti},
  {Spagna}, {Jancart} et~al.}]{case08}
{Casertano} S, {Lattanzi} MG, {Sozzetti} A, {Spagna} A, {Jancart} S, et~al.
  2008.
\newblock \textit{\aap} 482:699--729

\bibitem[{{Catala} et~al.(2007){Catala}, {Donati}, {Shkolnik}, {Bohlender} \&
  {Alecian}}]{cata07}
{Catala} C, {Donati} JF, {Shkolnik} E, {Bohlender} D, {Alecian} E. 2007.
\newblock \textit{\mnras} 374:L42--L46

\bibitem[{{Chabrier} et~al.(2014){Chabrier}, {Johansen}, {Janson} \&
  {Rafikov}}]{chab14}
{Chabrier} G, {Johansen} A, {Janson} M, {Rafikov} R. 2014.
\newblock \textit{Protostars and Planets VI} :619--642

\bibitem[{{Chambers}(2016)}]{cham16}
{Chambers} JE. 2016.
\newblock \textit{\apj} 825:63

\bibitem[{{Chambers}, {Wetherill} \& {Boss}(1996)}]{cham96}
{Chambers} JE, {Wetherill} GW, {Boss} AP. 1996.
\newblock \textit{\icarus} 119:261--268

\bibitem[{{Chang}, {Gu} \& {Bodenheimer}(2010)}]{chan10}
{Chang} SH, {Gu} PG, {Bodenheimer} PH. 2010.
\newblock \textit{\apj} 708:1692--1702

\bibitem[{{Charbonneau} et~al.(2000){Charbonneau}, {Brown}, {Latham} \&
  {Mayor}}]{char00}
{Charbonneau} D, {Brown} TM, {Latham} DW, {Mayor} M. 2000.
\newblock \textit{\apjl} 529:L45--L48

\bibitem[{{Chatterjee} et~al.(2012){Chatterjee}, {Ford}, {Geller} \&
  {Rasio}}]{chat12}
{Chatterjee} S, {Ford} EB, {Geller} AM, {Rasio} FA. 2012.
\newblock \textit{\mnras} 427:1587--1602

\bibitem[{{Chatterjee} et~al.(2008){Chatterjee}, {Ford}, {Matsumura} \&
  {Rasio}}]{chat08}
{Chatterjee} S, {Ford} EB, {Matsumura} S, {Rasio} FA. 2008.
\newblock \textit{\apj} 686:580--602

\bibitem[{{Chiang} \& {Laughlin}(2013)}]{chia13}
{Chiang} E, {Laughlin} G. 2013.
\newblock \textit{\mnras} 431:3444--3455

\bibitem[{{Ciardi} et~al.(2013){Ciardi}, {Fabrycky}, {Ford}, {Gautier},
  {Howell} et~al.}]{ciar13}
{Ciardi} DR, {Fabrycky} DC, {Ford} EB, {Gautier} III TN, {Howell} SB, et~al.
  2013.
\newblock \textit{\apj} 763:41

\bibitem[{{Coleman} \& {Nelson}(2016)}]{cole16}
{Coleman} GAL, {Nelson} RP. 2016.
\newblock \textit{\mnras} 460:2779--2795

\bibitem[{{Collier Cameron} et~al.(2010){Collier Cameron}, {Guenther},
  {Smalley}, {McDonald}, {Hebb} et~al.}]{coll10}
{Collier Cameron} A, {Guenther} E, {Smalley} B, {McDonald} I, {Hebb} L, et~al.
  2010.
\newblock \textit{\mnras} 407:507--514

\bibitem[{{Cumming} et~al.(2008){Cumming}, {Butler}, {Marcy}, {Vogt}, {Wright}
  \& {Fischer}}]{cumm08}
{Cumming} A, {Butler} RP, {Marcy} GW, {Vogt} SS, {Wright} JT, {Fischer} DA.
  2008.
\newblock \textit{\pasp} 120:531

\bibitem[{{Damiani} \& {D{\'{\i}}az}(2016)}]{dami16}
{Damiani} C, {D{\'{\i}}az} RF. 2016.
\newblock \textit{\aap} 589:A55

\bibitem[{{Damiani} \& {Lanza}(2015)}]{dami15}
{Damiani} C, {Lanza} AF. 2015.
\newblock \textit{\aap} 574:A39

\bibitem[{{Dawson}(2014)}]{daws14a}
{Dawson} RI. 2014.
\newblock \textit{\apjl} 790:L31

\bibitem[{{Dawson} \& {Chiang}(2014)}]{daws14b}
{Dawson} RI, {Chiang} E. 2014.
\newblock \textit{Science} 346:212--216

\bibitem[{{Dawson}, {Chiang} \& {Lee}(2015)}]{daws15b}
{Dawson} RI, {Chiang} E, {Lee} EJ. 2015.
\newblock \textit{\mnras} 453:1471--1483

\bibitem[{{Dawson} \& {Johnson}(2012)}]{daws12}
{Dawson} RI, {Johnson} JA. 2012.
\newblock \textit{\apj} 756:122

\bibitem[{{Dawson} et~al.(2014){Dawson}, {Johnson}, {Fabrycky},
  {Foreman-Mackey}, {Murray-Clay} et~al.}]{daws14}
{Dawson} RI, {Johnson} JA, {Fabrycky} DC, {Foreman-Mackey} D, {Murray-Clay} RA,
  et~al. 2014.
\newblock \textit{\apj} 791:89

\bibitem[{{Dawson} et~al.(2012){Dawson}, {Johnson}, {Morton}, {Crepp},
  {Fabrycky} et~al.}]{daws12a}
{Dawson} RI, {Johnson} JA, {Morton} TD, {Crepp} JR, {Fabrycky} DC, et~al. 2012.
\newblock \textit{\apj} 761:163

\bibitem[{{Dawson}, {Lee} \& {Chiang}(2016)}]{daws16}
{Dawson} RI, {Lee} EJ, {Chiang} E. 2016.
\newblock \textit{\apj} 822:54

\bibitem[{{Dawson} \& {Murray-Clay}(2013)}]{daws13}
{Dawson} RI, {Murray-Clay} RA. 2013.
\newblock \textit{\apjl} 767:L24

\bibitem[{{Dawson}, {Murray-Clay} \& {Johnson}(2015)}]{daws15}
{Dawson} RI, {Murray-Clay} RA, {Johnson} JA. 2015.
\newblock \textit{\apj} 798:66

\bibitem[{{Donati} et~al.(2016){Donati}, {Moutou}, {Malo}, {Baruteau}, {Yu}
  et~al.}]{dona16}
{Donati} JF, {Moutou} C, {Malo} L, {Baruteau} C, {Yu} L, et~al. 2016.
\newblock \textit{\nat} 534:662--666

\bibitem[{{Dong} \& {Dawson}(2016)}]{dong16}
{Dong} R, {Dawson} R. 2016.
\newblock \textit{\apj} 825:77

\bibitem[{{Dong}, {Katz} \& {Socrates}(2014)}]{dong14}
{Dong} S, {Katz} B, {Socrates} A. 2014.
\newblock \textit{\apjl} 781:L5

\bibitem[{{Dong} et~al.(2017){Dong}, {Xie}, {Zhou}, {Zheng} \& {Luo}}]{dong17}
{Dong} S, {Xie} JW, {Zhou} JL, {Zheng} Z, {Luo} A. 2017.
\newblock \textit{ArXiv e-prints} arXiv:1706.07807

\bibitem[{{Dressing} \& {Charbonneau}(2015)}]{dres15}
{Dressing} CD, {Charbonneau} D. 2015.
\newblock \textit{\apj} 807:45

\bibitem[{{Duffell} \& {Chiang}(2015)}]{duff15}
{Duffell} PC, {Chiang} E. 2015.
\newblock \textit{\apj} 812:94

\bibitem[{{Duffell} et~al.(2014){Duffell}, {Haiman}, {MacFadyen}, {D'Orazio} \&
  {Farris}}]{duff14}
{Duffell} PC, {Haiman} Z, {MacFadyen} AI, {D'Orazio} DJ, {Farris} BD. 2014.
\newblock \textit{\apjl} 792:L10

\bibitem[{{Durisen} et~al.(2007){Durisen}, {Boss}, {Mayer}, {Nelson}, {Quinn}
  \& {Rice}}]{duri07}
{Durisen} RH, {Boss} AP, {Mayer} L, {Nelson} AF, {Quinn} T, {Rice} WKM. 2007.
\newblock \textit{Protostars and Planets V} :607--622

\bibitem[{{D{\"u}rmann} \& {Kley}(2015)}]{durm15}
{D{\"u}rmann} C, {Kley} W. 2015.
\newblock \textit{\aap} 574:A52

\bibitem[{{Eggleton}, {Kiseleva} \& {Hut}(1998)}]{eggl98}
{Eggleton} PP, {Kiseleva} LG, {Hut} P. 1998.
\newblock \textit{\apj} 499:853--870

\bibitem[{{Eisner} et~al.(2005){Eisner}, {Hillenbrand}, {White}, {Akeson} \&
  {Sargent}}]{eisn05}
{Eisner} JA, {Hillenbrand} LA, {White} RJ, {Akeson} RL, {Sargent} AI. 2005.
\newblock \textit{\apj} 623:952--966

\bibitem[{{Endl} et~al.(2014){Endl}, {Caldwell}, {Barclay}, {Huber}, {Isaacson}
  et~al.}]{endl14}
{Endl} M, {Caldwell} DA, {Barclay} T, {Huber} D, {Isaacson} H, et~al. 2014.
\newblock \textit{\apj} 795:151

\bibitem[{{Evans} et~al.(2016){Evans}, {Southworth}, {Maxted}, {Skottfelt},
  {Hundertmark} et~al.}]{evan16}
{Evans} DF, {Southworth} J, {Maxted} PFL, {Skottfelt} J, {Hundertmark} M,
  et~al. 2016.
\newblock \textit{\aap} 589:A58

\bibitem[{{Faber}, {Rasio} \& {Willems}(2005)}]{fabe05}
{Faber} JA, {Rasio} FA, {Willems} B. 2005.
\newblock \textit{\icarus} 175:248--262

\bibitem[{{Fabrycky} \& {Tremaine}(2007)}]{fabr07}
{Fabrycky} D, {Tremaine} S. 2007.
\newblock \textit{\apj} 669:1298--1315

\bibitem[{{Fabrycky}(2010)}]{fabr10}
{Fabrycky} DC. 2010.
\newblock \textit{{Non-Keplerian Dynamics of Exoplanets}}.
\newblock  217--238

\bibitem[{{Fabrycky} \& {Winn}(2009)}]{fabr09}
{Fabrycky} DC, {Winn} JN. 2009.
\newblock \textit{\apj} 696:1230--1240

\bibitem[{{Fedele} et~al.(2010){Fedele}, {van den Ancker}, {Henning},
  {Jayawardhana} \& {Oliveira}}]{fede10}
{Fedele} D, {van den Ancker} ME, {Henning} T, {Jayawardhana} R, {Oliveira} JM.
  2010.
\newblock \textit{\aap} 510:A72

\bibitem[{{Fielding} et~al.(2015){Fielding}, {McKee}, {Socrates}, {Cunningham}
  \& {Klein}}]{fiel15}
{Fielding} DB, {McKee} CF, {Socrates} A, {Cunningham} AJ, {Klein} RI. 2015.
\newblock \textit{\mnras} 450:3306--3318

\bibitem[{{Fischer} \& {Valenti}(2005)}]{fisc05}
{Fischer} DA, {Valenti} J. 2005.
\newblock \textit{\apj} 622:1102--1117

\bibitem[{{Ford} \& {Rasio}(2006)}]{ford06}
{Ford} EB, {Rasio} FA. 2006.
\newblock \textit{\apjl} 638:L45--L48

\bibitem[{{Gaia Collaboration} et~al.(2016){Gaia Collaboration}, {Prusti}, {de
  Bruijne}, {Brown}, {Vallenari} et~al.}]{gaia16}
{Gaia Collaboration}, {Prusti} T, {de Bruijne} JHJ, {Brown} AGA, {Vallenari} A,
  et~al. 2016.
\newblock \textit{\aap} 595:A1

\bibitem[{{G{\'a}sp{\'a}r}, {Rieke} \& {Ballering}(2016)}]{gasp16}
{G{\'a}sp{\'a}r} A, {Rieke} GH, {Ballering} N. 2016.
\newblock \textit{\apj} 826:171

\bibitem[{{Gaudi}, {Seager} \& {Mallen-Ornelas}(2005)}]{gaud05}
{Gaudi} BS, {Seager} S, {Mallen-Ornelas} G. 2005.
\newblock \textit{\apj} 623:472--481

\bibitem[{{Ginzburg} \& {Sari}(2016)}]{ginz16}
{Ginzburg} S, {Sari} R. 2016.
\newblock \textit{\apj} 819:116

\bibitem[{{Goldreich}, {Lithwick} \& {Sari}(2004)}]{gold04}
{Goldreich} P, {Lithwick} Y, {Sari} R. 2004.
\newblock \textit{\araa} 42:549--601

\bibitem[{{Goldreich} \& {Sari}(2003)}]{gold03}
{Goldreich} P, {Sari} R. 2003.
\newblock \textit{\apj} 585:1024--1037

\bibitem[{{Goldreich} \& {Schlichting}(2014)}]{gold14}
{Goldreich} P, {Schlichting} HE. 2014.
\newblock \textit{\aj} 147:32

\bibitem[{{Goldreich} \& {Tremaine}(1980)}]{gold80}
{Goldreich} P, {Tremaine} S. 1980.
\newblock \textit{\apj} 241:425--441

\bibitem[{{Gonzalez}(1997)}]{gonz97}
{Gonzalez} G. 1997.
\newblock \textit{\mnras} 285:403--412

\bibitem[{{Greenzweig} \& {Lissauer}(1990)}]{gree90}
{Greenzweig} Y, {Lissauer} JJ. 1990.
\newblock \textit{\icarus} 87:40--77

\bibitem[{{Guillochon}, {Ramirez-Ruiz} \& {Lin}(2011)}]{guil11}
{Guillochon} J, {Ramirez-Ruiz} E, {Lin} D. 2011.
\newblock \textit{\apj} 732:74

\bibitem[{{Guillot}(2005)}]{guil05}
{Guillot} T. 2005.
\newblock \textit{Annual Review of Earth and Planetary Sciences} 33:493--530

\bibitem[{{Guillot} \& {Showman}(2002)}]{guil02}
{Guillot} T, {Showman} AP. 2002.
\newblock \textit{\aap} 385:156--165

\bibitem[{{Guo} et~al.(2017){Guo}, {Johnson}, {Mann}, {Kraus}, {Curtis} \&
  {Latham}}]{guo17}
{Guo} X, {Johnson} JA, {Mann} AW, {Kraus} AL, {Curtis} JL, {Latham} DW. 2017.
\newblock \textit{\apj} 838:25

\bibitem[{{Hamers} et~al.(2017){Hamers}, {Antonini}, {Lithwick}, {Perets} \&
  {Portegies Zwart}}]{hame16}
{Hamers} AS, {Antonini} F, {Lithwick} Y, {Perets} HB, {Portegies Zwart} SF.
  2017.
\newblock \textit{\mnras} 464:688--701

\bibitem[{{Hansen}(2010)}]{hans10}
{Hansen} BMS. 2010.
\newblock \textit{\apj} 723:285--299

\bibitem[{{Hansen} \& {Murray}(2013)}]{hans13}
{Hansen} BMS, {Murray} N. 2013.
\newblock \textit{\apj} 775:53

\bibitem[{{Hao}, {Kouwenhoven} \& {Spurzem}(2013)}]{hao13}
{Hao} W, {Kouwenhoven} MBN, {Spurzem} R. 2013.
\newblock \textit{\mnras} 433:867--877

\bibitem[{{Hartman} et~al.(2016){Hartman}, {Bakos}, {Bhatti}, {Penev},
  {Bieryla} et~al.}]{hart16}
{Hartman} JD, {Bakos} G{\'A}, {Bhatti} W, {Penev} K, {Bieryla} A, et~al. 2016.
\newblock \textit{\aj} 152:182

\bibitem[{{Hayashi}(1981)}]{haya81}
{Hayashi} C. 1981.
\newblock \textit{Progress of Theoretical Physics Supplement} 70:35--53

\bibitem[{{Heng} \& {Lyons}(2016)}]{heng16}
{Heng} K, {Lyons} JR. 2016.
\newblock \textit{\apj} 817:149

\bibitem[{{Henry} et~al.(2000){Henry}, {Marcy}, {Butler} \& {Vogt}}]{henr00}
{Henry} GW, {Marcy} GW, {Butler} RP, {Vogt} SS. 2000.
\newblock \textit{\apjl} 529:L41--L44

\bibitem[{{Howard} et~al.(2010){Howard}, {Marcy}, {Johnson}, {Fischer},
  {Wright} et~al.}]{howa10}
{Howard} AW, {Marcy} GW, {Johnson} JA, {Fischer} DA, {Wright} JT, et~al. 2010.
\newblock \textit{Science} 330:653

\bibitem[{{Huang}, {Wu} \& {Triaud}(2016)}]{huan16}
{Huang} C, {Wu} Y, {Triaud} AHMJ. 2016.
\newblock \textit{\apj} 825:98

\bibitem[{{Huber} et~al.(2013){Huber}, {Carter}, {Barbieri}, {Miglio}, {Deck}
  et~al.}]{hube13}
{Huber} D, {Carter} JA, {Barbieri} M, {Miglio} A, {Deck} KM, et~al. 2013.
\newblock \textit{Science} 342:331--334

\bibitem[{{Hut}(1981)}]{hut81}
{Hut} P. 1981.
\newblock \textit{\aap} 99:126--140

\bibitem[{{Ibgui}, {Spiegel} \& {Burrows}(2011)}]{igbu11}
{Ibgui} L, {Spiegel} DS, {Burrows} A. 2011.
\newblock \textit{\apj} 727:75

\bibitem[{{Ida} \& {Lin}(2008)}]{ida08}
{Ida} S, {Lin} DNC. 2008.
\newblock \textit{\apj} 673:487--501

\bibitem[{{Ida}, {Lin} \& {Nagasawa}(2013)}]{ida13}
{Ida} S, {Lin} DNC, {Nagasawa} M. 2013.
\newblock \textit{\apj} 775:42

\bibitem[{{Jackson} et~al.(2016){Jackson}, {Jensen}, {Peacock}, {Arras} \&
  {Penev}}]{jack16}
{Jackson} B, {Jensen} E, {Peacock} S, {Arras} P, {Penev} K. 2016.
\newblock \textit{Celestial Mechanics and Dynamical Astronomy} 126:227--248

\bibitem[{{Jenkins} et~al.(2017){Jenkins}, {Jones}, {Tuomi}, {D{\'{\i}}az},
  {Cordero} et~al.}]{jenk16}
{Jenkins} JS, {Jones} HRA, {Tuomi} M, {D{\'{\i}}az} M, {Cordero} JP, et~al.
  2017.
\newblock \textit{\mnras} 466:443--473

\bibitem[{{Johansen} \& {Lambrechts}(2017)}]{joha17}
{Johansen} A, {Lambrechts} M. 2017.
\newblock \textit{Annual Review of Earth and Planetary Sciences} 45:359--387

\bibitem[{{Johnson} et~al.(2010){Johnson}, {Aller}, {Howard} \&
  {Crepp}}]{john10}
{Johnson} JA, {Aller} KM, {Howard} AW, {Crepp} JR. 2010.
\newblock \textit{\pasp} 122:905--915

\bibitem[{{Jones} et~al.(2003){Jones}, {Butler}, {Tinney}, {Marcy}, {Penny}
  et~al.}]{jone03}
{Jones} HRA, {Butler} RP, {Tinney} CG, {Marcy} GW, {Penny} AJ, et~al. 2003.
\newblock \textit{\mnras} 341:948--952

\bibitem[{{Juri{\'c}} \& {Tremaine}(2008)}]{juri08}
{Juri{\'c}} M, {Tremaine} S. 2008.
\newblock \textit{\apj} 686:603--620

\bibitem[{{Katz}, {Dong} \& {Malhotra}(2011)}]{katz11}
{Katz} B, {Dong} S, {Malhotra} R. 2011.
\newblock \textit{Physical Review Letters} 107:181101

\bibitem[{{Knutson} et~al.(2014){Knutson}, {Fulton}, {Montet}, {Kao}, {Ngo}
  et~al.}]{knut14}
{Knutson} HA, {Fulton} BJ, {Montet} BT, {Kao} M, {Ngo} H, et~al. 2014.
\newblock \textit{\apj} 785:126

\bibitem[{{Kov{\'a}cs} et~al.(2014){Kov{\'a}cs}, {Hartman}, {Bakos}, {Quinn},
  {Penev} et~al.}]{kova14}
{Kov{\'a}cs} G, {Hartman} JD, {Bakos} G{\'A}, {Quinn} SN, {Penev} K, et~al.
  2014.
\newblock \textit{\mnras} 442:2081--2093

\bibitem[{{Kozai}(1962)}]{koza62}
{Kozai} Y. 1962.
\newblock \textit{\aj} 67:591

\bibitem[{{Kraft}(1967)}]{kraf67}
{Kraft} RP. 1967.
\newblock \textit{\apj} 150:551

\bibitem[{{Kreidberg} et~al.(2015){Kreidberg}, {Line}, {Bean}, {Stevenson},
  {D{\'e}sert} et~al.}]{krei15}
{Kreidberg} L, {Line} MR, {Bean} JL, {Stevenson} KB, {D{\'e}sert} JM, et~al.
  2015.
\newblock \textit{\apj} 814:66

\bibitem[{{Kuchner} \& {Lecar}(2002)}]{kuch02}
{Kuchner} MJ, {Lecar} M. 2002.
\newblock \textit{\apjl} 574:L87--L89

\bibitem[{{Lai}(2012)}]{lai12}
{Lai} D. 2012.
\newblock \textit{\mnras} 423:486--492

\bibitem[{{Laskar}(2008)}]{lask08}
{Laskar} J. 2008.
\newblock \textit{\icarus} 196:1--15

\bibitem[{{Latham} et~al.(2011){Latham}, {Rowe}, {Quinn}, {Batalha}, {Borucki}
  et~al.}]{lath11}
{Latham} DW, {Rowe} JF, {Quinn} SN, {Batalha} NM, {Borucki} WJ, et~al. 2011.
\newblock \textit{\apjl} 732:L24

\bibitem[{{Lecar} et~al.(2006){Lecar}, {Podolak}, {Sasselov} \&
  {Chiang}}]{leca06}
{Lecar} M, {Podolak} M, {Sasselov} D, {Chiang} E. 2006.
\newblock \textit{\apj} 640:1115--1118

\bibitem[{{Leconte} et~al.(2010){Leconte}, {Chabrier}, {Baraffe} \&
  {Levrard}}]{leco10}
{Leconte} J, {Chabrier} G, {Baraffe} I, {Levrard} B. 2010.
\newblock \textit{\aap} 516:A64

\bibitem[{{Lee} \& {Chiang}(2016)}]{lee16}
{Lee} EJ, {Chiang} E. 2016.
\newblock \textit{\apj} 817:90

\bibitem[{{Lee} \& {Chiang}(2017)}]{lee17}
{Lee} EJ, {Chiang} E. 2017.
\newblock \textit{\apj} 842:40

\bibitem[{{Lee}, {Chiang} \& {Ormel}(2014)}]{lee14}
{Lee} EJ, {Chiang} E, {Ormel} CW. 2014.
\newblock \textit{\apj} 797:95

\bibitem[{{Lee} \& {Peale}(2002)}]{lee02}
{Lee} MH, {Peale} SJ. 2002.
\newblock \textit{\apj} 567:596--609

\bibitem[{{Li} et~al.(2014{\natexlab{a}}){Li}, {Naoz}, {Holman} \&
  {Loeb}}]{li14}
{Li} G, {Naoz} S, {Holman} M, {Loeb} A. 2014{\natexlab{a}}.
\newblock \textit{\apj} 791:86

\bibitem[{{Li} et~al.(2014{\natexlab{b}}){Li}, {Naoz}, {Kocsis} \&
  {Loeb}}]{li14a}
{Li} G, {Naoz} S, {Kocsis} B, {Loeb} A. 2014{\natexlab{b}}.
\newblock \textit{\apj} 785:116

\bibitem[{{Li} et~al.(2014{\natexlab{c}}){Li}, {Naoz}, {Valsecchi}, {Johnson}
  \& {Rasio}}]{li14c}
{Li} G, {Naoz} S, {Valsecchi} F, {Johnson} JA, {Rasio} FA. 2014{\natexlab{c}}.
\newblock \textit{\apj} 794:131

\bibitem[{{Li} \& {Winn}(2016)}]{li16}
{Li} G, {Winn} JN. 2016.
\newblock \textit{\apj} 818:5

\bibitem[{{Lidov}(1962)}]{lido62}
{Lidov} ML. 1962.
\newblock \textit{\planss} 9:719--759

\bibitem[{{Lin}, {Bodenheimer} \& {Richardson}(1996)}]{lin96}
{Lin} DNC, {Bodenheimer} P, {Richardson} DC. 1996.
\newblock \textit{\nat} 380:606--607

\bibitem[{{Lin} \& {Papaloizou}(1986)}]{lin86}
{Lin} DNC, {Papaloizou} J. 1986.
\newblock \textit{\apj} 309:846--857

\bibitem[{{Lin} \& {Ogilvie}(2017)}]{lin17}
{Lin} Y, {Ogilvie} GI. 2017.
\newblock \textit{\mnras} 468:1387--1397

\bibitem[{{Lissauer}, {Dawson} \& {Tremaine}(2014)}]{liss14}
{Lissauer} JJ, {Dawson} RI, {Tremaine} S. 2014.
\newblock \textit{\nat} 513:336--344

\bibitem[{{Lopez} \& {Fortney}(2013)}]{lope13}
{Lopez} ED, {Fortney} JJ. 2013.
\newblock \textit{\apj} 776:2

\bibitem[{{Lopez} \& {Fortney}(2014)}]{lope14}
{Lopez} ED, {Fortney} JJ. 2014.
\newblock \textit{\apj} 792:1

\bibitem[{{Lopez} \& {Fortney}(2016)}]{lope16}
{Lopez} ED, {Fortney} JJ. 2016.
\newblock \textit{\apj} 818:4

\bibitem[{{Lopez}, {Fortney} \& {Miller}(2012)}]{lope12}
{Lopez} ED, {Fortney} JJ, {Miller} N. 2012.
\newblock \textit{\apj} 761:59

\bibitem[{{Madhusudhan}, {Amin} \& {Kennedy}(2014)}]{madh14}
{Madhusudhan} N, {Amin} MA, {Kennedy} GM. 2014.
\newblock \textit{\apjl} 794:L12

\bibitem[{{Malhotra}(1993)}]{malh93}
{Malhotra} R. 1993.
\newblock \textit{\nat} 365:819--821

\bibitem[{{Marcy} \& {Butler}(2000)}]{marc00}
{Marcy} GW, {Butler} RP. 2000.
\newblock \textit{\pasp} 112:137--140

\bibitem[{{Masset} \& {Papaloizou}(2003)}]{mass03}
{Masset} FS, {Papaloizou} JCB. 2003.
\newblock \textit{\apj} 588:494--508

\bibitem[{{Masuda}(2017)}]{masu17}
{Masuda} K. 2017.
\newblock \textit{\aj} 154:64

\bibitem[{{Matsakos} \& {K{\"o}nigl}(2015)}]{mats15}
{Matsakos} T, {K{\"o}nigl} A. 2015.
\newblock \textit{\apjl} 809:L20

\bibitem[{{Matsumura}, {Peale} \& {Rasio}(2010)}]{mats10}
{Matsumura} S, {Peale} SJ, {Rasio} FA. 2010.
\newblock \textit{\apj} 725:1995--2016

\bibitem[{{Mayor} et~al.(2011){Mayor}, {Marmier}, {Lovis}, {Udry},
  {S{\'e}gransan} et~al.}]{mayo11}
{Mayor} M, {Marmier} M, {Lovis} C, {Udry} S, {S{\'e}gransan} D, et~al. 2011.
\newblock \textit{ArXiv e-prints} arXiv:1109.2497

\bibitem[{{Mayor} \& {Queloz}(1995)}]{mayo95}
{Mayor} M, {Queloz} D. 1995.
\newblock \textit{\nat} 378:355--359

\bibitem[{{Mazeh} et~al.(2015){Mazeh}, {Perets}, {McQuillan} \&
  {Goldstein}}]{maze15}
{Mazeh} T, {Perets} HB, {McQuillan} A, {Goldstein} ES. 2015.
\newblock \textit{\apj} 801:3

\bibitem[{{McLaughlin}(1924)}]{mcla24}
{McLaughlin} DB. 1924.
\newblock \textit{\apj}

\bibitem[{{Miller} et~al.(2015){Miller}, {Gallo}, {Wright} \&
  {Pearson}}]{mill15}
{Miller} BP, {Gallo} E, {Wright} JT, {Pearson} EG. 2015.
\newblock \textit{\apj} 799:163

\bibitem[{{Millholland}, {Wang} \& {Laughlin}(2016)}]{mill16}
{Millholland} S, {Wang} S, {Laughlin} G. 2016.
\newblock \textit{\apjl} 823:L7

\bibitem[{{Minton} \& {Malhotra}(2011)}]{mint11}
{Minton} DA, {Malhotra} R. 2011.
\newblock \textit{\apj} 732:53

\bibitem[{{Moro-Mart{\'{\i}}n} et~al.(2015){Moro-Mart{\'{\i}}n}, {Marshall},
  {Kennedy}, {Sibthorpe}, {Matthews} et~al.}]{moro15}
{Moro-Mart{\'{\i}}n} A, {Marshall} JP, {Kennedy} G, {Sibthorpe} B, {Matthews}
  BC, et~al. 2015.
\newblock \textit{\apj} 801:143

\bibitem[{{Morton} \& {Johnson}(2011)}]{mort11}
{Morton} TD, {Johnson} JA. 2011.
\newblock \textit{\apj} 729:138

\bibitem[{{Mu{\~n}oz}, {Lai} \& {Liu}(2016)}]{muno16}
{Mu{\~n}oz} DJ, {Lai} D, {Liu} B. 2016.
\newblock \textit{\mnras} 460:1086--1093

\bibitem[{{Mustill}, {Davies} \& {Johansen}(2015)}]{must15}
{Mustill} AJ, {Davies} MB, {Johansen} A. 2015.
\newblock \textit{\apj} 808:14

\bibitem[{{Mustill}, {Davies} \& {Johansen}(2017)}]{must17}
{Mustill} AJ, {Davies} MB, {Johansen} A. 2017.
\newblock \textit{\mnras} 468:3000--3023

\bibitem[{{Naef} et~al.(2001){Naef}, {Latham}, {Mayor}, {Mazeh}, {Beuzit}
  et~al.}]{naef01}
{Naef} D, {Latham} DW, {Mayor} M, {Mazeh} T, {Beuzit} JL, et~al. 2001.
\newblock \textit{\aap} 375:L27--L30

\bibitem[{{Nagasawa} \& {Ida}(2011)}]{naga11}
{Nagasawa} M, {Ida} S. 2011.
\newblock \textit{\apj} 742:72

\bibitem[{{Naoz}(2016)}]{naoz16}
{Naoz} S. 2016.
\newblock \textit{\araa} 54:441--489

\bibitem[{{Naoz} et~al.(2011){Naoz}, {Farr}, {Lithwick}, {Rasio} \&
  {Teyssandier}}]{naoz11}
{Naoz} S, {Farr} WM, {Lithwick} Y, {Rasio} FA, {Teyssandier} J. 2011.
\newblock \textit{\nat} 473:187--189

\bibitem[{{Naoz}, {Farr} \& {Rasio}(2012)}]{naoz12}
{Naoz} S, {Farr} WM, {Rasio} FA. 2012.
\newblock \textit{\apjl} 754:L36

\bibitem[{{Nelson}, {Ford} \& {Rasio}(2017)}]{nels17}
{Nelson} BE, {Ford} EB, {Rasio} FA. 2017.
\newblock \textit{\aj} 154:106

\bibitem[{{Neveu-VanMalle} et~al.(2016){Neveu-VanMalle}, {Queloz}, {Anderson},
  {Brown}, {Collier Cameron} et~al.}]{neve16}
{Neveu-VanMalle} M, {Queloz} D, {Anderson} DR, {Brown} DJA, {Collier Cameron}
  A, et~al. 2016.
\newblock \textit{\aap} 586:A93

\bibitem[{{Ngo} et~al.(2017){Ngo}, {Knutson}, {Bryan}, {Blunt}, {Nielsen}
  et~al.}]{ngo17}
{Ngo} H, {Knutson} HA, {Bryan} ML, {Blunt} S, {Nielsen} EL, et~al. 2017.
\newblock \textit{\aj} 153:242

\bibitem[{{Ngo} et~al.(2016){Ngo}, {Knutson}, {Hinkley}, {Bryan}, {Crepp}
  et~al.}]{ngo16}
{Ngo} H, {Knutson} HA, {Hinkley} S, {Bryan} M, {Crepp} JR, et~al. 2016.
\newblock \textit{\apj} 827:8

\bibitem[{{{\"O}berg}, {Murray-Clay} \& {Bergin}(2011)}]{ober11}
{{\"O}berg} KI, {Murray-Clay} R, {Bergin} EA. 2011.
\newblock \textit{\apjl} 743:L16

\bibitem[{{Obermeier} et~al.(2016){Obermeier}, {Koppenhoefer}, {Saglia},
  {Henning}, {Bender} et~al.}]{ober16}
{Obermeier} C, {Koppenhoefer} J, {Saglia} RP, {Henning} T, {Bender} R, et~al.
  2016.
\newblock \textit{\aap} 587:A49

\bibitem[{{Ogihara}, {Inutsuka} \& {Kobayashi}(2013)}]{ogih13}
{Ogihara} M, {Inutsuka} Si, {Kobayashi} H. 2013.
\newblock \textit{\apjl} 778:L9

\bibitem[{{Ogihara}, {Kobayashi} \& {Inutsuka}(2014)}]{ogih14}
{Ogihara} M, {Kobayashi} H, {Inutsuka} Si. 2014.
\newblock \textit{\apj} 787:172

\bibitem[{{Otor} et~al.(2016){Otor}, {Montet}, {Johnson}, {Charbonneau},
  {Collier-Cameron} et~al.}]{otor16}
{Otor} OJ, {Montet} BT, {Johnson} JA, {Charbonneau} D, {Collier-Cameron} A,
  et~al. 2016.
\newblock \textit{\aj} 152:165

\bibitem[{{Paardekooper} \& {Mellema}(2006)}]{paar06}
{Paardekooper} SJ, {Mellema} G. 2006.
\newblock \textit{\aap} 459:L17--L20

\bibitem[{{Pascucci} et~al.(2016){Pascucci}, {Testi}, {Herczeg}, {Long},
  {Manara} et~al.}]{pasc16}
{Pascucci} I, {Testi} L, {Herczeg} GJ, {Long} F, {Manara} CF, et~al. 2016.
\newblock \textit{\apj} 831:125

\bibitem[{{Perri} \& {Cameron}(1974)}]{peri74}
{Perri} F, {Cameron} AGW. 1974.
\newblock \textit{\icarus} 22:416--425

\bibitem[{{Petrovich}(2015{\natexlab{a}})}]{petr15}
{Petrovich} C. 2015{\natexlab{a}}.
\newblock \textit{\apj} 805:75

\bibitem[{{Petrovich}(2015{\natexlab{b}})}]{petr15a}
{Petrovich} C. 2015{\natexlab{b}}.
\newblock \textit{\apj} 799:27

\bibitem[{{Petrovich} \& {Tremaine}(2016)}]{petr16}
{Petrovich} C, {Tremaine} S. 2016.
\newblock \textit{\apj} 829:132

\bibitem[{{Petrovich}, {Tremaine} \& {Rafikov}(2014)}]{petr14}
{Petrovich} C, {Tremaine} S, {Rafikov} R. 2014.
\newblock \textit{\apj} 786:101

\bibitem[{{Piskorz} et~al.(2015){Piskorz}, {Knutson}, {Ngo}, {Muirhead},
  {Batygin} et~al.}]{pisk15}
{Piskorz} D, {Knutson} HA, {Ngo} H, {Muirhead} PS, {Batygin} K, et~al. 2015.
\newblock \textit{\apj} 814:148

\bibitem[{{Piso} et~al.(2015){Piso}, {{\"O}berg}, {Birnstiel} \&
  {Murray-Clay}}]{piso15}
{Piso} AMA, {{\"O}berg} KI, {Birnstiel} T, {Murray-Clay} RA. 2015.
\newblock \textit{\apj} 815:109

\bibitem[{{Piso}, {Youdin} \& {Murray-Clay}(2015)}]{piso15b}
{Piso} AMA, {Youdin} AN, {Murray-Clay} RA. 2015.
\newblock \textit{\apj} 800:82

\bibitem[{{Plavchan} \& {Bilinski}(2013)}]{plav13}
{Plavchan} P, {Bilinski} C. 2013.
\newblock \textit{\apj} 769:86

\bibitem[{{Pollack} et~al.(1996){Pollack}, {Hubickyj}, {Bodenheimer},
  {Lissauer}, {Podolak} \& {Greenzweig}}]{poll96}
{Pollack} JB, {Hubickyj} O, {Bodenheimer} P, {Lissauer} JJ, {Podolak} M,
  {Greenzweig} Y. 1996.
\newblock \textit{\icarus} 124:62--85

\bibitem[{{Pont} et~al.(2011){Pont}, {Husnoo}, {Mazeh} \& {Fabrycky}}]{pont11}
{Pont} F, {Husnoo} N, {Mazeh} T, {Fabrycky} D. 2011.
\newblock \textit{\mnras} 414:1278--1284

\bibitem[{{Poppenhaeger} \& {Wolk}(2014)}]{popp14}
{Poppenhaeger} K, {Wolk} SJ. 2014.
\newblock \textit{\aap} 565:L1

\bibitem[{{Quinn} \& {White}(2016)}]{quin16}
{Quinn} SN, {White} RJ. 2016.
\newblock \textit{\apj} 833:173

\bibitem[{{Quinn} et~al.(2012){Quinn}, {White}, {Latham}, {Buchhave},
  {Cantrell} et~al.}]{quin12}
{Quinn} SN, {White} RJ, {Latham} DW, {Buchhave} LA, {Cantrell} JR, et~al. 2012.
\newblock \textit{\apjl} 756:L33

\bibitem[{{Quinn} et~al.(2014){Quinn}, {White}, {Latham}, {Buchhave}, {Torres}
  et~al.}]{quin14}
{Quinn} SN, {White} RJ, {Latham} DW, {Buchhave} LA, {Torres} G, et~al. 2014.
\newblock \textit{\apj} 787:27

\bibitem[{{Rafikov}(2004)}]{rafi04}
{Rafikov} RR. 2004.
\newblock \textit{\aj} 128:1348--1363

\bibitem[{{Rafikov}(2005)}]{rafi05}
{Rafikov} RR. 2005.
\newblock \textit{\apjl} 621:L69--L72

\bibitem[{{Rafikov}(2006)}]{rafi06}
{Rafikov} RR. 2006.
\newblock \textit{\apj} 648:666--682

\bibitem[{{Rasio} \& {Ford}(1996)}]{rasi96}
{Rasio} FA, {Ford} EB. 1996.
\newblock \textit{Science} 274:954--956

\bibitem[{{Rauer} et~al.(2014){Rauer}, {Catala}, {Aerts}, {Appourchaux}, {Benz}
  et~al.}]{raue14}
{Rauer} H, {Catala} C, {Aerts} C, {Appourchaux} T, {Benz} W, et~al. 2014.
\newblock \textit{Experimental Astronomy} 38:249--330

\bibitem[{{Raymond}, {Mandell} \& {Sigurdsson}(2006)}]{raym06}
{Raymond} SN, {Mandell} AM, {Sigurdsson} S. 2006.
\newblock \textit{Science} 313:1413--1416

\bibitem[{{Reffert} et~al.(2015){Reffert}, {Bergmann}, {Quirrenbach},
  {Trifonov} \& {K{\"u}nstler}}]{reff15}
{Reffert} S, {Bergmann} C, {Quirrenbach} A, {Trifonov} T, {K{\"u}nstler} A.
  2015.
\newblock \textit{\aap} 574:A116

\bibitem[{{Rice}, {Armitage} \& {Hogg}(2008)}]{rice08}
{Rice} WKM, {Armitage} PJ, {Hogg} DF. 2008.
\newblock \textit{\mnras} 384:1242--1248

\bibitem[{{Ricker} et~al.(2015){Ricker}, {Winn}, {Vanderspek}, {Latham},
  {Bakos} et~al.}]{rick15}
{Ricker} GR, {Winn} JN, {Vanderspek} R, {Latham} DW, {Bakos} G{\'A}, et~al.
  2015.
\newblock \textit{Journal of Astronomical Telescopes, Instruments, and Systems}
  1:014003

\bibitem[{{Rogers}, {Lin} \& {Lau}(2012)}]{roge12}
{Rogers} TM, {Lin} DNC, {Lau} HHB. 2012.
\newblock \textit{\apjl} 758:L6

\bibitem[{{Rossiter}(1924)}]{ross24}
{Rossiter} RA. 1924.
\newblock \textit{\apj}

\bibitem[{{Sanchis-Ojeda} et~al.(2014){Sanchis-Ojeda}, {Rappaport}, {Winn},
  {Kotson}, {Levine} \& {El Mellah}}]{sanc14}
{Sanchis-Ojeda} R, {Rappaport} S, {Winn} JN, {Kotson} MC, {Levine} A, {El
  Mellah} I. 2014.
\newblock \textit{\apj} 787:47

\bibitem[{{Sanchis-Ojeda} et~al.(2011){Sanchis-Ojeda}, {Winn}, {Holman},
  {Carter}, {Osip} \& {Fuentes}}]{sanc11}
{Sanchis-Ojeda} R, {Winn} JN, {Holman} MJ, {Carter} JA, {Osip} DJ, {Fuentes}
  CI. 2011.
\newblock \textit{\apj} 733:127

\bibitem[{{Santerne} et~al.(2016){Santerne}, {Moutou}, {Tsantaki}, {Bouchy},
  {H{\'e}brard} et~al.}]{sant16}
{Santerne} A, {Moutou} C, {Tsantaki} M, {Bouchy} F, {H{\'e}brard} G, et~al.
  2016.
\newblock \textit{\aap} 587:A64

\bibitem[{{Santos} et~al.(2004){Santos}, {Bouchy}, {Mayor}, {Pepe}, {Queloz}
  et~al.}]{sant04b}
{Santos} NC, {Bouchy} F, {Mayor} M, {Pepe} F, {Queloz} D, et~al. 2004.
\newblock \textit{\aap} 426:L19--L23

\bibitem[{{Santos}, {Israelian} \& {Mayor}(2001)}]{sant01}
{Santos} NC, {Israelian} G, {Mayor} M. 2001.
\newblock \textit{\aap} 373:1019--1031

\bibitem[{{Santos}, {Israelian} \& {Mayor}(2004)}]{sant04}
{Santos} NC, {Israelian} G, {Mayor} M. 2004.
\newblock \textit{\aap} 415:1153--1166

\bibitem[{{Santos} et~al.(2003){Santos}, {Israelian}, {Mayor}, {Rebolo} \&
  {Udry}}]{sant03}
{Santos} NC, {Israelian} G, {Mayor} M, {Rebolo} R, {Udry} S. 2003.
\newblock \textit{\aap} 398:363--376

\bibitem[{{Schlaufman}(2010)}]{schl10}
{Schlaufman} KC. 2010.
\newblock \textit{\apj} 719:602--611

\bibitem[{{Schlaufman} \& {Winn}(2016)}]{schl16}
{Schlaufman} KC, {Winn} JN. 2016.
\newblock \textit{\apj} 825:62

\bibitem[{{Schlichting}(2014)}]{schl14}
{Schlichting} HE. 2014.
\newblock \textit{\apjl} 795:L15

\bibitem[{{Schneider} et~al.(2011){Schneider}, {Dedieu}, {Le Sidaner},
  {Savalle} \& {Zolotukhin}}]{schn11}
{Schneider} J, {Dedieu} C, {Le Sidaner} P, {Savalle} R, {Zolotukhin} I. 2011.
\newblock \textit{\aap} 532:A79

\bibitem[{{Shabram} et~al.(2016){Shabram}, {Demory}, {Cisewski}, {Ford} \&
  {Rogers}}]{shab16}
{Shabram} M, {Demory} BO, {Cisewski} J, {Ford} EB, {Rogers} L. 2016.
\newblock \textit{\apj} 820:93

\bibitem[{{Shara}, {Hurley} \& {Mardling}(2016)}]{shar16}
{Shara} MM, {Hurley} JR, {Mardling} RA. 2016.
\newblock \textit{\apj} 816:59

\bibitem[{{Sing} et~al.(2016){Sing}, {Fortney}, {Nikolov}, {Wakeford},
  {Kataria} et~al.}]{sing16}
{Sing} DK, {Fortney} JJ, {Nikolov} N, {Wakeford} HR, {Kataria} T, et~al. 2016.
\newblock \textit{\nat} 529:59--62

\bibitem[{{Socrates}(2013)}]{socr13}
{Socrates} A. 2013.
\newblock \textit{ArXiv e-prints} arXiv:1304.4121

\bibitem[{{Socrates} et~al.(2012){Socrates}, {Katz}, {Dong} \&
  {Tremaine}}]{socr12}
{Socrates} A, {Katz} B, {Dong} S, {Tremaine} S. 2012.
\newblock \textit{\apj} 750:106

\bibitem[{{Sousa} et~al.(2011){Sousa}, {Santos}, {Israelian}, {Mayor} \&
  {Udry}}]{sous11}
{Sousa} SG, {Santos} NC, {Israelian} G, {Mayor} M, {Udry} S. 2011.
\newblock \textit{\aap} 533:A141

\bibitem[{{Spalding} \& {Batygin}(2015)}]{spal15}
{Spalding} C, {Batygin} K. 2015.
\newblock \textit{\apj} 811:82

\bibitem[{{Spiegel} \& {Burrows}(2012)}]{spie12}
{Spiegel} DS, {Burrows} A. 2012.
\newblock \textit{\apj} 745:174

\bibitem[{{Steffen} et~al.(2012){Steffen}, {Ragozzine}, {Fabrycky}, {Carter},
  {Ford} et~al.}]{stef12}
{Steffen} JH, {Ragozzine} D, {Fabrycky} DC, {Carter} JA, {Ford} EB, et~al.
  2012.
\newblock \textit{Proceedings of the National Academy of Science}
  109:7982--7987

\bibitem[{{Stevenson}(1982)}]{stev82}
{Stevenson} DJ. 1982.
\newblock \textit{\planss} 30:755--764

\bibitem[{{Storch}, {Anderson} \& {Lai}(2014)}]{stor14a}
{Storch} NI, {Anderson} KR, {Lai} D. 2014.
\newblock \textit{Science} 345:1317--1321

\bibitem[{{Struve}(1952)}]{struv52}
{Struve} O. 1952.
\newblock \textit{The Observatory} 72:199--200

\bibitem[{{Teyssandier} et~al.(2013){Teyssandier}, {Naoz}, {Lizarraga} \&
  {Rasio}}]{tesy13}
{Teyssandier} J, {Naoz} S, {Lizarraga} I, {Rasio} FA. 2013.
\newblock \textit{\apj} 779:166

\bibitem[{{Thommes}, {Duncan} \& {Levison}(1999)}]{thom99}
{Thommes} EW, {Duncan} MJ, {Levison} HF. 1999.
\newblock \textit{\nat} 402:635--638

\bibitem[{{Toomre}(1964)}]{toom64}
{Toomre} A. 1964.
\newblock \textit{\apj} 139:1217--1238

\bibitem[{{Triaud}(2011)}]{tria11}
{Triaud} AHMJ. 2011.
\newblock \textit{\aap} 534:L6

\bibitem[{{Triaud}(2017)}]{tria17}
{Triaud} AHMJ. 2017.
\newblock \textit{ArXiv e-prints} arXiv:1709.06376

\bibitem[{{Trilling} et~al.(1998){Trilling}, {Benz}, {Guillot}, {Lunine},
  {Hubbard} \& {Burrows}}]{tril98}
{Trilling} DE, {Benz} W, {Guillot} T, {Lunine} JI, {Hubbard} WB, {Burrows} A.
  1998.
\newblock \textit{\apj} 500:428--439

\bibitem[{{Tsang}, {Turner} \& {Cumming}(2014)}]{tsan14a}
{Tsang} D, {Turner} NJ, {Cumming} A. 2014.
\newblock \textit{\apj} 782:113

\bibitem[{{Udry}, {Mayor} \& {Santos}(2003)}]{udry03}
{Udry} S, {Mayor} M, {Santos} NC. 2003.
\newblock \textit{\aap} 407:369--376

\bibitem[{{Valsecchi} et~al.(2015){Valsecchi}, {Rappaport}, {Rasio}, {Marchant}
  \& {Rogers}}]{vals15}
{Valsecchi} F, {Rappaport} S, {Rasio} FA, {Marchant} P, {Rogers} LA. 2015.
\newblock \textit{\apj} 813:101

\bibitem[{{Valsecchi}, {Rasio} \& {Steffen}(2014)}]{vals14}
{Valsecchi} F, {Rasio} FA, {Steffen} JH. 2014.
\newblock \textit{\apjl} 793:L3

\bibitem[{{Wahl} et~al.(2017){Wahl}, {Hubbard}, {Militzer}, {Guillot}, {Miguel}
  et~al.}]{wahl17}
{Wahl} SM, {Hubbard} WB, {Militzer} B, {Guillot} T, {Miguel} Y, et~al. 2017.
\newblock \textit{GeoRL} 44:4649--4659

\bibitem[{{Ward}(1997)}]{ward97}
{Ward} WR. 1997.
\newblock \textit{\icarus} 126:261--281

\bibitem[{{Weidenschilling} \& {Marzari}(1996)}]{weid96}
{Weidenschilling} SJ, {Marzari} F. 1996.
\newblock \textit{\nat} 384:619--621

\bibitem[{{Weiss} et~al.(2013){Weiss}, {Marcy}, {Rowe}, {Howard}, {Isaacson}
  et~al.}]{weis13}
{Weiss} LM, {Marcy} GW, {Rowe} JF, {Howard} AW, {Isaacson} H, et~al. 2013.
\newblock \textit{\apj} 768:14

\bibitem[{{Winn} et~al.(2010){Winn}, {Fabrycky}, {Albrecht} \&
  {Johnson}}]{winn10}
{Winn} JN, {Fabrycky} D, {Albrecht} S, {Johnson} JA. 2010.
\newblock \textit{\apjl} 718:L145--L149

\bibitem[{{Winn} et~al.(2017{\natexlab{a}}){Winn}, {Petigura}, {Morton},
  {Weiss}, {Dai} et~al.}]{winn17b}
{Winn} JN, {Petigura} EA, {Morton} TD, {Weiss} LM, {Dai} F, et~al.
  2017{\natexlab{a}}.
\newblock \textit{\aj} 154:270

\bibitem[{{Winn} et~al.(2017{\natexlab{b}}){Winn}, {Sanchis-Ojeda}, {Rogers},
  {Petigura}, {Howard} et~al.}]{winn17}
{Winn} JN, {Sanchis-Ojeda} R, {Rogers} L, {Petigura} EA, {Howard} AW, et~al.
  2017{\natexlab{b}}.
\newblock \textit{\aj} 154:60

\bibitem[{{Wittenmyer} et~al.(2010){Wittenmyer}, {O'Toole}, {Jones}, {Tinney},
  {Butler} et~al.}]{witt10}
{Wittenmyer} RA, {O'Toole} SJ, {Jones} HRA, {Tinney} CG, {Butler} RP, et~al.
  2010.
\newblock \textit{\apj} 722:1854--1863

\bibitem[{{Wright} et~al.(2012){Wright}, {Marcy}, {Howard}, {Johnson}, {Morton}
  \& {Fischer}}]{wrig12}
{Wright} JT, {Marcy} GW, {Howard} AW, {Johnson} JA, {Morton} TD, {Fischer} DA.
  2012.
\newblock \textit{\apj} 753:160

\bibitem[{{Wu}(2017)}]{wu17}
{Wu} Y. 2017.
\newblock \textit{ArXiv e-prints} arXiv:1710.02542

\bibitem[{{Wu} \& {Lithwick}(2011)}]{wu11}
{Wu} Y, {Lithwick} Y. 2011.
\newblock \textit{\apj} 735:109

\bibitem[{{Wu} \& {Lithwick}(2013)}]{wu13}
{Wu} Y, {Lithwick} Y. 2013.
\newblock \textit{\apj} 763:13

\bibitem[{{Wu} \& {Murray}(2003)}]{wu03}
{Wu} Y, {Murray} N. 2003.
\newblock \textit{\apj} 589:605--614

\bibitem[{{Wu}, {Murray} \& {Ramsahai}(2007)}]{wu07}
{Wu} Y, {Murray} NW, {Ramsahai} JM. 2007.
\newblock \textit{\apj} 670:820--825

\bibitem[{{Xiang-Gruess}(2016)}]{xian16}
{Xiang-Gruess} M. 2016.
\newblock \textit{\mnras} 455:3086--3100

\bibitem[{{Xu} \& {Lai}(2016)}]{xu16}
{Xu} W, {Lai} D. 2016.
\newblock \textit{\mnras} 459:2925--2939

\bibitem[{{Youdin} \& {Mitchell}(2010)}]{youd10}
{Youdin} AN, {Mitchell} JL. 2010.
\newblock \textit{\apj} 721:1113--1126

\bibitem[{{Yu} et~al.(2017){Yu}, {Donati}, {H{\'e}brard}, {Moutou}, {Malo}
  et~al.}]{yu17}
{Yu} L, {Donati} JF, {H{\'e}brard} EM, {Moutou} C, {Malo} L, et~al. 2017.
\newblock \textit{\mnras} 467:1342--1359

\bibitem[{{Yu} \& {MaTYSSE Collaboration}(2017)}]{yu17b}
{Yu} L, {MaTYSSE Collaboration}. 2017.
\newblock In \textit{Living Around Active Stars}, eds. D~{Nandy}, A~{Valio},
  P~{Petit}, vol. 328 of \textit{IAU Symposium}

\bibitem[{{Zechmeister} et~al.(2013){Zechmeister}, {K{\"u}rster}, {Endl}, {Lo
  Curto}, {Hartman} et~al.}]{zeic13}
{Zechmeister} M, {K{\"u}rster} M, {Endl} M, {Lo Curto} G, {Hartman} H, et~al.
  2013.
\newblock \textit{\aap} 552:A78

\end{thebibliography}

\end{document}